%% file: Arxiv_1April.tex
\documentclass[aps,prx,secnumarabic]{revtex4-2}

\usepackage[utf8]{inputenc}   % <-- add this
\usepackage[T1]{fontenc}

\usepackage[english]{babel}
\usepackage{comment}
\usepackage{amsmath,amsthm,amssymb,dsfont,amscd,mathrsfs}
\usepackage{float}
\usepackage[centercolon=true]{mathtools}
\usepackage{graphicx,microtype,slashed,bigints}
\usepackage[usenames,dvipsnames]{xcolor}
\usepackage{simplewick,xspace}
\usepackage{enumitem}
\usepackage{csquotes}
\usepackage[normalem]{ulem}
\usepackage[colorlinks=true, linkcolor=blue, citecolor=blue, urlcolor=blue]{hyperref}
\usepackage[capitalize]{cleveref}

\usepackage{tikz}
\usetikzlibrary{arrows,arrows.meta,decorations.pathmorphing,decorations.markings,shapes,calc}
\input{tikzit.sty}
\input{styles-HOOPT.tikzstyles}
\input{style_CRF.tikzstyles}
\usepackage{bbm} % <-- add this if you keep \mathbbm{1}

\def\>{\rangle}
\def\<{\langle}

\def\T+{\mathsf{T}_+}

\usepackage{stackengine}

\DeclareSymbolFont{sfletters}{OML}{cmbrm}{m}{it}
\DeclareMathSymbol{\srho}{\mathord}{sfletters}{"1A}
\DeclareMathSymbol{\salpha}{\mathord}{sfletters}{"0B}
\DeclareMathSymbol{\sbeta}{\mathord}{sfletters}{"0C}
\DeclareMathSymbol{\sgamma}{\mathord}{sfletters}{"0D}
\DeclareMathSymbol{\sdelta}{\mathord}{sfletters}{"0E}
\DeclareMathSymbol{\sepsilon}{\mathord}{sfletters}{"0F}
\DeclareMathSymbol{\szeta}{\mathord}{sfletters}{"10}
\DeclareMathSymbol{\seta}{\mathord}{sfletters}{"11}
\DeclareMathSymbol{\stheta}{\mathord}{sfletters}{"12}
\DeclareMathSymbol{\siota}{\mathord}{sfletters}{"13}
\DeclareMathSymbol{\skappa}{\mathord}{sfletters}{"14}
\DeclareMathSymbol{\slambda}{\mathord}{sfletters}{"15}
\DeclareMathSymbol{\smu}{\mathord}{sfletters}{"16}
\DeclareMathSymbol{\snu}{\mathord}{sfletters}{"17}
\DeclareMathSymbol{\sxi}{\mathord}{sfletters}{"18}
\DeclareMathSymbol{\spi}{\mathord}{sfletters}{"19}
\DeclareMathSymbol{\srho}{\mathord}{sfletters}{"1A}
\DeclareMathSymbol{\ssigma}{\mathord}{sfletters}{"1B}
\DeclareMathSymbol{\stau}{\mathord}{sfletters}{"1C}
\DeclareMathSymbol{\supsilon}{\mathord}{sfletters}{"1D}
\DeclareMathSymbol{\sphi}{\mathord}{sfletters}{"1E}
\DeclareMathSymbol{\schi}{\mathord}{sfletters}{"1F}
\DeclareMathSymbol{\spsi}{\mathord}{sfletters}{"20}
\DeclareMathSymbol{\somega}{\mathord}{sfletters}{"21}
\DeclareMathSymbol{\svarepsilon}{\mathord}{sfletters}{"22}
\DeclareMathSymbol{\svartheta}{\mathord}{sfletters}{"23}
\DeclareMathSymbol{\svarpi}{\mathord}{sfletters}{"24}
\DeclareMathSymbol{\svarrho}{\mathord}{sfletters}{"25}
\DeclareMathSymbol{\svarsigma}{\mathord}{sfletters}{"26}
\DeclareMathSymbol{\svarphi}{\mathord}{sfletters}{"27}

\newcommand\Item[1][]{%
  \ifx\relax#1\relax  \item \else \item[#1] \fi
  \abovedisplayskip=0pt\abovedisplayshortskip=0pt~\vspace*{-\baselineskip}}
\newcommand{\bra}[1]{\langle#1|}
\newcommand{\ket}[1]{|#1\rangle}

\newcommand{\ketbra}[2]{{\ket{#1}\bra{#2}}}
\newcommand{\dbra}[1]{\langle \! \langle#1|}
\newcommand{\dket}[1]{|#1\rangle \! \rangle}

\newcommand{\set}[1]{{\sf #1}}

\begin{document}
\title{
%Reference frames for process matrices: 
Frame perspectives for process matrices:
\\ from coordinate parametrization to spacetime representation}

\author{Luca Apadula}
\email{luca.apadula@cea.fr}
%\affiliation{Universit\'e Paris-Saclay}
\affiliation{Commissariat \`a l'\'energie atomique et aux \'energies alternatives (CEA-Saclay), IRFU/LARSIM, 91190 Gif-sur-Yvette, France}
\author{Alexei Grinbaum}
\email{alexei.grinbaum@cea.fr}
\affiliation{Commissariat \`a l'\'energie atomique et aux \'energies alternatives (CEA-Saclay), IRFU/LARSIM, 91190 Gif-sur-Yvette, France}
\author{Časlav Brukner}
\email{caslav.brukner@unive.ac.at}
\affiliation{Vienna Center for Quantum Science and Technology (VCQ), Faculty of Physics,
University of Vienna, Boltzmanngasse 5, A-1090 Vienna, Austria}
\affiliation{Institute of Quantum Optics and Quantum Information (IQOQI),
Austrian Academy of Sciences, Boltzmanngasse 3, A-1090 Vienna, Austria}
\begin{abstract}
We study how to implement and transform \emph{frame perspectives} for quantum processes in the
process-matrix formalism. We argue that, for pure processes, the causal reference frames (CRF)
and time-delocalized subsystems (TDS) formalisms should be understood as
coordinate parametrizations of a single perspective-neutral higher-order object.
%, rather than as descriptions tied to physical frames.
A genuine  perspective arises
when one endows the process with additional \emph{frame data} by choosing an operational foliation
into circuit fragments (events). With this distinction, existing no-go results acquire a clear scope:
they rule out %\sout{slot-preserving}  \lucacom{unitaries preserving time fragmentation/ foliation-preserving / fragmentation preserving unitaries}\sout{, i.e.\ transformations that attempt to switch
%perspectives while keeping the fragment boundaries---hence the global past/future partition---fixed}
unitary transformations that preserve time foliation, attempting to switch
perspectives while keeping the fragment boundaries---hence the global past/future partition---fixed. Focusing on the quantum switch, we construct explicit  maps that transform perspectives  unitarily at the price of reshuffling the notions of past and future. We then show that unitary transformations between perspectives can also be achieved in a different way, namely by extending the process with subsystems that define quantum reference frames and provide a shared spatiotemporal scaffold. In this extended setting, complementary CRF/TDS perspectives become unitarily related while preserving  global past and future. We discuss how this frame-perspectival approach informs the broader question of empirical realizability of abstract process matrices.

\end{abstract}

\maketitle
%\lucacom{The modifications are minimal: 
%\begin{itemize}
%    \item terminology updated,
%    \item few lines of introduction at the beginning of section IV in order to remark the aim of establishing unitarity between frame perspectives,
%    \item titles of section IV and the corresponding subsections,
%    \item the section referred to the spacetime scaffold now is a sub section of IV.
%\end{itemize}}
\section{Introduction}

In recent years, two prominent research lines have developed complementary
frameworks for describing spacetime and causality in quantum theory:
quantum reference frames  (QRF)~\cite{aharonov_quantum_1984, Bartlett_2007, gour_resource_2008, Giacomini_spin, vanrietvelde2018change, vantrietvelde_switching_2018, H_hn_2020, castroruiz2019time, Streiter_2020, Castro_Ruiz_2017, giacomini2021einsteins, apadula2022quantum, Barbado_2020, delahamette2021perspectiveneutral, delaHamette2021falling, glowacki2024quantumreferenceframeshomogeneous, glowacki, carette2023operational, cepollaro2024, delaHamette2020, Kabel2022conformal, Kabel_2025, castroruiz2021relative, de_la_Hamette_2020, mattei2026quantumlimitsspacetimereference, garmier2025perspectivesnonidealquantumreference, delahamette2025quantumreferenceframesarbitrary, doat2025symmetryconstrainedperspectiveimportancetotal, cepollaro2024, castroruiz2025interpretingquantumreferenceframe}
and the process-matrix formalism~\cite{Oreshkov_2012, Araujo_2015, Branciard_2015, Oreshkov_2016, Baumeler_2016}.
Although developed in parallel, they meet on common ground:
both accommodate situations in which the causal relation between events can be indefinite.

A key conceptual distinction—relevant for the analysis of both QRFs and the process-matrix formalism—is that “coordinates” and “reference frames” can be understood in two distinct, yet standard, ways.
On the one hand, coordinates may be treated as \emph{abstract labels}:
one introduces a parametrization (times, positions, circuit lots or fragments) and
requires that the empirical content of the theory, i.e.\ observable predictions,
be independent of the chosen labels.
On the other hand, coordinates may be understood as \emph{physical instantiations}:
they arise from matter fields (rods and clocks) that define a
coordinate assignment. In this second sense, the coordinate fields themselves
are dynamical since matter fields can interact with other physical systems, and
backreaction from this interaction on the coordinates is in principle unavoidable.
In general relativity (GR), both viewpoints are routinely employed: we use coordinate transformations (diffeomorphisms) to express the laws in different charts, and we also compare descriptions tied to different physical frames that are built from matter fields. The physical content of the theory is diffeomorphism-invariant, i.e., it is unchanged under coordinate transformations on the spacetime manifold.
Spatiotemporal QRFs~\cite{Zych_2018_EEP, zych_2019, delahamette2022quantum, delaHamette2021falling, Kabel2022conformal, kabel2023quantum, apadula2022quantum, Barbado_2020, Kabel_2025, vilasini2025localization} push both viewpoints into the quantum regime: quantum coordinates and quantum reference fields can serve either as abstract or as physical “quantum rods” and “quantum clocks”, thereby providing a relational description of physical systems.
%degrees of freedom.
This framework naturally accommodates the possibility of superposing event locations and, consequently, superposing causal relations in both gravitational and non-gravitational settings.

On both viewpoints, however, a  question remains:
how does one transform between different descriptions, i.e., between descriptions associated with different QRFs or coordinate systems? In quantum theory, the changes of description that preserve physical content are
typically represented by unitary (or, more generally, isometric) maps.
This is the operational  meaning of  the expression ``same physics in different coordinates/frames'':
probabilities, operational predictions, and the structure of admissible
transformations must be preserved across  different descriptions.

The process-matrix framework characterizes the full space of correlations
compatible with quantum mechanics at each local laboratory without assuming a
 global causal order.
A process matrix is a higher-order quantum map~\cite{Higher-order, Apadula_2024, bavaresco2021strict, jencova2024structurehigherorderquantum, taranto2025higherorderquantumoperations}
encompassing causally separable processes, causally nonseparable processes
such as the quantum switch~\cite{Procopio:2015ab, Rubino_2017},
and also processes that violate causal inequalities~\cite{Oreshkov_2012, Branciard_2015, Wechs_2023, Baumeler_2016, 6874888, PhysRevA.90.042106},
often called noncausal.
A key open issue is \emph{empirical realizability}:
which abstract processes can arise from bona fide dynamics of physical systems
embedded in spacetime? In Ref.~\cite{Parker2022background} a background-independent formulation of the process-matrix framework was proposed, based on permutation invariance of the laboratory labels, and it was shown that under this requirement one loses the notion of distinct local operations unless a reference frame is used to distinguish them; in this work, we explicitly retain the identity of the laboratories.

The central concept of our study is that of an agent’s perspective. We identify it as the perspective in which the agent’s operation appears “local” and should therefore be represented by a “local box” in a diagrammatic representation of a process.
So far, two frameworks explored the operational meaning of perspectives within the process-matrix formalism.
Causal reference frames (CRF)~\cite{CausalReferenceFrames} provide event-dependent
parametrizations in which a chosen agent's operation appears local, while time-delocalized
subsystems (TDS)~\cite{Oreshkov_2019, Wechs_2021, Wechs_2023, wechs2024subsystemdecompositionsquantumevolutions}
reinterpret indefinite causal order as time delocalization of subsystems relative
to an agent’s time frame.
For pure processes, these representations are equivalent to the underlying
process matrix in the sense that they yield the same higher-order map,
i.e.\ the same operational statistics for all local instruments.

At this point, a tension emerges.
On the one hand, if CRF/TDS representations are ``just different descriptions''
of the same physical process, one expects that moving between them should be
implementable as a symmetry—in the quantum setting, as a unitary transformation.
On the other hand, it has been shown for the quantum switch that
natural candidates for such transformations fail:
one cannot unitarily map the circuit fragments associated with one agent perspective to the circuit fragments of another agent while keeping fixed the time-frame data, i.e.  the notions of ``past'' and ``future''~\cite{Oreshkov_2019}. In other words, the two descriptions are operationally equivalent, yet there is no unitary transformation that maps one circuit decomposition on the other preserving the boundaries of preparation and measurement. How can ``same physical content'' coexist with the apparent absence of a unitary change of perspective?

We resolve this tension by taking seriously the dual role of the coordinates.
Our first claim is that CRF and TDS representations are  naturally understood
as coordinate parametrizations of a perspective-neutral process:
they are rewritings of the same higher-order object, and their equivalence is a
statement of coordinate invariance at the level of the \emph{unfragmented} process. By a \emph{fragment} we mean a time-ordered piece of the circuit, delimited by chosen interface boundaries. 

Our second claim is that a \emph{frame perspective} enters only when one
promotes the coordinate assignment to frame data—most notably, when the
``time coordinate'' is treated as a reading of a physical clock.
This corresponds to an operational cut of the process into
circuit fragments (events), thereby endowing the description with a definite
time foliation relative to a chosen frame.
Once such frame data are included, a correct change of perspective must act, not
only on the system but also on frame data itself. 
The no-go result then acquires sharp meaning: it excludes unitary
maps that preserve the original time foliation, i.e.\ keep the original
time labels and boundary partition fixed, and therefore do not implement a
genuine change of time frame.

With this distinction in place, we proceed as follows.
We define a \emph{frame perspective} via the operational cut of a
CRF/TDS representation and construct explicit perspective-change maps.
We show that there exists a unitary transformation that takes one agent perspective (in which agent $A$ operations are local) to another agent perspective (in which agent $B$ operations are local)~\cite{grinbaum2025agencyindefinitecausalityoperational}\footnote{In  Ref.~\cite{grinbaum2025agencyindefinitecausalityoperational}, the author develops a derived notion of an agent--—emerging from a choice of subsystem factorization and from the grouping of input–output data—--that aligns with the present notion of agent perspective}. However, this unitary transformation necessarily reshuffles the boundaries, so that the two agent perspectives do not share common  past and
future. 

We then extend the process by introducing additional quantum reference systems that
provide a background circuit fragment—i.e.\ a spatiotemporal scaffold into which
the process is embedded~\cite{delahamette2022quantum, Kabel_2025}.
In this extended frame perspective, we construct a unitary perspective-change
transformation that maps the CRF/TDS associated with one spacetime viewpoint to the
complementary CRF/TDS associated with a different spatiotemporal frame, while
preserving common global past and future.

This last viewpoint does not require abandoning abstract coordinates.
Rather, abstract coordinates can be understood as \emph{gedanken coordinate fields} that could in principle be instantiated as rods and clocks, but need not be.
As in classical GR, one may express dynamical laws in arbitrary coordinates and
prove coordinate invariance of physical statements. 
However, when one moves between coordinate choices interpreted as 
potential frames, the transformation must also act on the frame data.
This is precisely where the quantum setting differs from the classical one:
the frame is itself quantum and entangled with the system, so the
change of perspective must account for that.

The remainder of this work is organized as follows.
In Section~\ref{Causal Reference Frames and Time-Delocalized Subsystems representations} we recall the CRF and TDS formalisms and recast their equivalence in the form of coordinate invariance of the perspective-neutral process.
In Section~\ref{Coordinate time parametrization and foliation into fragments} we introduce a frame perspective via an operational cut and construct explicit perspective-change transformations, clarifying the scope of the existing no-go results.
In Section~\ref{Spacetime frame for process matrices} we introduce extended frame perspectives for the quantum switch by extending its definition with a spatiotemporal scaffold built out of additional reference systems. Using this new definition, we show in Section~\ref{sect5} that complementary extended frame perspectives are unitarily related.

\section{Causal Reference Frames and Time-Delocalized Subsystems}\label{Causal Reference Frames and Time-Delocalized Subsystems representations}
%\subsection{Causal reference frames and cyclic composition}
In Ref.~\cite{CausalReferenceFrames} it was shown that any pure bipartite process matrix with global past and future can be expressed in equivalent decompositions called causal reference frames (CRF).
Let us consider a unitary  process $W$, where $P_S, P_C$ and $F_S, F_C$ belong to the  past and future target $(S)$ and control $(C)$ systems, while $(A_I,A_O)$ and $(B_I,B_O)$ are the input-output systems of Alice's and Bob's laboratories respectively. 
Choosing a causal reference frame amounts to decomposing the full process into the past and future of a given laboratory, so that the corresponding agent is effectively local within their own causal frame. In this setting, the equivalence between CFR decompositions is expressed as: 
\begin{align}\label{equivalence_CRF}
\dket{W}&=\sum_i\dbra{I}^{E_AE_A}\dket{\Pi_A(U_i)}^{E_AA_IP_SP_C}\dket{\Phi_A(U_i)}^{F_SF_CE_AA_O}\dket{U_i}^{B_OB_I}\\
&=\sum_i\dbra{I}^{E_BE_B}\dket{\Pi_B(V_i)}^{E_BB_IP_SP_C}\dket{\Phi_B(V_i)}^{F_SF_CE_BB_O}\dket{V_i}^{A_OA_I},  
\end{align}
 where $E_{A/B}$ are suitable  isomorphic ancillary systems and $\{U_i\}$ and $\{V_i\}$ are two orthonormal unitary bases for $B$ and $A$ respectively. Here, $\Pi_{A/B}$, $\Phi_{A/B}$ are ``future'' and ``past'' functions mapping unitaries to unitaries, which describe how the global process decomposes from the perspective of Alice’s or Bob’s causal frame.
Hence, for any choice of unitary operations $U_A$ of Alice and $U_B$ of Bob: 
\begin{equation}\label{equivalence_process_CRF}
\dbra{U_A}\dbra{U_B}\dket{W}=\dket{W(U_A,U_B)}=\scalebox{1.0}{\input{W_B_A.tikz}}=\scalebox{1.0}{\input{W_A_B.tikz}}.    \end{equation}

In Ref.~\cite{CausalReferenceFrames} the authors established a connection between
CRFs and the time-delocalized subsystems (TDS) decomposition~\cite{Oreshkov_2019}.
Specifically, the same unitary process in TDS is $\dket{W}=\dbra{I}^{EE}\dket{L_A}^{EA_IPB_O}\dket{R_A}^{FB_IEA_O}$. 
Expressed as a quantum circuit 
\begin{equation}
\dket{W}=\scalebox{1.0}{\input{causalframe_circuit_A.tikz}},
\end{equation}
this makes evident the time delocalization of 
Bob's output and input 
in the global past and future of Alice's localized operations.  Note that $E'_{A/B}$ are ancillary systems isomorphic to $E_{A/B}$, while the subscripts in $L_A$ and $R_A$ merely indicate that the process is being expressed in Alice's time frame. At this stage, no agent operation has been applied,  hence $L_A$ and $R_A$ do not depend on $U_A$ or $U_B$.

Therefore, the two decompositions are 
related:
\begin{align}
\dket{W(U_A,U_B)}&=\dbra{I}^{E_AE_A}\dbra{U_A}^{A_OA_I}\dket{\Phi_A(U_B)}^{E_AA_IP_SP_C}\dket{\Pi_A(U_B)}^{F_SF_CE_AA_O}\\
&=\dbra{U_B}^{B_OB_I}\dbra{U_A}^{A_OA_I}\dbra{I}^{E_A'E_A'}\dket{L_A}^{E_A'A_IP_SP_CB_O}\dket{R_A}^{F_SF_CB_IE_A'A_O}.  
\end{align}
Graphically, this yields
\begin{equation}\label{CRF=TDS}
 \dket{W(U_A,U_B)}=\scalebox{1.0}{\input{W_B_A.tikz}}=   \scalebox{1.0}{\input{time-delocalized_B_A.tikz}}.
\end{equation}
Therefore, Eq.~\eqref{equivalence_process_CRF} can be 
expressed in terms of the time-delocalized decomposition of $W$, namely 
\begin{equation}\label{equivalence_process_TDS}
\scalebox{1.0}{\input{time-delocalized_B_A.tikz}}=\scalebox{1.0}{\input{time-delocalized_A_B.tikz}}
\end{equation} 
where $L_A, R_A$ and $L_B, R_B$ are related by a change of subsystem factorization~\cite{Oreshkov_2019}.

In what follows, we recast the equivalences in Eqs.~\eqref{equivalence_process_CRF} and~\eqref{equivalence_process_TDS} as instances of coordinate invariance. More specifically, we point out that different CRF or TDS decompositions are alternative parametrizations of the abstract process rather than genuinely different physical perspectives.
%provided by a choice of reference frame. 

\section{No-Go Results for Unitaries Preserving Time Foliation}
%Fragmentation}
\label{Coordinate time parametrization and foliation into fragments}
%\subsection{Coordinate time parametrization and foliation into fragments}

For the purposes of this work, it is useful to distinguish two levels of description.
At the level of the \emph{perspective-neutral} (higher-order) process, there is no
preferred global time parameter: any ``time labels'' attached to boxes or wires merely
parametrize a chosen representation, and the operational content is invariant under such
reparametrizations.  A frame perspective only enters when one chooses an explicit
\emph{time foliation}  that breaks the circuit into  fragments, which ought to be interpreted as events in this perspective.

\paragraph*{Coordinate parametrization.} Let us start with the circuit in Alice's CRF/TDS
viewed purely as a \emph{coordinate parametrization} of the underlying
process.  We introduce labels $t_1$ and $t_2$ to mark the beginning and the end of Alice's
local gate, and we denote the global input and output boundaries by $t_0$ and $t_3$,
respectively.  In this parametrization Bob's gate is time-delocalized:

\begin{equation}\label{time_coordinates}
 \scalebox{1.2}{\input{time_coordinates.tikz}}
\end{equation}

We stress that the ``time delocalization'' displayed here is \emph{not} defined relative to
any clock.  Rather, it is an intrinsically
relational phenomenon that occurs as a consequence of choosing CRF/TDS: Alice operates locally by construction, and
Bob is represented as delocalized relative to that choice.
%Consequently, for a quantum-switch--like process there is, at this stage, no intrinsic temporal
%localization with respect to a third system that could serve as a background clock.
Symbols $t_0,t_1,t_2,t_3$ in Eq.~\eqref{time_coordinates} function only as
coordinate parameters---bookkeeping devices for a particular representation of the
same underlying higher-order object.  They should not be conflated with a frame
perspective, where ``time'' is a reading of a clock relative to which events are defined.
%and which is therefore, in principle, subject to backreaction and quantum correlations.

\paragraph*{Foliation into fragments.}
To pass from a mere parametrization to a frame perspective, one must
choose a concrete foliation of the process into circuit fragments that are to be regarded
as events. This corresponds to introducing an
\emph{operational cut}: we cut the target wires in the vicinity of the labels $t_1$ and 
$t_2$, thereby decomposing the process into Alice's local gate and Bob's time-delocalized
operation.  In the resulting picture, symbols $t_0,t_1,t_2,t_3$ label the
boundary interfaces between fragments and thus encode not only a parametrization but also a
specific foliation of the process.  Within this foliation,
Alice’s laboratory is local at step $(t_1,t_2)$, while Bob’s intervention is in a
superposition of temporal locations, i.e. delocalized at $(t_0,t_1)$ and $(t_2,t_3)$:
\begin{equation}\label{clock_A}
   % \scalebox{1.0}{\tikzfig{clock_A}}.  
\includegraphics[width=280pt]{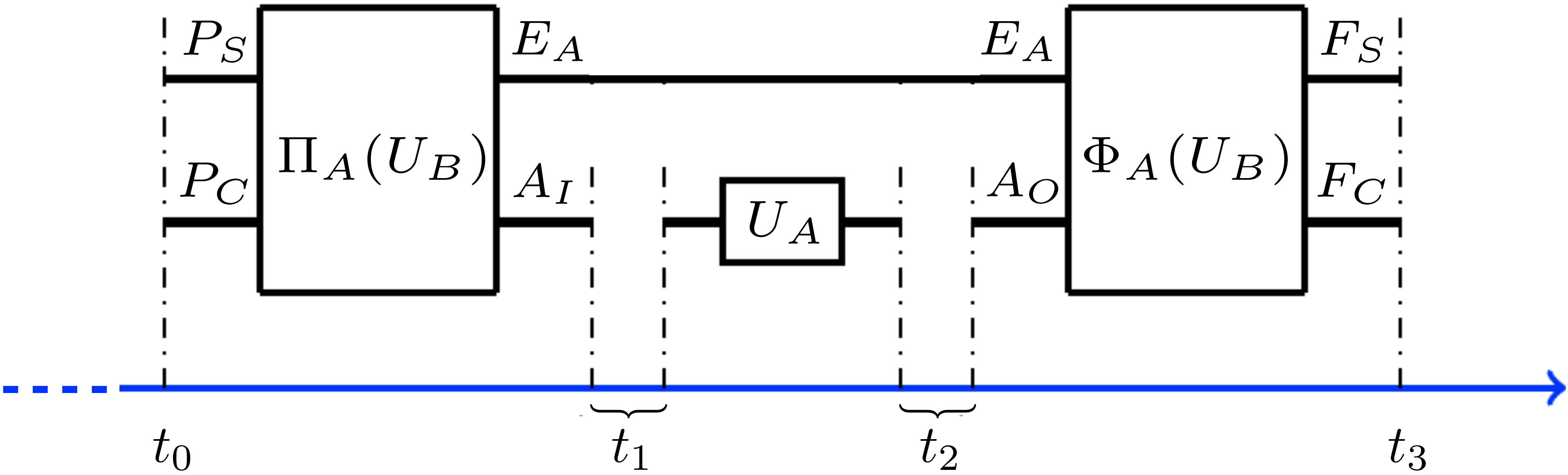}
\end{equation}
Here again, the operational cut does not imply that $t_i$  must be  readings of a
clock: in the absence of additional reference information, the interfaces are 
defined only relative to the chosen decomposition.  What has changed is that the
description now carries extra structure beyond the timeless process matrix, namely a
distinguished foliation into fragments---a choice of ``event boundaries''---where one may insert extra systems, gates, or implement transformations by acting on the corresponding boundary input--output systems.
%Such fragment boundaries are then operational as one can now insert additional system, gates or implementing a transformation by acting on the boundary input-output systems. 
%\lucacom{These fragment boundaries are indeed operationally meaningful: they define \emph{accessible} interfaces on which one may insert extra systems or gates, or implement transformations by acting on the corresponding boundary input--output systems.}
A  fully physical frame perspective---``$t_i$ is what a clock shows''---will further require an
explicit reference system whose distinguishable states would be correlated with these
interfaces, making the boundary data operationally accessible as frame information. This
will be the role of the background scaffold in Section~\ref{Spacetime frame for process matrices}.

A natural question is whether the equivalence in
Eqs.~\eqref{equivalence_process_CRF} and~\eqref{equivalence_process_TDS} can be realized at
the level of \emph{fragmented} descriptions, i.e.\ whether one can unitarily ``re-frame''
the fragments obtained from Alice's foliation in such a way that they coincide with fragments obtained by
cutting the process in Bob's CRF. In Refs.~\cite{Oreshkov_2019,wechs2024subsystemdecompositionsquantumevolutions} it was
shown that such a mapping is impossible for a specific class of unitaries: there exist no
maps $J$ (and its conjugate)
acting at the boundary interface $t_0$ and $t_3$ %, as shown in Fig.~\eqref{clock_A} 
such that 
\begin{equation}\label{eq:wrong_map_B}
    \scalebox{0.9}{\input{unitaries_fixedboundaries.tikz}}=
    \scalebox{0.9}{\input{fragments_CRF_B.tikz}},
\end{equation}
where $\phi$ and $\psi$ refer to the initial state preparation and final measurement respectively. 

In Ref.~\cite{wechs2024subsystemdecompositionsquantumevolutions} the authors considered  two other equivalent circuit representations of the quantum switch representing the causal perspectives of Alice and Bob with a finer resolution in time  
\begin{align}
    \scalebox{0.9}{\input{timecircuit_A.tikz}}, \;
     \scalebox{0.9}{\input{timecircuit_B.tikz}},
\end{align}
where  additional time steps---identified by $\scalebox{0.8}{\input{timestep_target.tikz}}$ and $\scalebox{0.8}{\input{timestep_control.tikz}}$---are placed between events.
%Analogously to the case discussed in~\ref{Coordinate time parametrization and foliation into fragments}, 
It is then argued that it is impossible to transform
the fragments of one circuit decomposition, for example
\begin{equation}\label{fragments_timecircuit}
\scalebox{1.0}{\input{fragments_timecircuit_A.tikz}},
\end{equation}
into the fragments of the other circuit decomposition through unitary maps.
Following our line of reasoning, the time data labeling the interfaces between the  circuit fragments is preserved by the unitary mapping in Eq.~\eqref{eq:wrong_map_B}, thereby preventing a proper 
transformation into a new time foliation. 
%As a matter of fact
We can reproduce the fragments %from the previous section 
%We can see the connection with the case in the previous section
%We can reproduce the fragments of the previous case 
by contracting over  $T_1, T_2, T_5, T_6$ and $C_1,\dots, C_6$, i.e erasing the corresponding boundaries and time data in Eq.~\eqref{fragments_timecircuit}.
We obtain: 
\begin{align}
    \scalebox{0.8}{\input{trace_fragments_timecircuit_A.tikz}}=\scalebox{0.8}{\input{fragments_causalframe_A.tikz}}.
\end{align}
The resulting fragment corresponds to Bob’s 
delocalized intervention as described in Alice’s causal 
frame. 

In this context, the no-go results of Refs.~\cite{Oreshkov_2019,wechs2024subsystemdecompositionsquantumevolutions} can be understood as follows: once particular fragmentation has been fixed, they rule out
transformations that preserve the fragments---i.e.  boundary partitions into
``past'', ``future'', and intermediate interfaces---while attempting to transform between  frame perspectives.
By contrast, adopting a new time  foliation amounts to
changing this boundary identification. In general, this requires a
global reshuffling of ``past'' and ``future''.
Accordingly, the unitaries in Eq.~\eqref{eq:wrong_map_B}---which
act only fixed  boundary interfaces---necessarily preserve the original interfaces, labels, and boundaries, and therefore cannot implement a genuine change of frame: they
may change the representation \emph{conditional} on a fixed foliation, but they do not
transform the foliation itself. 
%\subsection{Time circuit representation and time frame}

It is worth mentioning that these results are in agreement  with the findings of Ref.~\cite{Vilasini_2022,Vilasini_2024Realize, vilasini2025localization}.
The results there show that, starting from localized events in a definite acyclic spacetime, one cannot generate indefinite causal order. In our framework, this further suggests that if events are localized in one perspective, then, as long as the foliation is preserved, there is no perspective in which they become delocalized.

\section{Unitary perspective-change transformation for the quantum switch
}
\label{Time frame-change transformation for quantum switch}

In this section we construct unitary maps between different frame perspectives using the example of the quantum switch. We show how to recover unitarity among perspectives that can be interpreted as an \enquote{agent's point of view}, namely frames in which the agent’s operation is local in time and appears as a circuit fragment. For this, we examine the bare switch—composed only of target and control systems. When Alice's perspective is changed to Bob's, the two frames do not in general share the same notion of past and future.

We then show that unitarity can be recovered also for frame perspectives that do share past and future.
This requires extending the process with additional quantum reference systems that serve as a spatiotemporal scaffold for the frame perspectives. The unitary  equivalence between Alice's and Bob's complementary perspectives is then restored by accounting for the transformation of these additional systems.

\subsection{Perspective-change by reshuffling past and future}
\label{non-extended-scenario}

Having made the distinction between (i) a mere coordinate parametrization of a perspective-neutral
process and (ii) a foliation into operational fragments, we now construct an explicit
perspective-change transformation for the quantum switch.
Consider the foliation associated with Alice’s perspective in
Eq.~\ref{clock_A}.
Bob’s operation is time-delocalized:
\begin{equation}\label{switch_B|A}
  \dket{W(U_B)}=\scalebox{1.0}{\input{causalframe_A.tikz}}
  =\dket{00}^{F_CP_C}\dket{I}^{A_IP_S}\dket{U_B}^{F_SA_O}
  +\dket{11}^{F_CP_C}\dket{I}^{F_SA_O}\dket{U_B}^{A_IP_S} .
\end{equation}
Bob’s gate appears in a coherent superposition of two temporal locations:
one in the ``global past'' segment $(t_0,t_1)$ with input--output systems $(P_S,A_I)$, and one in the
``global future'' segment $(t_2,t_3)$ with input--output systems $(A_O,F_S)$.
By contrast, Alice’s own operation
$\dket{U_A}^{A_OA_I}=\scalebox{0.8}{\input{gate_A.tikz}}$
is local and confined to the interval $(t_1,t_2)$.

Our goal is to pass to Bob’s perspective.
%As emphasized in Section~\ref{Coordinate time parametrization and foliation into fragments}, 
This cannot be achieved by any
foliation-preserving transformation, i.e.\ any unitary that keeps Alice’s boundary interfaces---her past and future---fixed.
Indeed, Eq.~\eqref{switch_B|A} shows that localizing Bob by switching to his perspective necessarily requires a reassignment of the delocalized boundary subsystems $P_S$ and $F_S$, and therefore a concomitant re-identification of %which degrees of freedom constitute the
relevant boundary slices.
In particular, since $P_S$ and $F_S$ sit at the original boundaries, any such reassignment must
also transform the associated preparation
$\scalebox{0.7}{\input{state.tikz}}$
and final measurement
$\scalebox{0.7}{\input{measure.tikz}}$.

The key ingredient is therefore to implement a \emph{global} reshuffling of the foliation data.
Operationally, this is achieved by employing a coherently controlled unitary applied at a
single interface: because the control system is localized on one slice, the induced transformation can act nontrivially on the target system on other slices and thereby
reassign the global past/future partition.
This stands in contrast with the constructions ruled out by 
Eq.~\eqref{eq:wrong_map_B}, where transformations were distributed over
distinct interfaces and thus preserved the original time foliation.

A perspective-change transformation is %unitary that realizes such a time-frame change is
\begin{equation}\label{eq:J_AtoB}
\dket{J_{A\to B}}
=\dket{00}^{F_CF_C}\dket{I}^{B_OF_S}\dket{I}^{B_IA_O}\dket{I}^{F_S'A_I}\dket{I}^{P_S'P_S}
+\dket{11}^{F_CF_C}\dket{I}^{B_OA_O}\dket{I}^{B_IP_S}\dket{I}^{P_S'A_O}\dket{I}^{F_S'F_S}.
\end{equation}
Defining the collections of systems
\[
X \coloneqq F_CF_SA_IA_OP_S,
\qquad
Y \coloneqq F_CP_SA_OA_IF_S,
\]
the transformed fragments are obtained by
\begin{align}
&\dbra{I}^{XX}\dbra{I}^{YY}\,
\dket{J_{A\to B}}\,
\dket{W(U_B)}\,\dket{U_A}\,\dket{\phi}\,\dket{\psi}\,
\dket{J_{A\to B}^\dagger}
=\scalebox{1.0}{\input{transformation_switch.tikz}}
\label{QRF_J}\\[1mm]
&=\dket{U_B}^{B_OB_I}\Big(
\dket{00}^{F_CP_C}\dket{I}^{F_S'P_S'}\dket{U_A}^{B_IF_S'}\ket{\phi}^{P_S'}\ket\psi^{B_O}
+\dket{11}^{F_CP_C}\dket{I}^{F_S'P_S'}\dket{U_A}^{P_S'B_O}\ket{\phi}^{B_I}\ket{\psi}^{F_S'}
\Big)\nonumber\\
&=\scalebox{1.0}{\input{transformed_AtoB.tikz}}
=\dket{U_B}^{B_OB_I}\dket{\widetilde W}^{F_CF_S'P_S'B_IP_CP_S'F_S'B_O},
\label{tildeW}
\end{align}
where the line crossing the transformation diagram denotes a system that is invariant under the perspective change.
The resulting process $\widetilde W$ admits a circuit representation in which time-delocalization
of relevant target subsystems is explicit:
\begin{align}\label{tildeW1}
\dket{\widetilde W}
=\scalebox{1.0}{\input{timeframe_B.tikz}}
=\dbra{I}^{EE}\,\dket{\Phi_A(U_A,\phi,\psi)}\,\dket{\Pi_A(U_A,\phi,\psi)},
\end{align}
where, writing $E\coloneqq E_SE_C$, we have
\begin{align}
\dket{\Phi_A(U_A,\phi,\psi)}
&=\dket{00}^{E_CP_C}\dket{I}^{E_SP_S'}\dket{U_A}^{B_IF_S'}
+\dket{11}^{E_CP_C}\dket{I}^{E_SP_S'}\ket{\psi}^{B_I}\ket{\phi}^{P_S'},\\
\dket{\Pi_A(U_A,\phi,\psi)}
&=\dket{00}^{F_CE_C}\dket{I}^{F_S'E_S}\ket{\phi}^{P_S'}\ket{\psi}^{B_O}
+\dket{11}^{F_CE_C}\dket{I}^{F_S'E_S}\dket{U_A}^{P_S'B_O}.
\end{align}

Two structural features are worth stressing.
First, unlike in the  foliation-preserving scenario, both $J_{A\to B}$ and $J_{A\to B}^\dagger$ are
controlled by $F_C$ at a \emph{single} interface (Alice’s future slice in Eq.~\ref{clock_A}). Their action \emph{wraps the fragments around}, i.e. globally reassigns the temporal placement of the target subsystems; this is
precisely what enables a nontrivial change of foliation.
Second, because the transformation necessarily re-identifies the past/future partition, the
transformed fragments of the preparation, the final measurement, and Alice’s gate (collected into
$\widetilde W$) do \emph{not} coincide with those appearing in Bob’s CRF under the \emph{same} boundary
identification: the original preparation/measurement assignment does not agree with the new past/future partition in the new foliation and thus becomes time-delocalized.
Equivalently, in the absence of further reference structure, a unitary perspective change that
localizes Bob’s operation unavoidably alters the global notion of past and future.

This provides the methodological point that aligns with Section~\ref{Coordinate time parametrization and foliation into fragments}.
CRF/TDS representations remain equivalent as coordinate parametrizations of the underlying
perspective-neutral process.
However, once one had fixed a foliation into fragments and asked for a change of perspective, the
relevant symmetry notion must allow transformations that act not only on the system  within slots  but also on the foliation data itself (i.e.\ on the interface identities and the induced boundary partition).
The construction above achieves exactly this: it implements a unitary change of foliation that
renders Bob local at the price of reshuffling the global past and future.
At the next step, we will supply additional reference systems that provide a
background circuit fragment (a spatiotemporal scaffold), allowing complementary agent perspectives to be
unitarily related while preserving  shared  past and future. We call such perspectives \textit{extended frame perspectives}.

\subsection{Quantum frames for process matrices}\label{Spacetime frame for process matrices}
So far we have analyzed the quantum switch at the level of a chosen \emph{time foliation}.
By performing an operational cut, we have separated circuit fragments at the boundaries of the agents' interventions, preparations and measurements. We treat the resulting interfaces  as
\emph{frame data} specifying an agent-dependent set of time-ordered events.
Crucially, the bare switch process is defined only on the target and control systems.
It contains no additional subsystem that could serve as an \emph{external geometric
background}---no clock or rod relative to which events could be localized. In this sense, the previous time-foliated description fixes an ordering, but still lacks an embedding  into a spatiotemporal scaffold.

The aim of this section is to supply precisely this geometric structure by extending the process with explicit quantum reference systems (clocks and, where relevant, rods) that carry operationally accessible spatiotemporal information.
These additional reference systems provide a \emph{background circuit fragment} shared across
perspectives, i.e.\ a spatiotemporal scaffold into which the process embeds.
We will show that, in this extended setting, one can implement unitary changes of
spatiotemporal perspective that relate complementary CRF/TDS while preserving
global past and future.
This yields a  concrete handle on how the quantum switch-type correlations can be understood
as arising from bona fide dynamics of matter fields (rods and clocks) in spacetime.

\subsubsection{Quantum switch-like process in a general relativistic setting}
\begin{figure}[h]
\centering
\includegraphics[width=250pt]{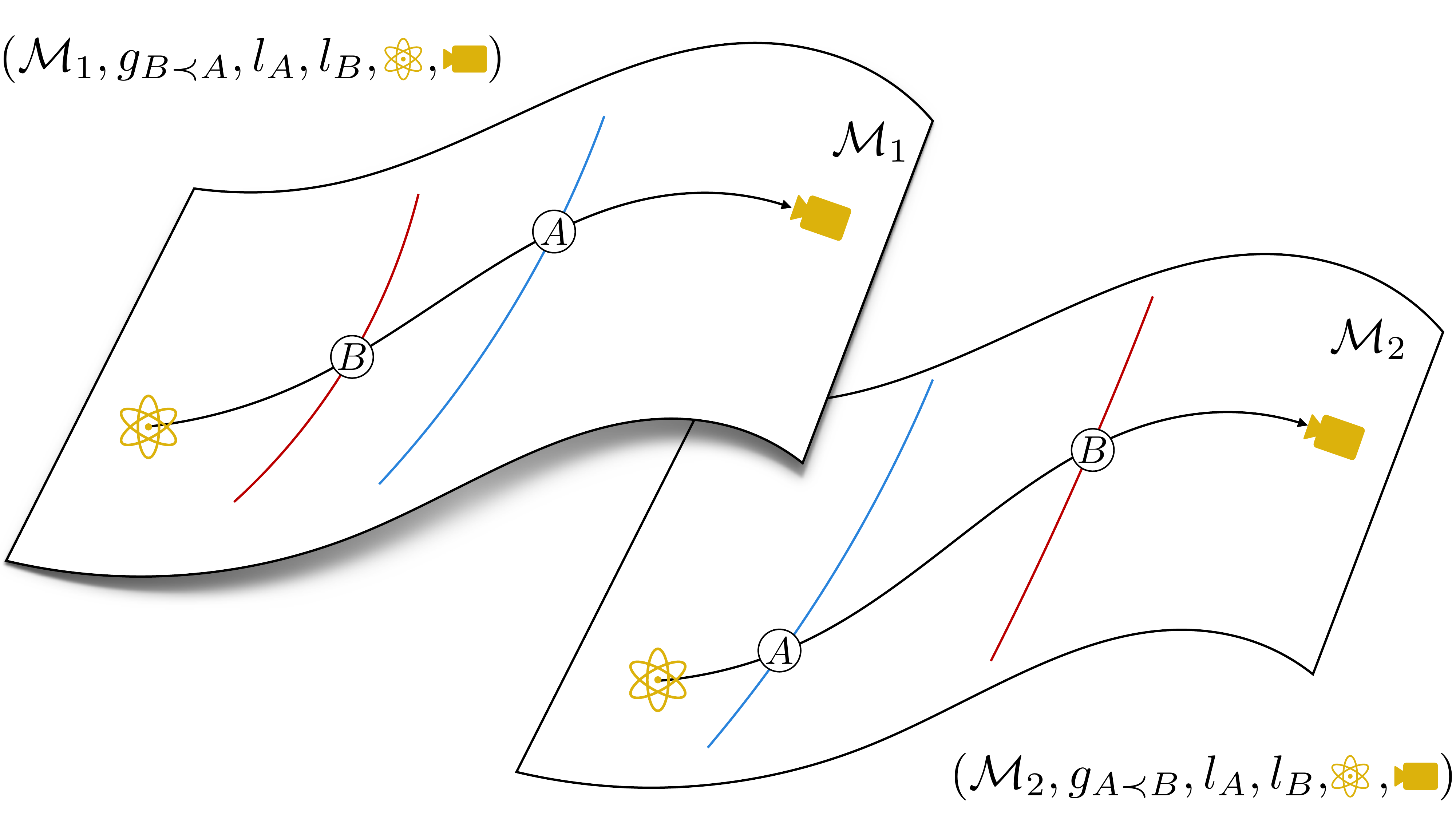}
\caption[Superposition of physically inequivalent configurations]{ \textbf{Superposition of physically inequivalent configurations}. Events 
$A$ and $B$ are defined as intersections of the world-lines of the target system (black arrow), Alice’s laboratory (blue line), and Bob’s laboratory (red line). The state preparation and the final measurement are represented by the yellow atom-shaped symbol and the detector symbol, respectively.}
\label{fig:sup_configurations}
\end{figure}

We consider a general relativistic representation of a quantum switch–like process.
This digression further clarifies the way to extend the process matrix in order to achieve a fully relational spatiotemporal description.
%This digression may cfurther clarify the preceding results and indicates how to extend the process structure, so that a relational spatiotemporal description can be employed.
%that will help to consolodate the results shown so far and pave the way toward an extension of the considered process that would allow to adopt a spacetime point of view.  
%To clarify the discussion and prepare for the 
%construction developed later, we now consider 
%a spacetime representation of a physical 
%scenario with indefinite causal structure. 
Specifically, we examine a superposition of 
two inequivalent physical configurations in 
which the causal order between two events, 
$A$ and $B$, is indefinite: in one branch of 
the superposition, $B$ lies in the past of 
$A$ $(B\prec A)$, while in the other 
branch, the order is reversed $(A\prec B)$. 
This scenario is illustrated in 
Fig.~\ref{fig:sup_configurations}.

\begin{figure}[h]
\centering
\includegraphics[width=400pt]{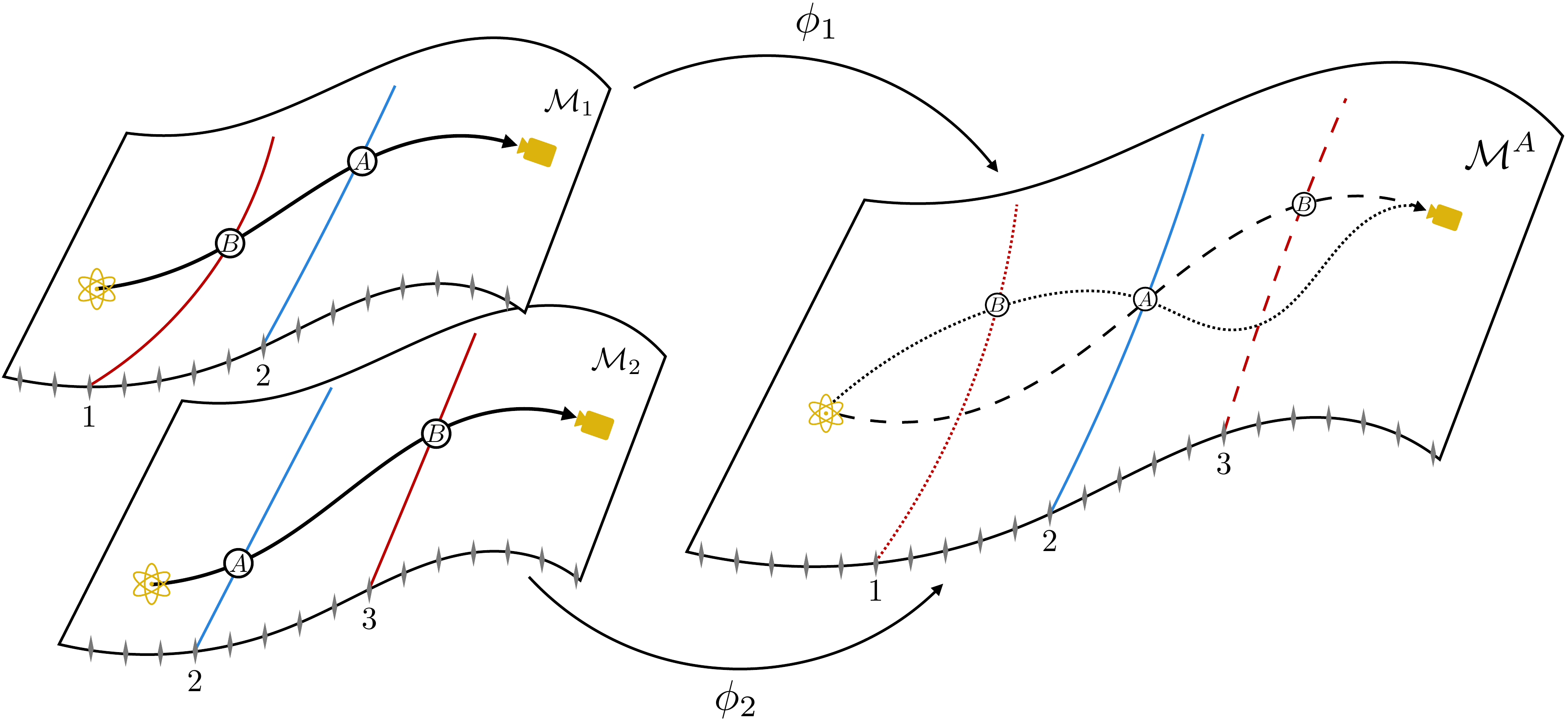}
\caption[Quantum coordinate transformation]{\textbf{Quantum coordinatization.} The quantum-controlled diffeomorphism allows one to take the perspective of the geometric scaffolding—i.e. quantum coordinates—depicted as the gray grid. 
The transformed state on the right is expressed in those quantum coordinates, such that Alice’s worldline is localized (configuration 2), whereas Bob’s worldline is spacetime delocalized (superposition of configurations 2 and 3).}
\label{transformation_perspectiveA}
\end{figure}

Each configuration is specified by the tuple of the type $\begin{array}{l} \raisebox{-0.3\height}{\includegraphics[width=100pt]{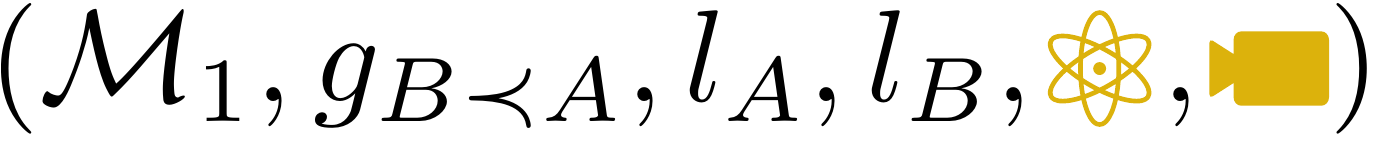}} \end{array}$, which encodes the space of parameters: the manifold 
$\mathcal M$ and the physical systems, here consisting of the 
gravitational field, the world-lines of 
Alice’s and Bob’s laboratories, the target 
system, and the detector performing the 
final measurement. The differentiable manifolds and the spacetime points comprising them have no physical significance in and by themselves since, without an implementation of coordinates  by matter fields, they lack observable properties through which they could be individuated. At this stage, only the source, the detection event, and the events associated with Alice's and Bob's operations are specified through intersection points within each manifold. No reference matter fields have yet been introduced that would permit the identification, or ``sewing together,'' of individual points across the two manifolds via equal values of the considered  fields, and thereby allow one to determine whether the events are local in the chosen coordinates or to establish their order (see Ref.~\cite{Kabel_2025} for a detailed discussion). More precisely, in the absence of such coordinates, there is no criterion for deciding whether the locations corresponding to event \(A\), rather than those corresponding to event \(B\), should be identified across the manifolds, and hence whether \(A\) or \(B\) is to be regarded as local, with the other remaining non-local. This underdetermination is similar in spirit to that encountered in complementary CRF/TDS descriptions in Section~\ref{Causal Reference Frames and Time-Delocalized Subsystems representations}.

The underlying physics is invariant under 
any change of coordinate system 
implemented by the group of 
diffeomorphisms on the manifold. In the 
regime considered, the standard notion of 
symmetry is extended to include quantum-controlled 
symmetry transformations: we 
refer to the resulting group as the 
\emph{quantum symmetry group}~\cite{Kabel2022conformal, delaHamette2021falling, Kabel_2025}. 
This extension allows one to adopt either an actual or a gedanken physical perspective by parametrizing spacetime through a suitable set of scalar fields, the configuration of which defines \emph{quantum coordinates} or, equivalently, \emph{quantum reference frames} (QRF)~\footnote{Ref.~\cite{Kabel_2025} constructs a relational parametrization of spacetime using scalar fields; the resulting physical local chart is termed ``quantum coordinates,'' to be distinguished from standard coordinate charts defined independently of any physical system.}.
A pictorial representation of the extended physical picture is given in the left-hand side of Fig.~\ref{transformation_perspectiveA}, where the new quantum coordinates correspond to the gray grid~$\raisebox{-0.3\height}{\includegraphics[width=4pt]{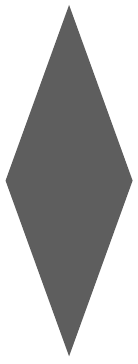}}$.   
In the chosen ``quantum coordinate chart''   Alice’s inputs and outputs are local, while Bob’s and the target’s worldlines become spacetime-delocalized. Nevertheless, event $A$ remains local because the superposed target worldlines intersect Alice’s worldline at the same point of the quantum coordinate chart.
Moreover, both branches of the 
superposition share the same notion of 
global past and future, making the initial 
state preparation and final measurement 
local as well. 

This relational description in terms of quantum coordinates can be be achieved via a 
suitable quantum-controlled diffeomorphism 
$\phi:=(\phi_1,\phi_2)$, where 
$\phi_{1}:\mathcal{M}_{1}\to\mathcal M_A$
and 
$\phi_{2}:\mathcal{M}_{2}\to\mathcal M_A$ 
act on the extended tuples 
$\raisebox{-0.3\height}{\includegraphics[width=100pt]{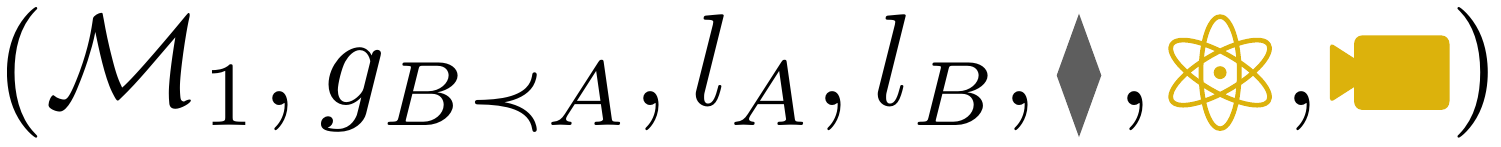}}$ and $\raisebox{-0.3\height}{\includegraphics[width=100pt]{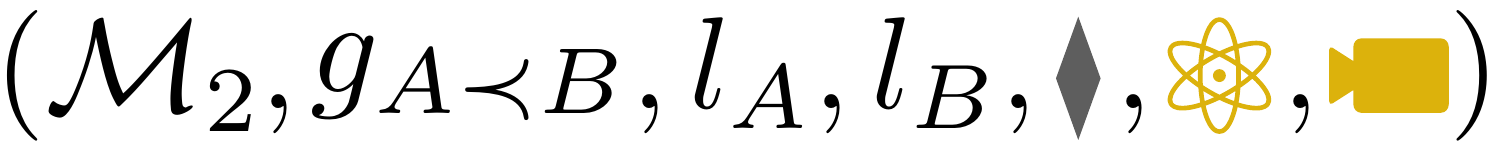}}$ 
respectively. 
%The physical configuration is now described  by the tuple$\begin{array}{l} \raisebox{-0.3\height}{\includegraphics[width=100pt]{tuple+frame.pdf}} \end{array}$, including also the reference degrees of freedom–i.e.$\begin{array}{l} \raisebox{-0.3\height}{\includegraphics[width=4pt]{frame.pdf}} \end{array}$–, which in Fig.~\ref{transformation_perspectiveA} is represented as the gray grid that provides the  spatiotemporal coordination.
The right-hand side of  Fig.~\ref{transformation_perspectiveA} shows the resulting physical configuration, where the chosen quantum coordinates provide the spatiotemporal scaffolding %coordinatization
for the coordinates of the events.
%and worldlines.

\begin{figure}[H]
\centering
\includegraphics[width=450pt]{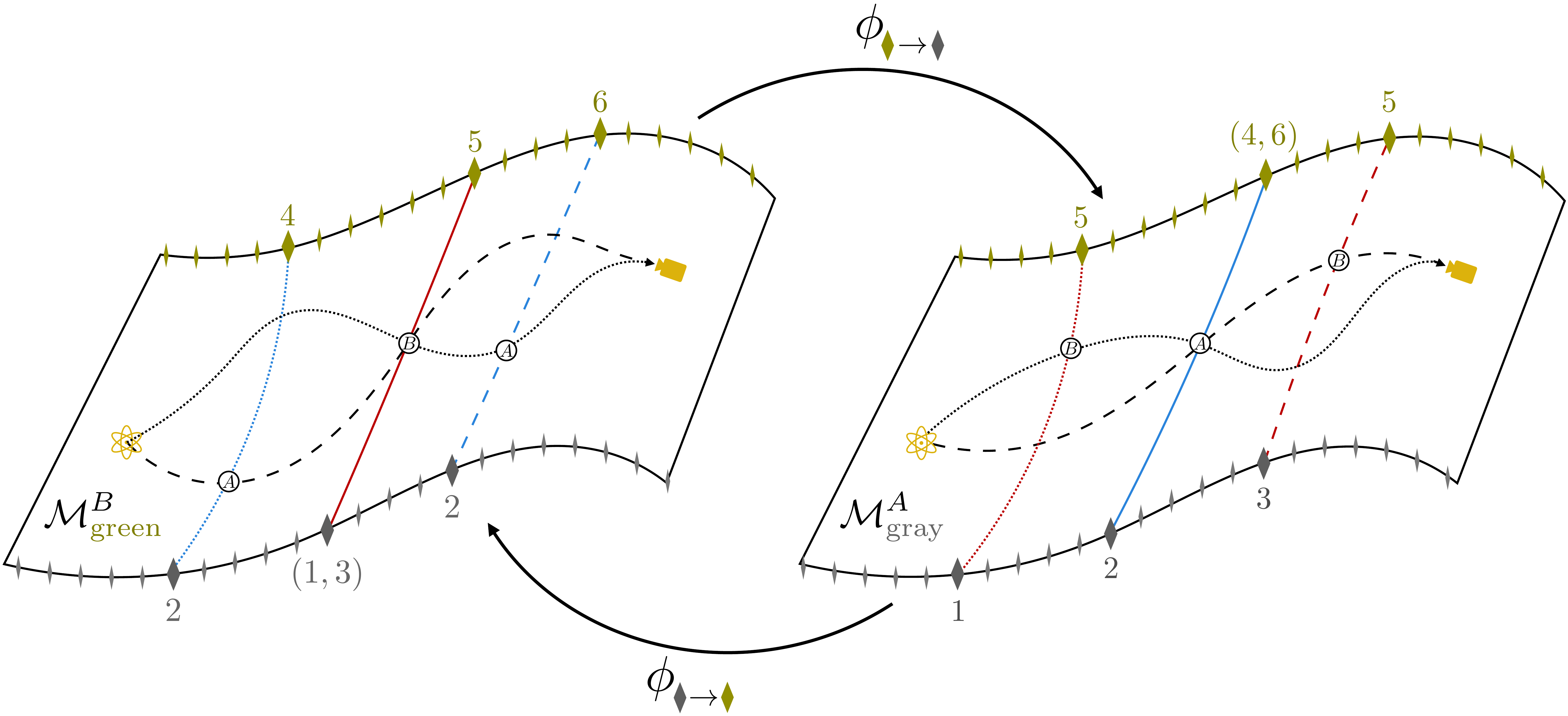}
\caption[Equivalent quantum coordinate representations] {\textbf{Quantum coordinate transformations.} Two  different coordinate representations of the same process: (left) Bob's causal reference frame; (right) Alice's  causal reference frame. One can change representation via a suitable quantum-controlled diffeomorphism.}
\label{diagram_coordianterepresentations}
\end{figure}

The ability to adopt Bob's agent perspective—in which event $B$ is local while $A$ is delocalized—depends solely on the availability of another frame prepared in an appropriate configuration. When such a frame is available—i.e. $\raisebox{-0.3\height}{\includegraphics[width=4pt]{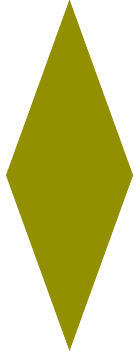}}$—both perspectives can be realized, as illustrated in Fig.~\ref{diagram_coordianterepresentations}.

%the physical scenario is also endowed of a 

%It is likewise possible to move to a 
%complementary coordinate representation in 
%which event $B$ is local while 
%$A$ is delocalized, or to a representation 
%in which both $A$ and 
%$B$ are local. These three configurations 
%are depicted in 
%Fig.~\ref{diagram_coordianterepresentations}.

The equivalence underlined by Fig.~\ref{diagram_coordianterepresentations} 
requires the notion of quantum symmetry and at least two quantum coordinates charts. It is therefore substantially different from the equivalence established in Eq.~\eqref{equivalence_process_TDS}, which relied only on relabeling in the perspective-neutral, unfragmented representation of the bare process. This leads us to 
extending the previous notion of frame perspective.
%replace the previous notion of agent perspective with a notion of \emph{frame perspective}.
%\subsubsection{\sout{Lack of spacetime  frame of reference}}
%The configurations displayed in Fig.\ref{diagram_coordianterepresentations} do not presuppose any underlying spatiotemporal reference frame.
In the extended setting, the possibility to switch between different perspectives is granted by the presence of suitable QRFs in form of quantum fields. 
%By introducing additional systems, however, one can adopt a description in which these systems serve as a QRFs, providing a relational coordinatization of the remaining ones. 
In particular, obtaining a QRF that matches one of the considered configurations depends entirely on the physical state of the QRF—specifically, on whether it is suitably correlated with the rest of the systems~\cite{Kabel_2025}.
More precisely, a QRF initialization in which $A$-event is local amounts to using a specific set of reference quantum fields to label and identify spacetime points across the superposed branches. A full QRF change then requires switching to a different set of reference quantum fields to maintain this identification, so that the comparison of points could now be defined relative to the new fields. In such a description, the newly chosen reference quantum fields are factorized out of branches, while the previously used reference fields are generically in superposition and become entangled with the other degrees of freedom, including the metric~\cite{delahamette2022quantum, Kabel_2025}.

Note that the non-extended scenario studied in Section~\ref{non-extended-scenario} includes, beyond the clock’s readings in Eq.~\eqref{clock_A}, only the target system and the control, the latter typically identified with the gravitational field in the general-relativistic setting. There are no QRFs. Consequently, this scenario describes time delocalization of a single gate, not the spacetime delocalization of the corresponding worldline. 
By contrast, the extended scenario  endows the frame perspective with an additional spacetime scaffold serving as a quantum reference frame.

%Namely, given a physical perspective on the quantum switch–like process that yields a circuit fragment consistent with one causal reference frame decomposition (Alice), there is no alternative QRF that can produce a decomposition coinciding with the complementary causal reference frame (Bob). 

\subsection{Spacetime scaffold for a switch-like process}

We now aim to bridge the gap between the diagrammatic description of the bare switch in Eq.~\eqref{clock_A} and its spacetime representation. As emphasized previously, the time foliation of the bare switch lacks a spatiotemporal background to act as a counterpart of the quantum coordinates in Fig.~\ref{diagram_coordianterepresentations}. To remedy this, the algebraic process can be extended by introducing additional systems beyond the target and control which would supply spatial dimensionality. This is achieved by the following circuit fragment that plays the role of the new spatiotemporal scaffold:

\begin{equation}\label{acyclicW_diagram}
\dket{W_{B|A}}=\scalebox{1.0}{\input{spacetime_process.tikz}},
\end{equation}
where the single boxes defining the time steps of the process 
are the following  controlled swaps
\begin{align}
    \dket{\set T^{0\to 1}}&=\dket{00}^{C_1C_0}\dket{I}^{S_1R_0}\dket{I}^{T_1S_0}+\dket{11}^{C_1C_0}\dket{I}^{T_1R_0}\dket{I}^{S_1S_0},\\
    \dket{\set T^{1\to 2}}&=\dket{00}^{C_3C_2}\dket{I}^{T_3T_2}\dket{I}^{S_3S_2}+\dket{11}^{C_3C_2}\dket{I}^{T_3S_2}\dket{I}^{S_3T_2},\\
    \dket{\set T^{4\to 5}}&=\dket{00}^{C_5C_4}\dket{I}^{S_5T_4}\dket{I}^{T_5S_4}+\dket{11}^{C_5C_4}\dket{I}^{T_5T_4}\dket{I}^{S_5S_4},\\
    \dket{\set T^{6\to7}}&=\dket{00}^{C_7C_6}\dket{I}^{S_7S_6}\dket{I}^{R_7T_6}+\dket{11}^{C_7C_6}\dket{I}^{S_7T_6}\dket{I}^{T_7S_6}.
\end{align}
%By linking the systems as in figure~\eqref{acyclicW_diagram}, we obtain the acyclic 
%process
%\begin{align}\dket{W_{B|A}}=\dket{00}^{C_7C_0}\dket{I}^{T_1S_0}\dket{I}^{T_5R_0}\dket{I}^{T_3T_2}\dket{I}^{S_7T_4}\dket{I}^{R_7T_6}+\dket{11}^{C_7C_0}\dket{I}^{T_1R_0}\dket{I}^{T_3S_0}\dket{I}^{R_7T_2}\dket{I}^{T_5T_4}\dket{I}^{S_7T_6}\end{align}
This circuit fragment presents three open slots 
corresponding
to the spacetime locations of Bob's and Alice's events.
In particular, by introducing an additional reference system, the scaffold~\eqref{acyclicW_diagram} can represent the superposition of worldlines described by the quantum coordinates in Fig.~\ref{transformation_perspectiveA}. 
Note that $S$ and $C$ label the target and control systems respectively, while $R$ denotes an additional reference system. 

We describe the agent's operations in the frame perspective in which Bob's operation is delocalized: 
\begin{equation}\label{Bob's operation|A}
   \dket{\mathbf{U}_{B|A}(U_B)}=\scalebox{1.0}{\input{timecircuit_fragment_B.tikz}}=\dket{00}^{C_0C_0}\dket{U_B}^{T_2T_1}\dket{I}^{T_6T_5}+\dket{11}^{C_0C_0}\dket{I}^{T_2T_1}\dket{U_B}^{T_6T_5}. 
\end{equation}
%which is represented by the circuit fragment 
%\begin{equation}
% \dket{\mathbf{U}_{B|A}(U_B)}=\scalebox{1.2}{\tikzfig{timecircuit_fragment_B}}.
%\end{equation}
Alice's operation occurs at time step $3\to 4$, namely $\dket{U_A}^{T_4T_3}=\scalebox{0.8}{\input{spacetime_gate_A.tikz}}$. 
Note that upon placing the operations in the corresponding slots of the circuit, one obtains
\begin{align}
\dbra{\mathbf{U}_{B|A}(U_B)}\dbra{U_A}\dket{W_{B|A}}=\dket{I}^{R_7R_0}\dket{W_{SW}(U_A,U_B)},
\end{align}
where $W$ corresponds to the quantum 
switch. Hence, discarding the extra system $R_7$ while plugging any deterministic state in $R_0$
does not cause any loss of coherence.
This guarantees statistical equivalence with the standard quantum 
switch. 

\subsection{Perspective-change for spatiotemporal representation of the process%Transformations  between spatiotemporal perspectives
}
\label{sect5}
%Now we discuss how to  change perspective.
We seek a frame perspective in which the background scaffold %fragment 
places Bob’s operation at a well-defined spacetime 
location, i.e. a single circuit slot.
In this frame, Bob’s intervention is represented by 
a single non-controlled gate, whereas Alice’s operation becomes delocalized and entangled with the control system. This unitary map transforms the fragments accordingly:
\begin{align}
    \dket{S_{A\to B}}=&\dket{00}^{C_0C_0}\dket{I}^{\widetilde T_4T_2}\dket{I}^{\widetilde T_5T_3}\dket{I}^{\widetilde T_6T_4}\dket{I}^{\widetilde T_3T_1}\dket{I}^{\widetilde T_1T_5}\dket{I}^{\widetilde T_2T_6}\\ 
    &+\dket{11}^{C_0C_0}\dket{I}^{\widetilde T_4T_6}\dket{I}^{\widetilde T_1T_3}\dket{I}^{\widetilde T_2T_4}\dket{I}^{\widetilde T_3T_5}\dket{I}^{\widetilde T_6T_2}\dket{I}^{\widetilde T_5T_1}.
\end{align}
Define $\set K:=C_0T_1T_2T_3T_4T_5T_6$. Alice's and Bob's fragments of the circuit transform as
\begin{align}
    \dbra{I}^{\set K\set K}\dket{S_{A\to B}}\dket{\mathbf{U}_{B|A}(U_B)}\dket{U_A}&=\scalebox{1.0}{\input{transformed_fragments_AB.tikz}}=\scalebox{1.0}{\input{transformed_fragments_AB_1.tikz}}\label{Alice's operation|B}\\
    &=\dket{\mathbf U_{A|B}(U_A)}\dket{U_B},
\end{align}
which in the current notation is
\begin{align}
    \dket{\mathbf U_{A|B}(U_A)}^{C_0 \widetilde T_2\widetilde T_4\widetilde T_6C_0\widetilde T_1 \widetilde T_3\widetilde T_5}\dket{U_B}^{\widetilde T_4\widetilde T_3}&=\left(\dket{00}^{C_0C_0}\dket{U_A}^{\widetilde T_2\widetilde T_1}\dket{I}^{\widetilde T_6\widetilde T_5}+\dket{11}^{C_0C_0}\dket{I}^{\widetilde T_2\widetilde T_1}\dket{U_A}^{\widetilde T_6\widetilde T_5}\right)\dket{U_B}^{\widetilde T_4\widetilde T_3}.
\end{align}
As expected, Bob’s operation is now disentangled from the 
control and localized at the time step $T_3\to T_4$, 
whereas Alice’s operation is delocalized between
$T_5\to T_6$ and $T_1\to T_2$, being entangled with $C_0$.
%The transformed operations coincides with the desiderata, as  Alice gate is delocalized in two time slots, while Bob is local according to his causal frame.

Now transform the fragment of the background circuit
\begin{align}
 \dbra{I}^{\set K\set K}\dket{S_{A\to B}^\dagger}\dket{W_{B|A}}=&\dket{00}^{C_7C_0}\dket{I}^{\widetilde T_3S_0}\dket{I}^{\widetilde T_1R_0}\dket{I}^{\widetilde T_5T_4}\dket{I}^{S_7\widetilde T_6}\dket{I}^{R_7\widetilde T_2}\nonumber\\
&+\dket{11}^{C_7C_0}\dket{I}^{\widetilde T_5R_0}\dket{I}^{\widetilde T_1S_0}\dket{I}^{R_7\widetilde T_6}\dket{I}^{\widetilde T_3\widetilde T_2}\dket{I}^{S_7\widetilde T_4}=\dket{W_{A|B}}.
\end{align}
The fragment of the circuit corresponding 
to the background scaffold transforms in such a way 
that the first and last time fragments correspond to 
Alice's lab, while Bob's one is localized between them: 
\begin{align}
\dket{ S_{A\to B}^\dagger (W_{B|A})}&=\scalebox{1.0}{\input{transformed_background.tikz}}\\
&=\scalebox{1.0}{\input{spacetime_process_B.tikz}}=\dket{W_{A|B}}.\label{Background|B}
%&=\scalebox{1.2}{\tikzfig{spacetime_process_B_relabelled}}
\end{align}
Hence, $\ket{W_{A|B}}$ describes the spacetime scaffold providing the spatiotemporal coordinatization of Bob's TDS decomposition.
This demonstrates the possibility to transform unitarily the fragments associated Alice's TDS (Eq.~\eqref{Bob's operation|A}) and the corresponding background (Eq.~\eqref{acyclicW_diagram}) into  those associated with Bob's TDS (Eq.~\eqref{Alice's operation|B}) and background (Eq.~\eqref{Background|B}).

\section{Conclusions}

The starting point of this work is a distinction between two conceptions of the coordinates: on one hand, they are understood as abstract labels in the perspective-neutral or unfragmented picture; on the other, they are conceived of as frame data that can be instantiated by matter fields, e.g. rods and clocks. We first argue that the CRF and TDS representations of a pure process are best understood as coordinate parameterizations of a perspective-neutral higher-order object. Their equivalence is therefore a statement of coordinate invariance at the level of the unfragmented process, rather than a relativization of the process with respect to a physical frame.

Our next contribution is methodological. We formalize a frame perspective by foliating a chosen CRF/TDS decomposition into circuit fragments. An operational cut helps us upgrade the interface labels from bookkeeping parameters to
frame data, namely clock-readings associated with the chosen time frame. In this setting, a genuine
perspective change must co-transform both the circuit fragments and the frame data---i.e. the
fragment interfaces---so as to effectively re-foliate the process. This requirement is precisely what underlies
the no-go result for  unitary
transformations that preserve time foliation  
~\cite{Oreshkov_2019, wechs2024subsystemdecompositionsquantumevolutions}: acting locally at two fragment interfaces  necessarily preserves their
original frame data (global past and future) and therefore cannot implement a change of perspective.

Focusing on the quantum switch, we show that although our perspective-change map unitarily transforms one perspective into another, it generically \emph{does not preserve}
common notions of past and future---instead, both become reshuffled and mixed. In this way, the time-fragmented TDS decomposition does not need to be unitarily related to its \emph{complementary} decomposition, which \emph{does} share the same global past and future. This may seem paradoxical, since the quantum switch can be empirically realized in both gravitational and optical settings~\cite{Rubino_2017, zych_2019}, where one can choose descriptions in which either agent’s operation is local while the other’s is nonlocal, yet both descriptions share the same global past and future. How, then, can two physically equivalent situations—each realizable and sharing a common global past and future—fail to be related by a unitary transformation? The question is naturally resolved once one realizes that, in contrast to physical realizations of the quantum switch that include quantum reference frames (QRFs), the bare switch-like process does not contain extra subsystems to instantiate distinct frame perspectives that agree on a global past and future. In other words, the process lacks the geometric background scaffold relative to which complementary perspectives could be related by a unitary change of frame.

To overcome this limitation, we extend the quantum switch
by explicitly adding QRFs,  thus supplying a shared
spatiotemporal scaffold in which the process is embedded. The extended process reproduces the same statistics as the abstract switch, but it now admits a genuinely relational spacetime description.
In this extended setting, we exhibit an explicit unitary QRF-change transformation that maps the fragments of one extended TDS---spacetime TDS---to the complementary one, while retaining shared
 past and future. 
%This restores, at the level of spatiotemporal frames, the perspective complementarity in TDS/CRF that is not
%unitarily accessible using a time frame only. 

We contend that developing a notion of frame perspective for processes and a corresponding perspective-change transformation is directly relevant to the question of empirical realizability. 
By “realizable” we mean that the coupled dynamics of spacetime and matter admits configurations which yield statistics that are operationally equivalent to the statistics of the abstract---bare---process. In Refs.~\cite{Tselentis_2023,tselentis2025mobiusgamequantuminspiredtest,tselentis2025mobiusgamequantuminspiredtest}, the authors use graph-theoretic techniques to derive admissibility criteria for causal structures, formulating explicit graph constraints that any spatiotemporal causal structure connecting local quantum laboratories  must satisfy. They also propose a general-relativistic generalization of a Bell-type test that can certify the presence of a dynamical causal structure between the parties involved. Processes with quantum-controlled causal orders can be realized in general relativistic settings~\cite{zych_2019, delahamette2022quantum, S_M_ller_2024}. However, such a realization of  unitary noncausal processes that violate causal inequalities, e.g. the Baumeler--Wolf process~\cite{baumeler2014perfect}, may appear to require spacetime solutions with closed timelike curves (CTCs). At the same time, recently it has been argued that there exits processes violating causal inequalities that nevertheless admit realizations on time-delocalised subsystems within standard quantum theory, i.e., without invoking exotic spacetime structures~\cite{Wechs_2023}.

We see our work as providing a concrete route toward assessing empirical realizability of process matrices through bona fide spacetime-and-matter dynamics. This naturally motivates the next step: extending the same perspective-based construction beyond switch-like processes. In particular, applying this framework to unitary noncausal processes would test whether
noncausality can be re-expressed as quantum spatiotemporal delocalization of laboratories  without invoking exotic geometric structures.

Finally, we return to our initial distinction between abstract coordinates and geometric scaffolding. We argue that the latter is crucial for implementing a proper transformation between perspectives. Does it imply that physical equivalence between two perspectives cannot be established without invoking  rods and clocks, i.e. specific particles and fields? If yes, this might contradict general relativity, where coordinates are conceptually prior to matter fields defined on a manifold. Our understanding is that the answer is negative: the geometric scaffolding in the frame-perspectival approach can be viewed as \emph{potential} or \emph{gedanken} matter fields, not necessarily the physically implemented ones. With this understanding, the situation is not very different from classical general relativity: one may choose different gedanken coordinate fields to express observables, then demonstrate that the observables are independent of that choice. However, the change in the gedanken fields must still be included in the overall transformation in order to preserve its unitarity.

%Overall, extending the QRF line of reasoning supplies the missing physical layer beneath CRF and TDS
%descriptions: when ``clocks and rods'' are treated as physical reference systems rather than mere parameters,
%indefinite causal structure can be represented and transformed \emph{relationally}. 

%\vspace{0.5cm}

\section*{Acknowledgments}

The authors are grateful to Julien Wechs, Ognyan Oreshkov, {\"A}min Baumeler, and Marco Erba for helpful discussions and comments. This research was funded by l’Agence Nationale de la Recherche (ANR), project ANR-22-CE47-0012. 
%For the purpose of open access, the authors have applied a CC-BY public copyright licence to any Author Accepted Manuscript (AAM) version arising from this submission.
This research was also funded in part by the Austrian Science Fund (FWF) [10.55776/F71] (BeyondC), and [10.55776/COE1]. This publication was also made possible through financial support of WOST (WithOutSpaceTime) grant from the John Templeton Foundation. The opinions expressed in this publication are those of the authors and do not necessarily reflect the views of the John Templeton Foundation. For open access purposes, the authors have applied a CC-BY public copyright license to any author-accepted manuscript arising from this submission.

\bibliographystyle{apsrev4-1}
\bibliography{bibliography}

\end{document}

%% file: styles-HOOPT.tikzstyles
\tikzstyle{small-square}=[fill=white, draw=black, shape=rectangle, line width=1 pt]
\tikzstyle{tall}=[fill=white, draw=black, shape=rectangle, minimum width=1.5 cm, minimum height=2.5 cm, line width=1 pt]
\tikzstyle{huge}=[fill=white, draw=black, shape=rectangle, minimum width=2 cm, minimum height=4.5 cm, line width=1 pt]

\tikzstyle{contr-wire}=[-, draw={rgb,255: red,0; green,53; blue,244}, line width=2 pt, densely dashdotted]
\tikzstyle{choi-wire2}=[-, draw={rgb,255: red,79; green,4; blue,134}, line width=2.5 pt]
\tikzstyle{choi-wire1}=[-, draw={rgb,255: red,0; green,98; blue,98}, line width=2.5 pt]
\tikzstyle{portal}=[-, draw={rgb,255: red,0; green,53; blue,244}, double, line width=1.5 pt]
\tikzstyle{syst-wire}=[-, line width=1.5 pt]
\tikzstyle{func-wire}=[-, draw={rgb,255: red,145; green,18; blue,18}, line width=2 pt, densely dashdotted]
\tikzstyle{choi-wire3}=[-, draw={rgb,255: red,0; green,53; blue,244}, line width=2.5 pt]
\tikzstyle{choi-wire4}=[-, draw={rgb,255: red,145; green,18; blue,18}, line width=2.5 pt]
\tikzstyle{cross-wire}=[-, draw={rgb,255: red,145; green,18; blue,18}, line width=1.5 pt]
\tikzstyle{box-wire}=[-, line width=1 pt]

%% file: style_CRF.tikzstyles
\tikzstyle{small-square}=[fill=white, draw=black, shape=rectangle, line width=1 pt]
\tikzstyle{tall}=[fill=white, draw=black, shape=rectangle, minimum width=1.5 cm, minimum height=2.5 cm, line width=1 pt]
\tikzstyle{huge}=[fill=white, draw=black, shape=rectangle, minimum width=2 cm, minimum height=4.5 cm, line width=1 pt]
\tikzstyle{new style 0}=[fill=white, draw=black, shape=circle, minimum size=0.1cm]
\tikzstyle{new style 1}=[fill=black, draw=black, shape=circle]
\tikzstyle{new style 2}=[fill=white, draw=black, shape=circle]

\tikzstyle{contr-wire}=[-, draw={rgb,255: red,0; green,53; blue,244}, line width=0.5 pt, dashdotted]
\tikzstyle{choi-wire2}=[-, draw={rgb,255: red,79; green,4; blue,134}, line width=2.5 pt]
\tikzstyle{choi-wire1}=[-, draw={rgb,255: red,0; green,98; blue,98}, line width=2.5 pt]
\tikzstyle{portal}=[-, draw={rgb,255: red,0; green,53; blue,244}, double, line width=1.5 pt]
\tikzstyle{syst-wire}=[-, line width=1.5 pt]
\tikzstyle{func-wire}=[-, draw={rgb,255: red,145; green,18; blue,18}, line width=2 pt, densely dashdotted]
\tikzstyle{choi-wire3}=[draw={rgb,255: red,0; green,53; blue,244}, line width=1 pt, ->]
\tikzstyle{choi-wire4}=[-, draw={rgb,255: red,145; green,18; blue,18}, line width=2.5 pt]
\tikzstyle{cross-wire}=[-, draw={rgb,255: red,145; green,18; blue,18}, line width=1.5 pt]
\tikzstyle{box-wire}=[-, line width=1 pt]
\tikzstyle{new edge style 1}=[line width=0.5  pt, ->]
\tikzstyle{new edge style 2}=[line width=1.5 pt, ->]
\tikzstyle{new edge style 3}=[-, line width=0.5 pt, dashdotted, draw=black]

%% file: W_B_A.tikz
\begin{tikzpicture}
	\begin{pgfonlayer}{nodelayer}
		\node [style=none] (0) at (-4.25, 1.5) {};
		\node [style=none] (1) at (-1.75, 1.5) {};
		\node [style=none] (2) at (-4.25, -1.5) {};
		\node [style=none] (3) at (-1.75, -1.5) {};
		\node [style=none] (4) at (1.75, 1.5) {};
		\node [style=none] (5) at (4.25, 1.5) {};
		\node [style=none] (6) at (1.75, -1.5) {};
		\node [style=none] (7) at (4.25, -1.5) {};
		\node [style=none] (8) at (-1.75, 0.75) {};
		\node [style=none] (9) at (1.75, 0.75) {};
		\node [style=none] (10) at (0.5, -0.75) {};
		\node [style=none] (11) at (1.75, -0.75) {};
		\node [style=none] (12) at (-1.75, -0.75) {};
		\node [style=none] (13) at (-0.5, -0.75) {};
		\node [style=none] (14) at (-5.25, 0.75) {};
		\node [style=none] (15) at (-4.25, 0.75) {};
		\node [style=none] (16) at (4.25, 0.75) {};
		\node [style=none] (17) at (5.25, 0.75) {};
		\node [style=none] (18) at (4.25, -0.75) {};
		\node [style=none] (19) at (5.25, -0.75) {};
		\node [style=none] (20) at (-5.25, -0.75) {};
		\node [style=none] (21) at (-4.25, -0.75) {};
		\node [style=none] (22) at (-3, 0) {};
		\node [style=none] (23) at (3, 0) {};
		\node [style=none] (24) at (-3, 0) {$\scriptstyle\Pi_A(U_B)$};
		\node [style=none] (25) at (3, 0) {$\scriptstyle\Phi_A(U_B)$};
		\node [style=none] (26) at (1.25, -0.25) {};
		\node [style=none] (27) at (-1.25, -0.25) {$\scriptstyle A_I$};
		\node [style=none] (28) at (1.25, -0.25) {$\scriptstyle A_O$};
		\node [style=none] (29) at (-1.2, 1.25) {$\scriptstyle E_A$};
        \node [style=none] (29) at (1.20, 1.25) {$\scriptstyle E_A$};
		\node [style=none] (30) at (4.75, 1.25) {$\scriptstyle F_S$};
		\node [style=none] (31) at (-4.75, 1.25) {$\scriptstyle P_S$};
		\node [style=none] (32) at (-4.75, -0.25) {$\scriptstyle P_C$};
		\node [style=none] (33) at (4.75, -0.25) {$\scriptstyle F_C$};
		\node [style=small-square] (34) at (0, -0.75) {$\scriptstyle U_A$};
	\end{pgfonlayer}
	\begin{pgfonlayer}{edgelayer}
		\draw [style=box-wire] (0.center) to (1.center);
		\draw [style=box-wire] (1.center) to (3.center);
		\draw [style=box-wire] (3.center) to (2.center);
		\draw [style=box-wire] (0.center) to (2.center);
		\draw [style=box-wire] (4.center) to (5.center);
		\draw [style=box-wire] (5.center) to (7.center);
		\draw [style=box-wire] (7.center) to (6.center);
		\draw [style=box-wire] (4.center) to (6.center);
		\draw [style=syst-wire] (8.center) to (9.center);
		\draw [style=syst-wire] (10.center) to (11.center);
		\draw [style=syst-wire] (12.center) to (13.center);
		\draw [style=syst-wire] (14.center) to (15.center);
		\draw [style=syst-wire] (16.center) to (17.center);
		\draw [style=syst-wire] (18.center) to (19.center);
		\draw [style=syst-wire] (20.center) to (21.center);
	\end{pgfonlayer}
\end{tikzpicture}

%% file: W_A_B.tikz
\begin{tikzpicture}
	\begin{pgfonlayer}{nodelayer}
		\node [style=none] (0) at (-4.25, 1.5) {};
		\node [style=none] (1) at (-1.75, 1.5) {};
		\node [style=none] (2) at (-4.25, -1.5) {};
		\node [style=none] (3) at (-1.75, -1.5) {};
		\node [style=none] (4) at (1.75, 1.5) {};
		\node [style=none] (5) at (4.25, 1.5) {};
		\node [style=none] (6) at (1.75, -1.5) {};
		\node [style=none] (7) at (4.25, -1.5) {};
		\node [style=none] (8) at (-1.75, 0.75) {};
		\node [style=none] (9) at (1.75, 0.75) {};
		\node [style=none] (10) at (0.5, -0.75) {};
		\node [style=none] (11) at (1.75, -0.75) {};
		\node [style=none] (12) at (-1.75, -0.75) {};
		\node [style=none] (13) at (-0.5, -0.75) {};
		\node [style=none] (14) at (-5.25, 0.75) {};
		\node [style=none] (15) at (-4.25, 0.75) {};
		\node [style=none] (16) at (4.25, 0.75) {};
		\node [style=none] (17) at (5.25, 0.75) {};
		\node [style=none] (18) at (4.25, -0.75) {};
		\node [style=none] (19) at (5.25, -0.75) {};
		\node [style=none] (20) at (-5.25, -0.75) {};
		\node [style=none] (21) at (-4.25, -0.75) {};
		\node [style=none] (22) at (-3, 0) {};
		\node [style=none] (23) at (3, 0) {};
		\node [style=none] (24) at (-3, 0) {$\scriptstyle\Pi_B(U_A)$};
		\node [style=none] (25) at (3, 0) {$\scriptstyle\Phi_B(U_A)$};
		\node [style=none] (26) at (1.25, -0.25) {};
		\node [style=none] (27) at (-1.25, -0.25) {$\scriptstyle B_I$};
		\node [style=none] (28) at (1.25, -0.25) {$\scriptstyle B_O$};
		\node [style=none] (29) at (-1.20, 1.25) {$\scriptstyle E_B$};
        \node [style=none] (29) at (1.20, 1.25) {$\scriptstyle E_B$};
		\node [style=none] (30) at (4.75, 1.25) {$\scriptstyle F_S$};
		\node [style=none] (31) at (-4.75, 1.25) {$\scriptstyle P_S$};
		\node [style=none] (32) at (-4.75, -0.25) {$\scriptstyle P_C$};
		\node [style=none] (33) at (4.75, -0.25) {$\scriptstyle F_C$};
		\node [style=small-square] (34) at (0, -0.75) {$\scriptstyle U_B$};
	\end{pgfonlayer}
	\begin{pgfonlayer}{edgelayer}
		\draw [style=box-wire] (0.center) to (1.center);
		\draw [style=box-wire] (1.center) to (3.center);
		\draw [style=box-wire] (3.center) to (2.center);
		\draw [style=box-wire] (0.center) to (2.center);
		\draw [style=box-wire] (4.center) to (5.center);
		\draw [style=box-wire] (5.center) to (7.center);
		\draw [style=box-wire] (7.center) to (6.center);
		\draw [style=box-wire] (4.center) to (6.center);
		\draw [style=syst-wire] (8.center) to (9.center);
		\draw [style=syst-wire] (10.center) to (11.center);
		\draw [style=syst-wire] (12.center) to (13.center);
		\draw [style=syst-wire] (14.center) to (15.center);
		\draw [style=syst-wire] (16.center) to (17.center);
		\draw [style=syst-wire] (18.center) to (19.center);
		\draw [style=syst-wire] (20.center) to (21.center);
	\end{pgfonlayer}
\end{tikzpicture}

%% file: causalframe_circuit_A.tikz
\begin{tikzpicture}
	\begin{pgfonlayer}{nodelayer}
		\node [style=none] (0) at (-3.75, 1.5) {};
		\node [style=none] (1) at (-1.75, 1.5) {};
		\node [style=none] (2) at (-3.75, -1.5) {};
		\node [style=none] (3) at (-1.75, -1.5) {};
		\node [style=none] (4) at (1.75, 1.5) {};
		\node [style=none] (5) at (3.75, 1.5) {};
		\node [style=none] (6) at (1.75, -1.5) {};
		\node [style=none] (7) at (3.75, -1.5) {};
		\node [style=none] (8) at (-1.75, 1) {};
		\node [style=none] (9) at (1.75, 1) {};
		\node [style=none] (10) at (0.75, -0.75) {};
		\node [style=none] (11) at (1.75, -0.75) {};
		\node [style=none] (12) at (-1.75, -0.75) {};
		\node [style=none] (13) at (-0.75, -0.75) {};
		\node [style=none] (14) at (-4.75, 0) {};
		\node [style=none] (15) at (-3.75, 0) {};
		\node [style=none] (16) at (3.75, 1) {};
		\node [style=none] (17) at (4.75, 1) {};
		\node [style=none] (18) at (3.75, -1) {};
		\node [style=none] (19) at (4.75, -1) {};
		\node [style=none] (20) at (-4.75, -1) {};
		\node [style=none] (21) at (-3.75, -1) {};
		\node [style=none] (22) at (-2.75, 0) {};
		\node [style=none] (23) at (2.75, 0) {};
		\node [style=none] (24) at (-2.75, 0) {$\scriptstyle L_A$};
		\node [style=none] (25) at (2.75, 0) {$\scriptstyle R_A$};
		\node [style=none] (26) at (1.25, -0.25) {};
		\node [style=none] (27) at (-1.25, -0.25) {$\scriptstyle A_I$};
		\node [style=none] (28) at (1.25, -0.25) {$\scriptstyle A_O$};
		\node [style=none] (29) at (-1.25, 1.5) {$\scriptstyle E_A'$};
        \node [style=none] (29) at (1.25, 1.5) {$\scriptstyle E_A'$};
		\node [style=none] (30) at (4.25, 0.5) {$\scriptstyle F_S$};
		\node [style=none] (31) at (-4.25, 0.5) {$\scriptstyle P_S$};
		\node [style=none] (32) at (-4.25, -0.5) {$\scriptstyle P_C$};
		\node [style=none] (33) at (4.25, -0.5) {$\scriptstyle F_C$};
		\node [style=none] (34) at (-4.75, 1) {};
		\node [style=none] (35) at (-3.75, 1) {};
		\node [style=none] (36) at (3.75, 0) {};
		\node [style=none] (37) at (4.75, 0) {};
		\node [style=none] (38) at (-4.25, 1.5) {$\scriptstyle B_O$};
		\node [style=none] (39) at (4.25, 1.5) {$\scriptstyle B_I$};
	\end{pgfonlayer}
	\begin{pgfonlayer}{edgelayer}
		\draw [style=box-wire] (0.center) to (1.center);
		\draw [style=box-wire] (1.center) to (3.center);
		\draw [style=box-wire] (3.center) to (2.center);
		\draw [style=box-wire] (0.center) to (2.center);
		\draw [style=box-wire] (4.center) to (5.center);
		\draw [style=box-wire] (5.center) to (7.center);
		\draw [style=box-wire] (7.center) to (6.center);
		\draw [style=box-wire] (4.center) to (6.center);
		\draw [style=syst-wire] (8.center) to (9.center);
		\draw [style=syst-wire] (10.center) to (11.center);
		\draw [style=syst-wire] (12.center) to (13.center);
		\draw [style=syst-wire] (14.center) to (15.center);
		\draw [style=syst-wire] (16.center) to (17.center);
		\draw [style=syst-wire] (18.center) to (19.center);
		\draw [style=syst-wire] (20.center) to (21.center);
		\draw [style=syst-wire] (34.center) to (35.center);
		\draw [style=syst-wire] (36.center) to (37.center);
	\end{pgfonlayer}
\end{tikzpicture}

%% file: time-delocalized_B_A.tikz
\begin{tikzpicture}
	\begin{pgfonlayer}{nodelayer}
		\node [style=none] (0) at (-3.75, 1.5) {};
		\node [style=none] (1) at (-1.75, 1.5) {};
		\node [style=none] (2) at (-3.75, -1.5) {};
		\node [style=none] (3) at (-1.75, -1.5) {};
		\node [style=none] (4) at (1.75, 1.5) {};
		\node [style=none] (5) at (3.75, 1.5) {};
		\node [style=none] (6) at (1.75, -1.5) {};
		\node [style=none] (7) at (3.75, -1.5) {};
		\node [style=none] (8) at (-1.75, 1) {};
		\node [style=none] (9) at (1.75, 1) {};
		\node [style=none] (10) at (0.5, -0.75) {};
		\node [style=none] (11) at (1.75, -0.75) {};
		\node [style=none] (12) at (-1.75, -0.75) {};
		\node [style=none] (13) at (-0.5, -0.75) {};
		\node [style=none] (14) at (-4.75, 0) {};
		\node [style=none] (15) at (-3.75, 0) {};
		\node [style=none] (16) at (3.75, 1) {};
		\node [style=none] (17) at (4.75, 1) {};
		\node [style=none] (18) at (3.75, -1) {};
		\node [style=none] (19) at (4.75, -1) {};
		\node [style=none] (20) at (-4.75, -1) {};
		\node [style=none] (21) at (-3.75, -1) {};
		\node [style=none] (22) at (-2.75, 0) {};
		\node [style=none] (23) at (2.75, 0) {};
		\node [style=none] (24) at (-2.75, 0) {$\scriptstyle L_A$};
		\node [style=none] (25) at (2.75, 0) {$\scriptstyle R_A$};
		\node [style=none] (26) at (1.25, -0.25) {};
		\node [style=none] (27) at (-1.25, -0.25) {$\scriptstyle A_I$};
		\node [style=none] (28) at (1.25, -0.25) {$\scriptstyle A_O$};
		\node [style=none] (29) at (-1.25, 1.5) {$\scriptstyle E_A'$};
        \node [style=none] (29) at (1.25, 1.5) {$\scriptstyle E_A'$};
		\node [style=none] (30) at (4.25, 0.5) {$\scriptstyle F_S$};
		\node [style=none] (31) at (-4.25, 0.5) {$\scriptstyle P_S$};
		\node [style=none] (32) at (-4.25, -0.5) {$\scriptstyle P_C$};
		\node [style=none] (33) at (4.25, -0.5) {$\scriptstyle F_C$};
		\node [style=none] (34) at (-5, 1) {};
		\node [style=none] (35) at (-3.75, 1) {};
		\node [style=none] (36) at (3.75, 0) {};
		\node [style=none] (37) at (4.75, 0) {};
		\node [style=none] (38) at (-4.25, 1.5) {$\scriptstyle B_O$};
		\node [style=none] (39) at (4.25, 1.5) {$\scriptstyle B_I$};
		\node [style=small-square] (40) at (-5.5, 1) {$\scriptstyle U_B$};
		\node [style=none] (41) at (-6.5, 1) {};
		\node [style=none] (42) at (-6.5, 2.25) {};
		\node [style=none] (43) at (4.75, 2.25) {};
		\node [style=none] (44) at (4.75, 1) {};
		\node [style=none] (45) at (4.75, 2.25) {};
		\node [style=none] (46) at (-6.5, 1) {};
		\node [style=none] (47) at (-6, 1) {};
		\node [style=small-square] (48) at (0, -0.75) {$\scriptstyle U_A$};
	\end{pgfonlayer}
	\begin{pgfonlayer}{edgelayer}
		\draw [style=box-wire] (0.center) to (1.center);
		\draw [style=box-wire] (1.center) to (3.center);
		\draw [style=box-wire] (3.center) to (2.center);
		\draw [style=box-wire] (0.center) to (2.center);
		\draw [style=box-wire] (4.center) to (5.center);
		\draw [style=box-wire] (5.center) to (7.center);
		\draw [style=box-wire] (7.center) to (6.center);
		\draw [style=box-wire] (4.center) to (6.center);
		\draw [style=syst-wire] (8.center) to (9.center);
		\draw [style=syst-wire] (10.center) to (11.center);
		\draw [style=syst-wire] (12.center) to (13.center);
		\draw [style=syst-wire] (14.center) to (15.center);
		\draw [style=syst-wire] (16.center) to (17.center);
		\draw [style=syst-wire] (18.center) to (19.center);
		\draw [style=syst-wire] (20.center) to (21.center);
		\draw [style=syst-wire] (34.center) to (35.center);
		\draw [style=syst-wire] (36.center) to (37.center);
		\draw [style=syst-wire] (42.center) to (43.center);
		\draw [style=syst-wire, bend left=90, looseness=1.50] (41.center) to (42.center);
		\draw [style=syst-wire, bend right=90, looseness=1.50] (44.center) to (45.center);
		\draw [style=syst-wire] (46.center) to (47.center);
	\end{pgfonlayer}
\end{tikzpicture}

%% file: time-delocalized_A_B.tikz
\begin{tikzpicture}
	\begin{pgfonlayer}{nodelayer}
		\node [style=none] (0) at (-3.75, 1.5) {};
		\node [style=none] (1) at (-1.75, 1.5) {};
		\node [style=none] (2) at (-3.75, -1.5) {};
		\node [style=none] (3) at (-1.75, -1.5) {};
		\node [style=none] (4) at (1.75, 1.5) {};
		\node [style=none] (5) at (3.75, 1.5) {};
		\node [style=none] (6) at (1.75, -1.5) {};
		\node [style=none] (7) at (3.75, -1.5) {};
		\node [style=none] (8) at (-1.75, 1) {};
		\node [style=none] (9) at (1.75, 1) {};
		\node [style=none] (10) at (0.5, -0.75) {};
		\node [style=none] (11) at (1.75, -0.75) {};
		\node [style=none] (12) at (-1.75, -0.75) {};
		\node [style=none] (13) at (-0.5, -0.75) {};
		\node [style=none] (14) at (-4.75, 0) {};
		\node [style=none] (15) at (-3.75, 0) {};
		\node [style=none] (16) at (3.75, 1) {};
		\node [style=none] (17) at (4.75, 1) {};
		\node [style=none] (18) at (3.75, -1) {};
		\node [style=none] (19) at (4.75, -1) {};
		\node [style=none] (20) at (-4.75, -1) {};
		\node [style=none] (21) at (-3.75, -1) {};
		\node [style=none] (22) at (-2.75, 0) {};
		\node [style=none] (23) at (2.75, 0) {};
		\node [style=none] (24) at (-2.75, 0) {$\scriptstyle L_B$};
		\node [style=none] (25) at (2.75, 0) {$\scriptstyle R_B$};
		\node [style=none] (26) at (1.25, -0.25) {};
		\node [style=none] (27) at (-1.25, -0.25) {$\scriptstyle B_I$};
		\node [style=none] (28) at (1.25, -0.25) {$\scriptstyle B_O$};
		\node [style=none] (29) at (-1.25, 1.5) {$\scriptstyle E_B'$};
        \node [style=none] (29) at (1.25, 1.5) {$\scriptstyle E_B'$};
        \node [style=none] (30) at (4.25, 0.5) {$\scriptstyle F_S$};
		\node [style=none] (31) at (-4.25, 0.5) {$\scriptstyle P_S$};
		\node [style=none] (32) at (-4.25, -0.5) {$\scriptstyle P_C$};
		\node [style=none] (33) at (4.25, -0.5) {$\scriptstyle F_C$};
		\node [style=none] (34) at (-5, 1) {};
		\node [style=none] (35) at (-3.75, 1) {};
		\node [style=none] (36) at (3.75, 0) {};
		\node [style=none] (37) at (4.75, 0) {};
		\node [style=none] (38) at (-4.25, 1.5) {$\scriptstyle A_O$};
		\node [style=none] (39) at (4.25, 1.5) {$\scriptstyle A_I$};
		\node [style=small-square] (40) at (-5.5, 1) {$\scriptstyle U_A$};
		\node [style=none] (41) at (-6.5, 1) {};
		\node [style=none] (42) at (-6.5, 2.25) {};
		\node [style=none] (43) at (4.75, 2.25) {};
		\node [style=none] (44) at (4.75, 1) {};
		\node [style=none] (45) at (4.75, 2.25) {};
		\node [style=none] (46) at (-6.5, 1) {};
		\node [style=none] (47) at (-6, 1) {};
		\node [style=small-square] (48) at (0, -0.75) {$\scriptstyle U_B$};
	\end{pgfonlayer}
	\begin{pgfonlayer}{edgelayer}
		\draw [style=box-wire] (0.center) to (1.center);
		\draw [style=box-wire] (1.center) to (3.center);
		\draw [style=box-wire] (3.center) to (2.center);
		\draw [style=box-wire] (0.center) to (2.center);
		\draw [style=box-wire] (4.center) to (5.center);
		\draw [style=box-wire] (5.center) to (7.center);
		\draw [style=box-wire] (7.center) to (6.center);
		\draw [style=box-wire] (4.center) to (6.center);
		\draw [style=syst-wire] (8.center) to (9.center);
		\draw [style=syst-wire] (10.center) to (11.center);
		\draw [style=syst-wire] (12.center) to (13.center);
		\draw [style=syst-wire] (14.center) to (15.center);
		\draw [style=syst-wire] (16.center) to (17.center);
		\draw [style=syst-wire] (18.center) to (19.center);
		\draw [style=syst-wire] (20.center) to (21.center);
		\draw [style=syst-wire] (34.center) to (35.center);
		\draw [style=syst-wire] (36.center) to (37.center);
		\draw [style=syst-wire] (42.center) to (43.center);
		\draw [style=syst-wire, bend left=90, looseness=1.50] (41.center) to (42.center);
		\draw [style=syst-wire, bend right=90, looseness=1.50] (44.center) to (45.center);
		\draw [style=syst-wire] (46.center) to (47.center);
	\end{pgfonlayer}
\end{tikzpicture}

%% file: time_coordinates.tikz
\begin{tikzpicture}
	\begin{pgfonlayer}{nodelayer}
		\node [style=none] (0) at (-4.5, 1.25) {};
		\node [style=none] (1) at (-2, 1.25) {};
		\node [style=none] (2) at (-4.5, -1.75) {};
		\node [style=none] (3) at (-2, -1.75) {};
		\node [style=none] (4) at (2, 1.25) {};
		\node [style=none] (5) at (4.5, 1.25) {};
		\node [style=none] (6) at (2, -1.75) {};
		\node [style=none] (7) at (4.5, -1.75) {};
		\node [style=none] (8) at (-2, 0.5) {};
		\node [style=none] (9) at (2, 0.5) {};
		\node [style=none] (10) at (0.5, -1) {};
		\node [style=none] (11) at (2, -1) {};
		\node [style=none] (12) at (-2, -1) {};
		\node [style=none] (13) at (-0.5, -1) {};
		\node [style=none] (14) at (-5.5, 0.5) {};
		\node [style=none] (15) at (-4.5, 0.5) {};
		\node [style=none] (16) at (4.5, 0.5) {};
		\node [style=none] (17) at (5.5, 0.5) {};
		\node [style=none] (18) at (4.5, -1) {};
		\node [style=none] (19) at (5.5, -1) {};
		\node [style=none] (20) at (-5.5, -1) {};
		\node [style=none] (21) at (-4.5, -1) {};
		\node [style=none] (22) at (-3.25, -0.25) {};
		\node [style=none] (23) at (3.25, -0.25) {};
		\node [style=none] (24) at (-3.25, -0.25) {$\scriptstyle\Pi_A(U_B)$};
		\node [style=none] (25) at (3.25, -0.25) {$\scriptstyle\Phi_A(U_B)$};
		\node [style=none] (26) at (1.5, -0.5) {};
		\node [style=none] (27) at (-1.5, -0.5) {$\scriptstyle A_I$};
		\node [style=none] (28) at (1.5, -0.5) {$\scriptstyle A_O$};
		\node [style=none] (29) at (0, 1) {$\scriptstyle E$};
		\node [style=none] (30) at (5, 1) {$\scriptstyle F_S$};
		\node [style=none] (31) at (-5, 1) {$\scriptstyle P_S$};
		\node [style=none] (32) at (-5, -0.5) {$\scriptstyle P_C$};
		\node [style=none] (33) at (5, -0.5) {$\scriptstyle F_C$};
		\node [style=small-square] (34) at (0, -1) {$\scriptstyle U_A$};
		\node [style=none] (35) at (-1, 1.75) {};
		\node [style=none] (36) at (-1, -2.25) {};
		\node [style=none] (37) at (1, 1.75) {};
		\node [style=none] (38) at (1, -2.25) {};
		\node [style=none] (39) at (5.5, 1.75) {};
		\node [style=none] (40) at (5.5, -2.25) {};
		\node [style=none] (41) at (-5.5, 1.75) {};
		\node [style=none] (42) at (-5.5, -2.25) {};
		\node [style=none] (45) at (5.5, 2.25) {$\scriptstyle t_3$};
		\node [style=none] (47) at (-5.5, 2.25) {$\scriptstyle t_0$};
		\node [style=none] (48) at (-1, 2.25) {$\scriptstyle t_1$};
		\node [style=none] (49) at (1, 2.25) {$\scriptstyle t_2$};
	\end{pgfonlayer}
	\begin{pgfonlayer}{edgelayer}
		\draw [style=box-wire] (0.center) to (1.center);
		\draw [style=box-wire] (1.center) to (3.center);
		\draw [style=box-wire] (3.center) to (2.center);
		\draw [style=box-wire] (0.center) to (2.center);
		\draw [style=box-wire] (4.center) to (5.center);
		\draw [style=box-wire] (5.center) to (7.center);
		\draw [style=box-wire] (7.center) to (6.center);
		\draw [style=box-wire] (4.center) to (6.center);
		\draw [style=syst-wire] (8.center) to (9.center);
		\draw [style=syst-wire] (10.center) to (11.center);
		\draw [style=syst-wire] (12.center) to (13.center);
		\draw [style=syst-wire] (14.center) to (15.center);
		\draw [style=syst-wire] (16.center) to (17.center);
		\draw [style=syst-wire] (18.center) to (19.center);
		\draw [style=syst-wire] (20.center) to (21.center);
		\draw [style=contr-wire] (41.center) to (42.center);
		\draw [style=contr-wire] (35.center) to (36.center);
		\draw [style=contr-wire] (37.center) to (38.center);
		\draw [style=contr-wire] (39.center) to (40.center);
	\end{pgfonlayer}
\end{tikzpicture}

%% file: unitaries_fixedboundaries.tikz
\begin{tikzpicture}
	\begin{pgfonlayer}{nodelayer}
		\node [style=none] (0) at (-4.25, 3) {};
		\node [style=none] (1) at (-1.75, 3) {};
		\node [style=none] (2) at (-4.25, 0) {};
		\node [style=none] (3) at (-1.75, 0) {};
		\node [style=none] (4) at (1.75, 3) {};
		\node [style=none] (5) at (4.25, 3) {};
		\node [style=none] (6) at (1.75, 0) {};
		\node [style=none] (7) at (4.25, 0) {};
		\node [style=none] (8) at (-1.75, 2.25) {};
		\node [style=none] (9) at (1.75, 2.25) {};
		\node [style=none] (10) at (0.75, 0.75) {};
		\node [style=none] (11) at (1.75, 0.75) {};
		\node [style=none] (12) at (-1.75, 0.75) {};
		\node [style=none] (13) at (-0.75, 0.75) {};
		\node [style=none] (14) at (-5.25, 2.25) {};
		\node [style=none] (15) at (-4.25, 2.25) {};
		\node [style=none] (16) at (4.25, 2.25) {};
		\node [style=none] (17) at (5.25, 2.25) {};
		\node [style=none] (18) at (4.25, 0.75) {};
		\node [style=none] (19) at (5.25, 0.75) {};
		\node [style=none] (20) at (-5.25, 0.75) {};
		\node [style=none] (21) at (-4.25, 0.75) {};
		\node [style=none] (22) at (-3, 1.5) {};
		\node [style=none] (23) at (3, 1.5) {};
		\node [style=none] (24) at (-3, 1.5) {$\scriptstyle\Pi_A(U_B)$};
		\node [style=none] (25) at (3, 1.5) {$\scriptstyle\Phi_A(U_B)$};
		\node [style=none] (26) at (1.25, 1.25) {};
		\node [style=none] (27) at (-1.25, 1.25) {$\scriptstyle A_I$};
		\node [style=none] (28) at (1.25, 1.25) {$\scriptstyle A_O$};
		\node [style=none] (29) at (1.25, 2.75) {$\scriptstyle E_A$};
		\node [style=none] (30) at (-1.25, 2.75) {$\scriptstyle E_A$};
		\node [style=none] (31) at (4.75, 2.75) {$\scriptstyle F_S$};
		\node [style=none] (32) at (-4.75, 2.75) {$\scriptstyle P_S$};
		\node [style=none] (33) at (-4.75, 1.25) {$\scriptstyle P_C$};
		\node [style=none] (34) at (4.75, 1.25) {$\scriptstyle F_C$};
		\node [style=none] (35) at (-0.75, 0.75) {};
		\node [style=none] (36) at (0.75, 0.75) {};
		\node [style=none] (37) at (-0.75, -0.75) {};
		\node [style=none] (38) at (0.75, -0.75) {};
		\node [style=none] (39) at (5.25, -0.75) {};
		\node [style=none] (40) at (-5.25, -0.75) {};
		\node [style=none] (41) at (4.75, -0.25) {$\scriptstyle A_I$};
		\node [style=none] (42) at (-4.75, -0.25) {};
		\node [style=none] (43) at (-4.75, -0.25) {$\scriptstyle A_O$};
		\node [style=none] (44) at (5.25, 5.5) {};
		\node [style=none] (45) at (7.25, 5.5) {};
		\node [style=none] (46) at (5.25, -2.5) {};
		\node [style=none] (47) at (7.25, -2.5) {};
		\node [style=none] (48) at (6.25, 1.5) {$\scriptstyle J^\dagger$};
		\node [style=none] (49) at (-7.25, 2.5) {};
		\node [style=none] (50) at (-7.25, 1) {};
		\node [style=none] (51) at (-7.25, -0.5) {};
		\node [style=none] (52) at (-7.25, 2.5) {};
		\node [style=none] (53) at (-5.25, 2.5) {};
		\node [style=none] (54) at (-7.25, -5.5) {};
		\node [style=none] (55) at (-5.25, -5.5) {};
		\node [style=none] (56) at (-6.25, -1.5) {$\scriptstyle J$};
		\node [style=none] (57) at (-8.25, 2.25) {};
		\node [style=none] (58) at (-7.25, 2.25) {};
		\node [style=none] (59) at (-8.25, 0.75) {};
		\node [style=none] (60) at (-7.25, 0.75) {};
		\node [style=none] (61) at (-8.25, -0.75) {};
		\node [style=none] (62) at (-7.25, -0.75) {};
		\node [style=none] (63) at (7.25, 2.25) {};
		\node [style=none] (64) at (8.25, 2.25) {};
		\node [style=none] (65) at (7.25, 0.75) {};
		\node [style=none] (66) at (8.25, 0.75) {};
		\node [style=none] (67) at (7.25, -0.75) {};
		\node [style=none] (68) at (8.25, -0.75) {};
		\node [style=none] (69) at (-7.75, 2.75) {$\scriptstyle P_S'$};
		\node [style=none] (70) at (-7.75, 1.25) {$\scriptstyle P_C$};
		\node [style=none] (71) at (-7.75, -0.25) {$\scriptstyle B_I$};
		\node [style=none] (72) at (7.75, 2.75) {$\scriptstyle F_S'$};
		\node [style=none] (73) at (7.75, 1.25) {$\scriptstyle F_C$};
		\node [style=none] (74) at (7.75, -0.25) {$\scriptstyle B_O$};
		\node [style=small-square] (75) at (0, -2.25) {$\scriptstyle U_A$};
		\node [style=none] (76) at (0.5, -2.25) {};
		\node [style=none] (77) at (5.25, -2.25) {};
		\node [style=none] (78) at (-5.25, -2.25) {};
		\node [style=none] (79) at (-0.5, -2.25) {};
		\node [style=none] (80) at (5.25, -2.25) {};
		\node [style=none] (81) at (-1.25, -1.75) {$\scriptstyle A_I$};
		\node [style=none] (82) at (1.25, -1.75) {$\scriptstyle A_O$};
		\node [style=none] (83) at (7.25, -2.25) {};
		\node [style=none] (84) at (8.25, -2.25) {};
		\node [style=none] (85) at (-8.25, -2.25) {};
		\node [style=none] (86) at (-7.25, -2.25) {};
		\node [style=none] (87) at (-7.75, -1.75) {$\scriptstyle B_O$};
		\node [style=none] (88) at (7.75, -1.75) {$\scriptstyle B_I$};
		\node [style=none] (89) at (-4.75, -1.75) {$\scriptstyle A_I$};
		\node [style=none] (90) at (4.25, 5) {};
		\node [style=none] (91) at (4.25, 4) {};
		\node [style=none] (92) at (3.75, 3.5) {};
		\node [style=none] (93) at (3.75, 5.5) {};
		\node [style=none] (94) at (4.25, 4) {};
		\node [style=none] (95) at (4.25, 5) {};
		\node [style=none] (96) at (3.75, 4.5) {$\phi$};
		\node [style=none] (97) at (4.25, 5) {};
		\node [style=none] (98) at (4.25, 5.5) {};
		\node [style=none] (99) at (4.25, 3.5) {};
		\node [style=none] (100) at (-4.25, -3.75) {};
		\node [style=none] (101) at (-4.25, -5.25) {};
		\node [style=none] (102) at (-4.25, -3.75) {};
		\node [style=none] (103) at (-4.25, -5.25) {};
		\node [style=none] (104) at (-3.75, -3.5) {};
		\node [style=none] (105) at (-3.75, -5.5) {};
		\node [style=none] (106) at (-4.25, -5.25) {};
		\node [style=none] (107) at (-3.75, -4.5) {$\psi$};
		\node [style=none] (108) at (-4.25, -5.25) {};
		\node [style=none] (109) at (-4.25, -5.25) {};
		\node [style=none] (110) at (-4.25, -5.5) {};
		\node [style=none] (111) at (-4.25, -3.5) {};
		\node [style=none] (112) at (-4.25, -3.75) {};
		\node [style=none] (113) at (-5.25, -3.75) {};
		\node [style=none] (114) at (-4.25, -5.25) {};
		\node [style=none] (115) at (-4.25, -5.25) {};
		\node [style=none] (116) at (-4.25, -5.25) {};
		\node [style=none] (117) at (-5.25, -5.25) {};
		\node [style=none] (118) at (5.25, 5.25) {};
		\node [style=none] (119) at (5.25, 3.75) {};
		\node [style=none] (120) at (5.25, 5.25) {};
		\node [style=none] (121) at (5.25, 3.75) {};
		\node [style=none] (122) at (5.25, 3.75) {};
		\node [style=none] (123) at (5.25, 3.75) {};
		\node [style=none] (124) at (5.25, 3.75) {};
		\node [style=none] (125) at (5.25, 5.25) {};
		\node [style=none] (126) at (4.25, 5.25) {};
		\node [style=none] (127) at (5.25, 3.75) {};
		\node [style=none] (128) at (5.25, 3.75) {};
		\node [style=none] (129) at (5.25, 3.75) {};
		\node [style=none] (130) at (4.25, 3.75) {};
		\node [style=none] (131) at (4.75, 5.75) {$\scriptstyle P_S$};
		\node [style=none] (132) at (4.75, 4.25) {$\scriptstyle P_C$};
		\node [style=none] (133) at (8.25, 5.5) {};
		\node [style=none] (134) at (8.25, 4) {};
		\node [style=none] (135) at (7.25, 5.25) {};
		\node [style=none] (136) at (8.25, 5.25) {};
		\node [style=none] (137) at (7.25, 3.75) {};
		\node [style=none] (138) at (8.25, 3.75) {};
		\node [style=none] (139) at (7.75, 5.75) {$\scriptstyle P_S'$};
		\node [style=none] (140) at (7.75, 4.25) {$\scriptstyle P_C$};
		\node [style=none] (141) at (-8.25, -3.75) {};
		\node [style=none] (142) at (-7.25, -3.75) {};
		\node [style=none] (143) at (-8.25, -5.25) {};
		\node [style=none] (144) at (-7.25, -5.25) {};
		\node [style=none] (145) at (-7.75, -3.25) {$\scriptstyle F_S'$};
		\node [style=none] (146) at (-7.75, -4.75) {$\scriptstyle F_C$};
		\node [style=none] (147) at (-4.75, -3.25) {$\scriptstyle F_S$};
		\node [style=none] (148) at (-4.75, -4.75) {$\scriptstyle F_C$};
	\end{pgfonlayer}
	\begin{pgfonlayer}{edgelayer}
		\draw [style=box-wire] (0.center) to (1.center);
		\draw [style=box-wire] (1.center) to (3.center);
		\draw [style=box-wire] (3.center) to (2.center);
		\draw [style=box-wire] (0.center) to (2.center);
		\draw [style=box-wire] (4.center) to (5.center);
		\draw [style=box-wire] (5.center) to (7.center);
		\draw [style=box-wire] (7.center) to (6.center);
		\draw [style=box-wire] (4.center) to (6.center);
		\draw [style=syst-wire] (8.center) to (9.center);
		\draw [style=syst-wire] (10.center) to (11.center);
		\draw [style=syst-wire] (12.center) to (13.center);
		\draw [style=syst-wire] (14.center) to (15.center);
		\draw [style=syst-wire] (16.center) to (17.center);
		\draw [style=syst-wire] (18.center) to (19.center);
		\draw [style=syst-wire] (20.center) to (21.center);
		\draw [style=syst-wire, in=180, out=0, looseness=0.75] (37.center) to (36.center);
		\draw [style=syst-wire, in=0, out=-180, looseness=0.75] (38.center) to (35.center);
		\draw [style=syst-wire] (40.center) to (37.center);
		\draw [style=syst-wire] (38.center) to (39.center);
		\draw [style=box-wire] (44.center) to (45.center);
		\draw [style=box-wire] (45.center) to (47.center);
		\draw [style=box-wire] (47.center) to (46.center);
		\draw [style=box-wire] (44.center) to (46.center);
		\draw [style=box-wire] (52.center) to (53.center);
		\draw [style=box-wire] (53.center) to (55.center);
		\draw [style=box-wire] (55.center) to (54.center);
		\draw [style=box-wire] (52.center) to (54.center);
		\draw [style=syst-wire] (57.center) to (58.center);
		\draw [style=syst-wire] (59.center) to (60.center);
		\draw [style=syst-wire] (61.center) to (62.center);
		\draw [style=syst-wire] (63.center) to (64.center);
		\draw [style=syst-wire] (65.center) to (66.center);
		\draw [style=syst-wire] (67.center) to (68.center);
		\draw [style=syst-wire] (76.center) to (77.center);
		\draw [style=syst-wire] (78.center) to (79.center);
		\draw [style=syst-wire] (83.center) to (84.center);
		\draw [style=syst-wire] (85.center) to (86.center);
		\draw [style=box-wire, bend right=90] (93.center) to (92.center);
		\draw [style=box-wire] (93.center) to (98.center);
		\draw [style=box-wire] (98.center) to (99.center);
		\draw [style=box-wire] (99.center) to (92.center);
		\draw [style=box-wire, bend left=90] (104.center) to (105.center);
		\draw [style=box-wire] (104.center) to (111.center);
		\draw [style=box-wire] (111.center) to (110.center);
		\draw [style=box-wire] (110.center) to (105.center);
		\draw [style=syst-wire] (113.center) to (112.center);
		\draw [style=syst-wire] (117.center) to (116.center);
		\draw [style=syst-wire] (126.center) to (125.center);
		\draw [style=syst-wire] (130.center) to (129.center);
		\draw [style=syst-wire] (135.center) to (136.center);
		\draw [style=syst-wire] (137.center) to (138.center);
		\draw [style=syst-wire] (141.center) to (142.center);
		\draw [style=syst-wire] (143.center) to (144.center);
	\end{pgfonlayer}
\end{tikzpicture}

%% file: fragments_CRF_B.tikz
\begin{tikzpicture}
	\begin{pgfonlayer}{nodelayer}
		\node [style=none] (0) at (-4.25, 3) {};
		\node [style=none] (1) at (-1.75, 3) {};
		\node [style=none] (2) at (-4.25, 0) {};
		\node [style=none] (3) at (-1.75, 0) {};
		\node [style=none] (4) at (1.75, 3) {};
		\node [style=none] (5) at (4.25, 3) {};
		\node [style=none] (6) at (1.75, 0) {};
		\node [style=none] (7) at (4.25, 0) {};
		\node [style=none] (8) at (-1.75, 2.25) {};
		\node [style=none] (9) at (1.75, 2.25) {};
		\node [style=none] (10) at (0.75, 0.75) {};
		\node [style=none] (11) at (1.75, 0.75) {};
		\node [style=none] (12) at (-1.75, 0.75) {};
		\node [style=none] (13) at (-0.75, 0.75) {};
		\node [style=none] (14) at (-5.25, 2.25) {};
		\node [style=none] (15) at (-4.25, 2.25) {};
		\node [style=none] (16) at (4.25, 2.25) {};
		\node [style=none] (17) at (5.25, 2.25) {};
		\node [style=none] (18) at (4.25, 0.75) {};
		\node [style=none] (19) at (5.25, 0.75) {};
		\node [style=none] (20) at (-5.25, 0.75) {};
		\node [style=none] (21) at (-4.25, 0.75) {};
		\node [style=none] (22) at (-3, 1.5) {};
		\node [style=none] (23) at (3, 1.5) {};
		\node [style=none] (24) at (-3, 1.5) {$\scriptstyle\Pi_B(U_A)$};
		\node [style=none] (25) at (3, 1.5) {$\scriptstyle\Phi_B(U_A)$};
		\node [style=none] (26) at (1.25, 1.25) {};
		\node [style=none] (27) at (-1.25, 1.25) {$\scriptstyle B_I$};
		\node [style=none] (28) at (1.25, 1.25) {$\scriptstyle B_O$};
		\node [style=none] (29) at (1.25, 2.75) {$\scriptstyle E_B$};
		\node [style=none] (29) at (-1.25, 2.75) {$\scriptstyle E_B$};
		\node [style=none] (30) at (4.75, 2.75) {$\scriptstyle F_S'$};
		\node [style=none] (31) at (-4.75, 2.75) {$\scriptstyle P_S'$};
		\node [style=none] (32) at (-4.75, 1.25) {$\scriptstyle P_C'$};
		\node [style=none] (33) at (4.75, 1.25) {$\scriptstyle F_C'$};
		\node [style=small-square] (34) at (0, -2.25) {$\scriptstyle U_B$};
		\node [style=none] (35) at (0.5, -2.25) {};
		\node [style=none] (36) at (5.25, -2.25) {};
		\node [style=none] (37) at (-5.25, -2.25) {};
		\node [style=none] (38) at (-0.5, -2.25) {};
		\node [style=none] (39) at (4.75, -1.75) {};
		\node [style=none] (40) at (-4.75, -1.75) {$\scriptstyle B_I$};
		\node [style=none] (41) at (4.75, -1.75) {$\scriptstyle B_O$};
		\node [style=none] (42) at (0.75, 0.75) {};
		\node [style=none] (43) at (-0.75, 0.75) {};
		\node [style=none] (44) at (-0.75, 0.75) {};
		\node [style=none] (45) at (0.75, 0.75) {};
		\node [style=none] (46) at (-0.75, -0.75) {};
		\node [style=none] (47) at (0.75, -0.75) {};
		\node [style=none] (48) at (-0.75, -0.75) {};
		\node [style=none] (49) at (0.75, -0.75) {};
		\node [style=none] (50) at (5.25, -0.75) {};
		\node [style=none] (51) at (-5.25, -0.75) {};
		\node [style=none] (52) at (4.75, -0.25) {$\scriptstyle B_I$};
		\node [style=none] (53) at (-4.75, -0.25) {};
		\node [style=none] (54) at (-4.75, -0.25) {$\scriptstyle B_O$};
		\node [style=none] (55) at (-1.25, -1.75) {$\scriptstyle B_I$};
		\node [style=none] (56) at (1.25, -1.75) {};
		\node [style=none] (57) at (1.25, -1.75) {$\scriptstyle B_O$};
		\node [style=none] (58) at (5.25, 5.5) {};
		\node [style=none] (59) at (4.25, 5) {};
		\node [style=none] (60) at (4.25, 4) {};
		\node [style=none] (61) at (3.75, 3.5) {};
		\node [style=none] (62) at (3.75, 5.5) {};
		\node [style=none] (63) at (4.25, 4) {};
		\node [style=none] (64) at (4.25, 5) {};
		\node [style=none] (65) at (3.75, 4.5) {$\phi$};
		\node [style=none] (66) at (4.25, 5) {};
		\node [style=none] (67) at (4.25, 5.5) {};
		\node [style=none] (68) at (4.25, 3.5) {};
		\node [style=none] (69) at (5.25, 5.25) {};
		\node [style=none] (70) at (5.25, 3.75) {};
		\node [style=none] (71) at (5.25, 5.25) {};
		\node [style=none] (72) at (5.25, 3.75) {};
		\node [style=none] (73) at (5.25, 3.75) {};
		\node [style=none] (74) at (5.25, 3.75) {};
		\node [style=none] (75) at (5.25, 3.75) {};
		\node [style=none] (76) at (5.25, 5.25) {};
		\node [style=none] (77) at (4.25, 5.25) {};
		\node [style=none] (78) at (5.25, 3.75) {};
		\node [style=none] (79) at (5.25, 3.75) {};
		\node [style=none] (80) at (5.25, 3.75) {};
		\node [style=none] (81) at (4.25, 3.75) {};
		\node [style=none] (82) at (4.75, 5.75) {$\scriptstyle P_S'$};
		\node [style=none] (83) at (4.75, 4.25) {$\scriptstyle P_C'$};
		\node [style=none] (84) at (-5.25, -5.5) {};
		\node [style=none] (85) at (-4.25, -3.75) {};
		\node [style=none] (86) at (-4.25, -5.25) {};
		\node [style=none] (87) at (-4.25, -3.75) {};
		\node [style=none] (88) at (-4.25, -5.25) {};
		\node [style=none] (89) at (-3.75, -3.5) {};
		\node [style=none] (90) at (-3.75, -5.5) {};
		\node [style=none] (91) at (-4.25, -5.25) {};
		\node [style=none] (92) at (-3.75, -4.5) {$\psi$};
		\node [style=none] (93) at (-4.25, -5.25) {};
		\node [style=none] (94) at (-4.25, -5.25) {};
		\node [style=none] (95) at (-4.25, -5.5) {};
		\node [style=none] (96) at (-4.25, -3.5) {};
		\node [style=none] (97) at (-4.25, -3.75) {};
		\node [style=none] (98) at (-5.25, -3.75) {};
		\node [style=none] (99) at (-4.25, -5.25) {};
		\node [style=none] (100) at (-4.25, -5.25) {};
		\node [style=none] (101) at (-4.25, -5.25) {};
		\node [style=none] (102) at (-5.25, -5.25) {};
		\node [style=none] (103) at (-4.75, -3.25) {$\scriptstyle F_S'$};
		\node [style=none] (104) at (-4.75, -4.75) {$\scriptstyle F_C'$};
	\end{pgfonlayer}
	\begin{pgfonlayer}{edgelayer}
		\draw [style=box-wire] (0.center) to (1.center);
		\draw [style=box-wire] (1.center) to (3.center);
		\draw [style=box-wire] (3.center) to (2.center);
		\draw [style=box-wire] (0.center) to (2.center);
		\draw [style=box-wire] (4.center) to (5.center);
		\draw [style=box-wire] (5.center) to (7.center);
		\draw [style=box-wire] (7.center) to (6.center);
		\draw [style=box-wire] (4.center) to (6.center);
		\draw [style=syst-wire] (8.center) to (9.center);
		\draw [style=syst-wire] (10.center) to (11.center);
		\draw [style=syst-wire] (12.center) to (13.center);
		\draw [style=syst-wire] (14.center) to (15.center);
		\draw [style=syst-wire] (16.center) to (17.center);
		\draw [style=syst-wire] (18.center) to (19.center);
		\draw [style=syst-wire] (20.center) to (21.center);
		\draw [style=syst-wire] (35.center) to (36.center);
		\draw [style=syst-wire] (37.center) to (38.center);
		\draw [style=syst-wire, in=180, out=0, looseness=0.75] (46.center) to (45.center);
		\draw [style=syst-wire, in=0, out=-180, looseness=0.75] (47.center) to (44.center);
		\draw [style=syst-wire] (51.center) to (48.center);
		\draw [style=syst-wire] (49.center) to (50.center);
		\draw [style=box-wire, bend right=90] (62.center) to (61.center);
		\draw [style=box-wire] (62.center) to (67.center);
		\draw [style=box-wire] (67.center) to (68.center);
		\draw [style=box-wire] (68.center) to (61.center);
		\draw [style=syst-wire] (77.center) to (76.center);
		\draw [style=syst-wire] (81.center) to (80.center);
		\draw [style=box-wire, bend left=90] (89.center) to (90.center);
		\draw [style=box-wire] (89.center) to (96.center);
		\draw [style=box-wire] (96.center) to (95.center);
		\draw [style=box-wire] (95.center) to (90.center);
		\draw [style=syst-wire] (98.center) to (97.center);
		\draw [style=syst-wire] (102.center) to (101.center);
	\end{pgfonlayer}
\end{tikzpicture}

%% file: timecircuit_A.tikz
\begin{tikzpicture}
	\begin{pgfonlayer}{nodelayer}
		\node [style=small-square] (3) at (-6, 0.75) {$\mathbf I$};
		\node [style=none] (4) at (-7.5, 0.75) {};
		\node [style=none] (5) at (-6.25, 0.75) {};
		\node [style=none] (6) at (-5.75, 0.75) {};
		\node [style=none] (7) at (-4.5, 0.75) {};
		\node [style=none] (8) at (-7, 1.25) {$\scriptstyle T_0$};
		\node [style=none] (9) at (-5, 1.25) {$\scriptstyle T_1$};
		\node [style=small-square] (10) at (-2, 0.75) {$\mathbf I$};
		\node [style=none] (11) at (-3.5, 0.75) {};
		\node [style=none] (12) at (-2.25, 0.75) {};
		\node [style=none] (13) at (-1.75, 0.75) {};
		\node [style=none] (14) at (-0.5, 0.75) {};
		\node [style=none] (15) at (-3, 1.25) {$\scriptstyle T_2$};
		\node [style=none] (16) at (-1, 1.25) {$\scriptstyle T_3$};
		\node [style=small-square] (17) at (2, 0.75) {$\mathbf I$};
		\node [style=none] (18) at (0.5, 0.75) {};
		\node [style=none] (19) at (1.75, 0.75) {};
		\node [style=none] (20) at (2.25, 0.75) {};
		\node [style=none] (21) at (3.5, 0.75) {};
		\node [style=none] (22) at (1, 1.25) {$\scriptstyle T_4$};
		\node [style=none] (23) at (3, 1.25) {$\scriptstyle T_5$};
		\node [style=small-square] (24) at (6, 0.75) {$\mathbf I$};
		\node [style=none] (25) at (4.5, 0.75) {};
		\node [style=none] (26) at (5.75, 0.75) {};
		\node [style=none] (27) at (6.25, 0.75) {};
		\node [style=none] (28) at (7.5, 0.75) {};
		\node [style=none] (29) at (5, 1.25) {$\scriptstyle T_6$};
		\node [style=none] (30) at (7, 1.25) {$\scriptstyle T_7$};
		\node [style=small-square] (31) at (-4, 0.75) {$\scriptstyle U_B$};
		\node [style=small-square] (32) at (0, 0.75) {$\scriptstyle U_A$};
		\node [style=small-square] (33) at (4, 0.75) {$\scriptstyle U_B$};
		\node [style=small-square] (34) at (-6, -0.75) {$\mathbf I$};
		\node [style=none] (35) at (-7.5, -0.75) {};
		\node [style=none] (36) at (-6.25, -0.75) {};
		\node [style=none] (37) at (-5.75, -0.75) {};
		\node [style=none] (38) at (-4.25, -0.75) {};
		\node [style=small-square] (39) at (-2, -0.75) {$\mathbf I$};
		\node [style=none] (40) at (-3.75, -0.75) {};
		\node [style=none] (41) at (-2.25, -0.75) {};
		\node [style=none] (42) at (-1.75, -0.75) {};
		\node [style=none] (43) at (-0.25, -0.75) {};
		\node [style=small-square] (44) at (2, -0.75) {$\mathbf I$};
		\node [style=none] (45) at (0.25, -0.75) {};
		\node [style=none] (46) at (1.75, -0.75) {};
		\node [style=none] (47) at (2.25, -0.75) {};
		\node [style=none] (48) at (3.75, -0.75) {};
		\node [style=small-square] (49) at (6, -0.75) {$\mathbf I$};
		\node [style=none] (50) at (4.25, -0.75) {};
		\node [style=none] (51) at (5.75, -0.75) {};
		\node [style=none] (52) at (6.25, -0.75) {};
		\node [style=none] (53) at (7.5, -0.75) {};
		\node [style=small-square] (55) at (0, -0.75) {$\mathbf I$};
		\node [style=none] (57) at (-7, -0.25) {$\scriptstyle C_0$};
		\node [style=none] (58) at (-5, -0.25) {$\scriptstyle C_1$};
		\node [style=none] (59) at (-3, -0.25) {$\scriptstyle C_2$};
		\node [style=none] (60) at (-1, -0.25) {$\scriptstyle C_3$};
		\node [style=none] (61) at (1, -0.25) {$\scriptstyle C_4$};
		\node [style=none] (62) at (3, -0.25) {$\scriptstyle C_5$};
		\node [style=none] (63) at (5, -0.25) {$\scriptstyle C_6$};
		\node [style=none] (64) at (7, -0.25) {$\scriptstyle C_7$};
		\node [style=none] (67) at (-8, -1.25) {};
		\node [style=none] (68) at (-8, 1.25) {};
		\node [style=none] (69) at (-7.5, -0.75) {};
		\node [style=none] (70) at (-7.5, 0.75) {};
		\node [style=none] (71) at (-8, 0) {$\phi$};
		\node [style=none] (72) at (-7.5, 0.75) {};
		\node [style=none] (73) at (7.5, 0.75) {};
		\node [style=none] (74) at (7.5, -0.75) {};
		\node [style=none] (75) at (8, 1.25) {};
		\node [style=none] (76) at (8, -1.25) {};
		\node [style=none] (79) at (7.5, -0.75) {};
		\node [style=none] (80) at (8, 0) {$\psi$};
		\node [style=none] (81) at (7.5, -0.75) {};
		\node [style=none] (82) at (7.5, -0.75) {};
		\node [style=none] (83) at (-7.5, 1.25) {};
		\node [style=none] (84) at (-7.5, -1.25) {};
		\node [style=none] (85) at (7.5, -1.25) {};
		\node [style=none] (86) at (7.5, 1.25) {};
		\node [style=new style 0] (87) at (-4, -0.75) {};
		\node [style=none] (88) at (-4, 0.5) {};
		\node [style=new style 1] (89) at (4, -0.75) {};
	\end{pgfonlayer}
	\begin{pgfonlayer}{edgelayer}
		\draw [style=syst-wire] (4.center) to (5.center);
		\draw [style=syst-wire] (6.center) to (7.center);
		\draw [style=syst-wire] (11.center) to (12.center);
		\draw [style=syst-wire] (13.center) to (14.center);
		\draw [style=syst-wire] (18.center) to (19.center);
		\draw [style=syst-wire] (20.center) to (21.center);
		\draw [style=syst-wire] (25.center) to (26.center);
		\draw [style=syst-wire] (27.center) to (28.center);
		\draw [style=syst-wire] (35.center) to (36.center);
		\draw [style=syst-wire] (37.center) to (38.center);
		\draw [style=syst-wire] (40.center) to (41.center);
		\draw [style=syst-wire] (42.center) to (43.center);
		\draw [style=syst-wire] (45.center) to (46.center);
		\draw [style=syst-wire] (47.center) to (48.center);
		\draw [style=syst-wire] (50.center) to (51.center);
		\draw [style=syst-wire] (52.center) to (53.center);
		\draw [style=box-wire, bend right=90] (68.center) to (67.center);
		\draw [style=box-wire] (68.center) to (83.center);
		\draw [style=box-wire] (83.center) to (84.center);
		\draw [style=box-wire] (84.center) to (67.center);
		\draw [style=box-wire, bend left=90] (75.center) to (76.center);
		\draw [style=box-wire] (75.center) to (86.center);
		\draw [style=box-wire] (86.center) to (85.center);
		\draw [style=box-wire] (85.center) to (76.center);
		\draw [style=syst-wire] (87) to (88.center);
		\draw [style=syst-wire] (33) to (89);
	\end{pgfonlayer}
\end{tikzpicture}

%% file: timecircuit_B.tikz
\begin{tikzpicture}
	\begin{pgfonlayer}{nodelayer}
		\node [style=small-square] (0) at (-6, 0.75) {$\mathbf I$};
		\node [style=none] (2) at (-6.25, 0.75) {};
		\node [style=none] (3) at (-5.75, 0.75) {};
		\node [style=none] (4) at (-4.5, 0.75) {};
		\node [style=none] (5) at (-7, 1.25) {$\scriptstyle T_0$};
		\node [style=none] (6) at (-5, 1.25) {$\scriptstyle T_1$};
		\node [style=small-square] (7) at (-2, 0.75) {$\mathbf I$};
		\node [style=none] (8) at (-3.5, 0.75) {};
		\node [style=none] (9) at (-2.25, 0.75) {};
		\node [style=none] (10) at (-1.75, 0.75) {};
		\node [style=none] (11) at (-0.5, 0.75) {};
		\node [style=none] (12) at (-3, 1.25) {$\scriptstyle T_2$};
		\node [style=none] (13) at (-1, 1.25) {$\scriptstyle T_3$};
		\node [style=small-square] (14) at (2, 0.75) {$\mathbf I$};
		\node [style=none] (15) at (0.5, 0.75) {};
		\node [style=none] (16) at (1.75, 0.75) {};
		\node [style=none] (17) at (2.25, 0.75) {};
		\node [style=none] (18) at (3.5, 0.75) {};
		\node [style=none] (19) at (1, 1.25) {$\scriptstyle T_4$};
		\node [style=none] (20) at (3, 1.25) {$\scriptstyle T_5$};
		\node [style=small-square] (21) at (6, 0.75) {$\mathbf I$};
		\node [style=none] (22) at (4.5, 0.75) {};
		\node [style=none] (23) at (5.75, 0.75) {};
		\node [style=none] (24) at (6.25, 0.75) {};
		\node [style=none] (26) at (5, 1.25) {$\scriptstyle T_6$};
		\node [style=none] (27) at (7, 1.25) {$\scriptstyle T_7$};
		\node [style=small-square] (28) at (-4, 0.75) {$\scriptstyle U_A$};
		\node [style=small-square] (29) at (0, 0.75) {$\scriptstyle U_B$};
		\node [style=small-square] (30) at (4, 0.75) {$\scriptstyle U_A$};
		\node [style=small-square] (31) at (-6, -0.75) {$\mathbf I$};
		\node [style=none] (33) at (-6.25, -0.75) {};
		\node [style=none] (34) at (-5.75, -0.75) {};
		\node [style=none] (35) at (-4.25, -0.75) {};
		\node [style=small-square] (36) at (-2, -0.75) {$\mathbf I$};
		\node [style=none] (37) at (-3.75, -0.75) {};
		\node [style=none] (38) at (-2.25, -0.75) {};
		\node [style=none] (39) at (-1.75, -0.75) {};
		\node [style=none] (40) at (-0.25, -0.75) {};
		\node [style=small-square] (41) at (2, -0.75) {$\mathbf I$};
		\node [style=none] (42) at (0.25, -0.75) {};
		\node [style=none] (43) at (1.75, -0.75) {};
		\node [style=none] (44) at (2.25, -0.75) {};
		\node [style=none] (45) at (3.75, -0.75) {};
		\node [style=small-square] (46) at (6, -0.75) {$\mathbf I$};
		\node [style=none] (47) at (4.25, -0.75) {};
		\node [style=none] (48) at (5.75, -0.75) {};
		\node [style=none] (49) at (6.25, -0.75) {};
		\node [style=small-square] (52) at (0, -0.75) {$\mathbf I$};
		\node [style=none] (54) at (-7, -0.25) {$\scriptstyle C_0$};
		\node [style=none] (55) at (-5, -0.25) {$\scriptstyle C_1$};
		\node [style=none] (56) at (-3, -0.25) {$\scriptstyle C_2$};
		\node [style=none] (57) at (-1, -0.25) {$\scriptstyle C_3$};
		\node [style=none] (58) at (1, -0.25) {$\scriptstyle C_4$};
		\node [style=none] (59) at (3, -0.25) {$\scriptstyle C_5$};
		\node [style=none] (60) at (5, -0.25) {$\scriptstyle C_6$};
		\node [style=none] (61) at (7, -0.25) {$\scriptstyle C_7$};
		\node [style=none] (80) at (-7.5, 0.75) {};
		\node [style=none] (81) at (-7.5, -0.75) {};
		\node [style=none] (82) at (-8, -1.25) {};
		\node [style=none] (83) at (-8, 1.25) {};
		\node [style=none] (84) at (-7.5, -0.75) {};
		\node [style=none] (85) at (-7.5, 0.75) {};
		\node [style=none] (86) at (-8, 0) {$\phi$};
		\node [style=none] (87) at (-7.5, 0.75) {};
		\node [style=none] (88) at (-7.5, 1.25) {};
		\node [style=none] (89) at (-7.5, -1.25) {};
		\node [style=none] (90) at (7.5, 0.75) {};
		\node [style=none] (91) at (7.5, -0.75) {};
		\node [style=none] (92) at (7.5, 0.75) {};
		\node [style=none] (93) at (7.5, -0.75) {};
		\node [style=none] (94) at (8, 1.25) {};
		\node [style=none] (95) at (8, -1.25) {};
		\node [style=none] (96) at (7.5, -0.75) {};
		\node [style=none] (97) at (8, 0) {$\psi$};
		\node [style=none] (98) at (7.5, -0.75) {};
		\node [style=none] (99) at (7.5, -0.75) {};
		\node [style=none] (100) at (7.5, -1.25) {};
		\node [style=none] (101) at (7.5, 1.25) {};
		\node [style=new style 0] (102) at (-4, -0.75) {};
		\node [style=none] (103) at (-4, 0.5) {};
		\node [style=new style 1] (104) at (4, -0.75) {};
	\end{pgfonlayer}
	\begin{pgfonlayer}{edgelayer}
		\draw [style=syst-wire] (3.center) to (4.center);
		\draw [style=syst-wire] (8.center) to (9.center);
		\draw [style=syst-wire] (10.center) to (11.center);
		\draw [style=syst-wire] (15.center) to (16.center);
		\draw [style=syst-wire] (17.center) to (18.center);
		\draw [style=syst-wire] (22.center) to (23.center);
		\draw [style=syst-wire] (34.center) to (35.center);
		\draw [style=syst-wire] (37.center) to (38.center);
		\draw [style=syst-wire] (39.center) to (40.center);
		\draw [style=syst-wire] (42.center) to (43.center);
		\draw [style=syst-wire] (44.center) to (45.center);
		\draw [style=syst-wire] (47.center) to (48.center);
		\draw [style=box-wire, bend right=90] (83.center) to (82.center);
		\draw [style=box-wire] (83.center) to (88.center);
		\draw [style=box-wire] (88.center) to (89.center);
		\draw [style=box-wire] (89.center) to (82.center);
		\draw [style=syst-wire] (87.center) to (2.center);
		\draw [style=syst-wire] (84.center) to (33.center);
		\draw [style=box-wire, bend left=90] (94.center) to (95.center);
		\draw [style=box-wire] (94.center) to (101.center);
		\draw [style=box-wire] (101.center) to (100.center);
		\draw [style=box-wire] (100.center) to (95.center);
		\draw [style=syst-wire] (24.center) to (92.center);
		\draw [style=syst-wire] (49.center) to (99.center);
		\draw [style=syst-wire] (102) to (103.center);
		\draw [style=syst-wire] (104) to (30);
	\end{pgfonlayer}
\end{tikzpicture}

%% file: timestep_target.tikz
\begin{tikzpicture}
	\begin{pgfonlayer}{nodelayer}
		\node [style=small-square] (0) at (0, 0) {$\mathbf I$};
		\node [style=none] (1) at (-1.5, 0) {};
		\node [style=none] (2) at (-0.25, 0) {};
		\node [style=none] (3) at (0.25, 0) {};
		\node [style=none] (4) at (1.75, 0) {};
		\node [style=none] (5) at (-1, 0.5) {$\scriptstyle T_{i}$};
		\node [style=none] (6) at (1.25, 0.5) {$\scriptstyle T_{i+1}$};
		\node [style=none] (15) at (-1.5, 0) {};
		\node [style=none] (16) at (-1.5, 0) {};
		\node [style=none] (17) at (-1.5, 0.5) {};
	\end{pgfonlayer}
	\begin{pgfonlayer}{edgelayer}
		\draw [style=syst-wire] (1.center) to (2.center);
		\draw [style=syst-wire] (3.center) to (4.center);
	\end{pgfonlayer}
\end{tikzpicture}

%% file: timestep_control.tikz
\begin{tikzpicture}
	\begin{pgfonlayer}{nodelayer}
		\node [style=small-square] (0) at (0, 0) {$\mathbf I$};
		\node [style=none] (1) at (-1.5, 0) {};
		\node [style=none] (2) at (-0.25, 0) {};
		\node [style=none] (3) at (0.25, 0) {};
		\node [style=none] (4) at (1.75, 0) {};
		\node [style=none] (5) at (-1, 0.5) {$\scriptstyle C_i$};
		\node [style=none] (6) at (1.25, 0.5) {$\scriptstyle C_{i+1}$};
		\node [style=none] (7) at (-1.5, 0) {};
	\end{pgfonlayer}
	\begin{pgfonlayer}{edgelayer}
		\draw [style=syst-wire] (1.center) to (2.center);
		\draw [style=syst-wire] (3.center) to (4.center);
	\end{pgfonlayer}
\end{tikzpicture}

%% file: fragments_timecircuit_A.tikz
\begin{tikzpicture}
	\begin{pgfonlayer}{nodelayer}
		\node [style=small-square] (0) at (-9, 0.75) {$\mathbf I$};
		\node [style=none] (1) at (-11.5, 0.75) {};
		\node [style=none] (2) at (-10.75, 0.75) {};
		\node [style=none] (3) at (-7.25, 0.75) {};
		\node [style=none] (4) at (-6.5, 0.75) {};
		\node [style=none] (5) at (-11, 1.25) {$\scriptstyle T_0$};
		\node [style=none] (6) at (-7, 1.25) {$\scriptstyle T_1$};
		\node [style=none] (8) at (-5.5, 0.75) {};
		\node [style=none] (9) at (-4.75, 0.75) {};
		\node [style=none] (10) at (-1.25, 0.75) {};
		\node [style=none] (11) at (-0.5, 0.75) {};
		\node [style=none] (12) at (-5, 1.25) {$\scriptstyle T_2$};
		\node [style=none] (13) at (-1, 1.25) {$\scriptstyle T_3$};
		\node [style=none] (15) at (0.5, 0.75) {};
		\node [style=none] (16) at (1.25, 0.75) {};
		\node [style=none] (17) at (4.75, 0.75) {};
		\node [style=none] (18) at (5.5, 0.75) {};
		\node [style=none] (19) at (1, 1.25) {$\scriptstyle T_4$};
		\node [style=none] (20) at (5, 1.25) {$\scriptstyle T_5$};
		\node [style=none] (22) at (6.5, 0.75) {};
		\node [style=none] (23) at (7.25, 0.75) {};
		\node [style=none] (24) at (10.75, 0.75) {};
		\node [style=none] (25) at (11.5, 0.75) {};
		\node [style=none] (26) at (7, 1.25) {$\scriptstyle T_6$};
		\node [style=none] (27) at (11, 1.25) {$\scriptstyle T_7$};
		\node [style=small-square] (28) at (-6, 0.75) {$\scriptstyle U_B$};
		\node [style=small-square] (29) at (0, 0.75) {$\scriptstyle U_A$};
		\node [style=small-square] (30) at (6, 0.75) {$\scriptstyle U_B$};
		\node [style=small-square] (31) at (-9, -0.75) {$\mathbf I$};
		\node [style=none] (32) at (-11.5, -0.75) {};
		\node [style=none] (33) at (-10.75, -0.75) {};
		\node [style=none] (34) at (-7.25, -0.75) {};
		\node [style=none] (35) at (-6.25, -0.75) {};
		\node [style=none] (37) at (-5.75, -0.75) {};
		\node [style=none] (38) at (-4.75, -0.75) {};
		\node [style=none] (39) at (-1.25, -0.75) {};
		\node [style=none] (40) at (-0.25, -0.75) {};
		\node [style=none] (42) at (0.25, -0.75) {};
		\node [style=none] (43) at (1.25, -0.75) {};
		\node [style=none] (44) at (4.75, -0.75) {};
		\node [style=none] (45) at (5.75, -0.75) {};
		\node [style=none] (47) at (6.25, -0.75) {};
		\node [style=none] (48) at (7.25, -0.75) {};
		\node [style=none] (49) at (10.75, -0.75) {};
		\node [style=none] (50) at (11.5, -0.75) {};
		\node [style=small-square] (51) at (0, -0.75) {$\mathbf I$};
		\node [style=none] (52) at (-11, -0.25) {$\scriptstyle C_0$};
		\node [style=none] (53) at (-7, -0.25) {$\scriptstyle C_1$};
		\node [style=none] (54) at (-5, -0.25) {$\scriptstyle C_2$};
		\node [style=none] (55) at (-1, -0.25) {$\scriptstyle C_3$};
		\node [style=none] (56) at (1, -0.25) {$\scriptstyle C_4$};
		\node [style=none] (57) at (5, -0.25) {$\scriptstyle C_5$};
		\node [style=none] (58) at (7, -0.25) {$\scriptstyle C_6$};
		\node [style=none] (59) at (11, -0.25) {$\scriptstyle C_7$};
		\node [style=none] (60) at (-12, -1.25) {};
		\node [style=none] (61) at (-12, 1.25) {};
		\node [style=none] (62) at (-11.5, -0.75) {};
		\node [style=none] (63) at (-11.5, 0.75) {};
		\node [style=none] (64) at (-12, 0) {$\phi$};
		\node [style=none] (65) at (-11.5, 0.75) {};
		\node [style=none] (66) at (11.5, 0.75) {};
		\node [style=none] (67) at (11.5, -0.75) {};
		\node [style=none] (68) at (12, 1.25) {};
		\node [style=none] (69) at (12, -1.25) {};
		\node [style=none] (70) at (11.5, -0.75) {};
		\node [style=none] (71) at (12, 0) {$\psi$};
		\node [style=none] (72) at (11.5, -0.75) {};
		\node [style=none] (73) at (11.5, -0.75) {};
		\node [style=none] (74) at (-11.5, 1.25) {};
		\node [style=none] (75) at (-11.5, -1.25) {};
		\node [style=none] (76) at (11.5, -1.25) {};
		\node [style=none] (77) at (11.5, 1.25) {};
		\node [style=new style 0] (78) at (-6, -0.75) {};
		\node [style=none] (79) at (-6, 0.5) {};
		\node [style=new style 1] (80) at (6, -0.75) {};
		\node [style=none] (81) at (-10.25, 0.75) {};
		\node [style=none] (82) at (-7.75, 0.75) {};
		\node [style=none] (83) at (-10.25, -0.75) {};
		\node [style=none] (84) at (-7.75, -0.75) {};
		\node [style=small-square] (85) at (-3, 0.75) {$\mathbf I$};
		\node [style=small-square] (86) at (-3, -0.75) {$\mathbf I$};
		\node [style=none] (87) at (-4.25, 0.75) {};
		\node [style=none] (88) at (-1.75, 0.75) {};
		\node [style=none] (89) at (-4.25, -0.75) {};
		\node [style=none] (90) at (-1.75, -0.75) {};
		\node [style=small-square] (91) at (3, 0.75) {$\mathbf I$};
		\node [style=small-square] (92) at (3, -0.75) {$\mathbf I$};
		\node [style=none] (93) at (1.75, 0.75) {};
		\node [style=none] (94) at (4.25, 0.75) {};
		\node [style=none] (95) at (1.75, -0.75) {};
		\node [style=none] (96) at (4.25, -0.75) {};
		\node [style=small-square] (97) at (9, 0.75) {$\mathbf I$};
		\node [style=small-square] (98) at (9, -0.75) {$\mathbf I$};
		\node [style=none] (99) at (7.75, 0.75) {};
		\node [style=none] (100) at (10.25, 0.75) {};
		\node [style=none] (101) at (7.75, -0.75) {};
		\node [style=none] (102) at (10.25, -0.75) {};
		\node [style=none] (103) at (-10.5, 2) {};
		\node [style=none] (104) at (-10.5, -1.5) {};
		\node [style=none] (105) at (-7.5, 2) {};
		\node [style=none] (106) at (-7.5, -1.5) {};
		\node [style=none] (107) at (-4.5, 2) {};
		\node [style=none] (108) at (-4.5, -1.5) {};
		\node [style=none] (109) at (-1.5, 2) {};
		\node [style=none] (110) at (-1.5, -1.5) {};
		\node [style=none] (111) at (1.5, 2) {};
		\node [style=none] (112) at (1.5, -1.5) {};
		\node [style=none] (113) at (4.5, 2) {};
		\node [style=none] (114) at (4.5, -1.5) {};
		\node [style=none] (115) at (7.5, 2) {};
		\node [style=none] (116) at (7.5, -1.5) {};
		\node [style=none] (117) at (10.5, 2) {};
		\node [style=none] (118) at (10.5, -1.5) {};
		\node [style=none] (119) at (-10, 1.25) {$\scriptstyle T_0$};
		\node [style=none] (120) at (-8, 1.25) {$\scriptstyle T_1$};
		\node [style=none] (121) at (-4, 1.25) {$\scriptstyle T_2$};
		\node [style=none] (122) at (-2, 1.25) {$\scriptstyle T_3$};
		\node [style=none] (123) at (-8, -0.25) {$\scriptstyle C_1$};
		\node [style=none] (124) at (-4, -0.25) {$\scriptstyle C_2$};
		\node [style=none] (125) at (-10, -0.25) {$\scriptstyle C_0$};
		\node [style=none] (126) at (-2, -0.25) {$\scriptstyle C_3$};
		\node [style=none] (127) at (2, -0.25) {$\scriptstyle C_4$};
		\node [style=none] (128) at (4, -0.25) {$\scriptstyle C_5$};
		\node [style=none] (129) at (8, -0.25) {$\scriptstyle C_6$};
		\node [style=none] (130) at (10, -0.25) {$\scriptstyle C_7$};
		\node [style=none] (131) at (2, 1.25) {$\scriptstyle T_4$};
		\node [style=none] (132) at (4, 1.25) {$\scriptstyle T_5$};
		\node [style=none] (133) at (8, 1.25) {$\scriptstyle T_6$};
		\node [style=none] (134) at (10, 1.25) {$\scriptstyle T_7$};
		\node [style=none] (135) at (-10.5, -2) {$t_0$};
		\node [style=none] (136) at (-7.5, -2) {$t_1$};
		\node [style=none] (137) at (-4.5, -2) {$t_2$};
		\node [style=none] (138) at (-1.5, -2) {$t_3$};
		\node [style=none] (139) at (1.5, -2) {$t_4$};
		\node [style=none] (140) at (4.5, -2) {$t_5$};
		\node [style=none] (141) at (7.5, -2) {$t_6$};
		\node [style=none] (142) at (10.5, -2) {$t_7$};
	\end{pgfonlayer}
	\begin{pgfonlayer}{edgelayer}
		\draw [style=syst-wire] (1.center) to (2.center);
		\draw [style=syst-wire] (3.center) to (4.center);
		\draw [style=syst-wire] (8.center) to (9.center);
		\draw [style=syst-wire] (10.center) to (11.center);
		\draw [style=syst-wire] (15.center) to (16.center);
		\draw [style=syst-wire] (17.center) to (18.center);
		\draw [style=syst-wire] (22.center) to (23.center);
		\draw [style=syst-wire] (24.center) to (25.center);
		\draw [style=syst-wire] (32.center) to (33.center);
		\draw [style=syst-wire] (34.center) to (35.center);
		\draw [style=syst-wire] (37.center) to (38.center);
		\draw [style=syst-wire] (39.center) to (40.center);
		\draw [style=syst-wire] (42.center) to (43.center);
		\draw [style=syst-wire] (44.center) to (45.center);
		\draw [style=syst-wire] (47.center) to (48.center);
		\draw [style=syst-wire] (49.center) to (50.center);
		\draw [style=box-wire, bend right=90] (61.center) to (60.center);
		\draw [style=box-wire] (61.center) to (74.center);
		\draw [style=box-wire] (74.center) to (75.center);
		\draw [style=box-wire] (75.center) to (60.center);
		\draw [style=box-wire, bend left=90] (68.center) to (69.center);
		\draw [style=box-wire] (68.center) to (77.center);
		\draw [style=box-wire] (77.center) to (76.center);
		\draw [style=box-wire] (76.center) to (69.center);
		\draw [style=syst-wire] (78) to (79.center);
		\draw [style=syst-wire] (30) to (80);
		\draw [style=syst-wire] (81.center) to (82.center);
		\draw [style=syst-wire] (83.center) to (84.center);
		\draw [style=syst-wire] (87.center) to (88.center);
		\draw [style=syst-wire] (89.center) to (90.center);
		\draw [style=syst-wire] (93.center) to (94.center);
		\draw [style=syst-wire] (95.center) to (96.center);
		\draw [style=syst-wire] (99.center) to (100.center);
		\draw [style=syst-wire] (101.center) to (102.center);
		\draw [style=contr-wire] (104.center) to (103.center);
		\draw [style=contr-wire] (106.center) to (105.center);
		\draw [style=contr-wire] (108.center) to (107.center);
		\draw [style=contr-wire] (110.center) to (109.center);
		\draw [style=contr-wire] (112.center) to (111.center);
		\draw [style=contr-wire] (114.center) to (113.center);
		\draw [style=contr-wire] (116.center) to (115.center);
		\draw [style=contr-wire] (118.center) to (117.center);
	\end{pgfonlayer}
\end{tikzpicture}

%% file: trace_fragments_timecircuit_A.tikz
\begin{tikzpicture}
	\begin{pgfonlayer}{nodelayer}
		\node [style=small-square] (0) at (-6.5, 0.75) {$\mathbf I$};
		\node [style=none] (1) at (-9, 0.75) {};
		\node [style=none] (2) at (-8.25, 0.75) {};
		\node [style=none] (3) at (-6, 0.75) {};
		\node [style=none] (4) at (-5.25, 0.75) {};
		\node [style=none] (5) at (-8.5, 1.25) {$\scriptstyle T_0$};
		\node [style=none] (7) at (-4.25, 0.75) {};
		\node [style=none] (8) at (-4, 0.75) {};
		\node [style=none] (9) at (-1.25, 0.75) {};
		\node [style=none] (10) at (-0.5, 0.75) {};
		\node [style=none] (12) at (-1, 1.25) {$\scriptstyle T_3$};
		\node [style=none] (13) at (0.5, 0.75) {};
		\node [style=none] (14) at (1.25, 0.75) {};
		\node [style=none] (15) at (3.5, 0.75) {};
		\node [style=none] (16) at (4.25, 0.75) {};
		\node [style=none] (17) at (1, 1.25) {$\scriptstyle T_4$};
		\node [style=none] (19) at (5.25, 0.75) {};
		\node [style=none] (20) at (6.5, 0.75) {};
		\node [style=none] (21) at (8.25, 0.75) {};
		\node [style=none] (22) at (9, 0.75) {};
		\node [style=none] (24) at (8.5, 1.25) {$\scriptstyle T_7$};
		\node [style=small-square] (25) at (-4.75, 0.75) {$\scriptstyle U_B$};
		\node [style=small-square] (26) at (0, 0.75) {$\scriptstyle U_A$};
		\node [style=small-square] (27) at (4.75, 0.75) {$\scriptstyle U_B$};
		\node [style=small-square] (28) at (-6.5, -0.75) {$\mathbf I$};
		\node [style=none] (29) at (-9, -0.75) {};
		\node [style=none] (30) at (-8.25, -0.75) {};
		\node [style=none] (31) at (-6, -0.75) {};
		\node [style=none] (32) at (-5, -0.75) {};
		\node [style=none] (33) at (-4.5, -0.75) {};
		\node [style=none] (34) at (-4, -0.75) {};
		\node [style=none] (35) at (-1.25, -0.75) {};
		\node [style=none] (36) at (-0.25, -0.75) {};
		\node [style=none] (37) at (0.25, -0.75) {};
		\node [style=none] (38) at (1.75, -0.75) {};
		\node [style=none] (39) at (3.5, -0.75) {};
		\node [style=none] (40) at (4.5, -0.75) {};
		\node [style=none] (41) at (5, -0.75) {};
		\node [style=none] (42) at (6.5, -0.75) {};
		\node [style=none] (43) at (8.25, -0.75) {};
		\node [style=none] (44) at (9, -0.75) {};
		\node [style=small-square] (45) at (0, -0.75) {$\mathbf I$};
		\node [style=none] (46) at (-8.5, -0.25) {$\scriptstyle C_0$};
		\node [style=none] (53) at (8.5, -0.25) {$\scriptstyle C_7$};
		\node [style=none] (54) at (-9.5, -1.25) {};
		\node [style=none] (55) at (-9.5, 1.25) {};
		\node [style=none] (56) at (-9, -0.75) {};
		\node [style=none] (57) at (-9, 0.75) {};
		\node [style=none] (58) at (-9.5, 0) {$\phi$};
		\node [style=none] (59) at (-9, 0.75) {};
		\node [style=none] (60) at (9, 0.75) {};
		\node [style=none] (61) at (9, -0.75) {};
		\node [style=none] (62) at (9.5, 1.25) {};
		\node [style=none] (63) at (9.5, -1.25) {};
		\node [style=none] (64) at (9, -0.75) {};
		\node [style=none] (65) at (9.5, 0) {$\psi$};
		\node [style=none] (66) at (9, -0.75) {};
		\node [style=none] (67) at (9, -0.75) {};
		\node [style=none] (68) at (-9, 1.25) {};
		\node [style=none] (69) at (-9, -1.25) {};
		\node [style=none] (70) at (9, -1.25) {};
		\node [style=none] (71) at (9, 1.25) {};
		\node [style=new style 0] (72) at (-4.75, -0.75) {};
		\node [style=none] (73) at (-4.75, 0.5) {};
		\node [style=new style 1] (74) at (4.75, -0.75) {};
		\node [style=none] (75) at (-7.75, 0.75) {};
		\node [style=none] (76) at (-6, 0.75) {};
		\node [style=none] (77) at (-7.75, -0.75) {};
		\node [style=none] (78) at (-6, -0.75) {};
		\node [style=small-square] (79) at (-3, 0.75) {$\mathbf I$};
		\node [style=small-square] (80) at (-3, -0.75) {$\mathbf I$};
		\node [style=none] (81) at (-4, 0.75) {};
		\node [style=none] (82) at (-1.75, 0.75) {};
		\node [style=none] (83) at (-4, -0.75) {};
		\node [style=none] (84) at (-1.25, -0.75) {};
		\node [style=small-square] (85) at (3, 0.75) {$\mathbf I$};
		\node [style=small-square] (86) at (3, -0.75) {$\mathbf I$};
		\node [style=none] (87) at (1.75, 0.75) {};
		\node [style=none] (88) at (3.5, 0.75) {};
		\node [style=none] (89) at (1.75, -0.75) {};
		\node [style=none] (90) at (3.5, -0.75) {};
		\node [style=small-square] (91) at (6.5, 0.75) {$\mathbf I$};
		\node [style=small-square] (92) at (6.5, -0.75) {$\mathbf I$};
		\node [style=none] (93) at (6.5, 0.75) {};
		\node [style=none] (94) at (7.75, 0.75) {};
		\node [style=none] (95) at (6.5, -0.75) {};
		\node [style=none] (96) at (7.75, -0.75) {};
		\node [style=none] (97) at (-8, 2) {};
		\node [style=none] (98) at (-8, -1.5) {};
		\node [style=none] (103) at (-1.5, 2) {};
		\node [style=none] (104) at (-1.5, -1.5) {};
		\node [style=none] (105) at (1.5, 2) {};
		\node [style=none] (106) at (1.5, -1.5) {};
		\node [style=none] (111) at (8, 2) {};
		\node [style=none] (112) at (8, -1.5) {};
		\node [style=none] (113) at (-7.5, 1.25) {$\scriptstyle T_0$};
		\node [style=none] (116) at (-2, 1.25) {$\scriptstyle T_3$};
		\node [style=none] (119) at (-7.5, -0.25) {$\scriptstyle C_0$};
		\node [style=none] (124) at (7.5, -0.25) {$\scriptstyle C_7$};
		\node [style=none] (125) at (2, 1.25) {$\scriptstyle T_4$};
		\node [style=none] (128) at (7.5, 1.25) {$\scriptstyle T_7$};
		\node [style=none] (129) at (-8, -2) {$t_0$};
		\node [style=none] (132) at (-1.5, -2) {$t_3$};
		\node [style=none] (133) at (1.5, -2) {$t_4$};
		\node [style=none] (136) at (8, -2) {$t_7$};
	\end{pgfonlayer}
	\begin{pgfonlayer}{edgelayer}
		\draw [style=syst-wire] (1.center) to (2.center);
		\draw [style=syst-wire] (3.center) to (4.center);
		\draw [style=syst-wire] (7.center) to (8.center);
		\draw [style=syst-wire] (9.center) to (10.center);
		\draw [style=syst-wire] (13.center) to (14.center);
		\draw [style=syst-wire] (15.center) to (16.center);
		\draw [style=syst-wire] (19.center) to (20.center);
		\draw [style=syst-wire] (21.center) to (22.center);
		\draw [style=syst-wire] (29.center) to (30.center);
		\draw [style=syst-wire] (31.center) to (32.center);
		\draw [style=syst-wire] (33.center) to (34.center);
		\draw [style=syst-wire] (35.center) to (36.center);
		\draw [style=syst-wire] (37.center) to (38.center);
		\draw [style=syst-wire] (39.center) to (40.center);
		\draw [style=syst-wire] (41.center) to (42.center);
		\draw [style=syst-wire] (43.center) to (44.center);
		\draw [style=box-wire, bend right=90] (55.center) to (54.center);
		\draw [style=box-wire] (55.center) to (68.center);
		\draw [style=box-wire] (68.center) to (69.center);
		\draw [style=box-wire] (69.center) to (54.center);
		\draw [style=box-wire, bend left=90] (62.center) to (63.center);
		\draw [style=box-wire] (62.center) to (71.center);
		\draw [style=box-wire] (71.center) to (70.center);
		\draw [style=box-wire] (70.center) to (63.center);
		\draw [style=syst-wire] (72) to (73.center);
		\draw [style=syst-wire] (27) to (74);
		\draw [style=syst-wire] (75.center) to (76.center);
		\draw [style=syst-wire] (77.center) to (78.center);
		\draw [style=syst-wire] (81.center) to (82.center);
		\draw [style=syst-wire] (83.center) to (84.center);
		\draw [style=syst-wire] (87.center) to (88.center);
		\draw [style=syst-wire] (89.center) to (90.center);
		\draw [style=syst-wire] (93.center) to (94.center);
		\draw [style=syst-wire] (95.center) to (96.center);
		\draw [style=contr-wire] (98.center) to (97.center);
		\draw [style=contr-wire] (104.center) to (103.center);
		\draw [style=contr-wire] (106.center) to (105.center);
		\draw [style=contr-wire] (112.center) to (111.center);
	\end{pgfonlayer}
\end{tikzpicture}

%% file: fragments_causalframe_A.tikz
\begin{tikzpicture}
	\begin{pgfonlayer}{nodelayer}
		\node [style=none] (0) at (-7.5, 0.75) {};
		\node [style=none] (1) at (-6.75, 0.75) {};
		\node [style=none] (2) at (-7, 1.25) {$\scriptstyle T_0$};
		\node [style=none] (3) at (-1.25, -0.75) {};
		\node [style=none] (4) at (-0.5, -0.75) {};
		\node [style=none] (5) at (-1, -0.25) {$\scriptstyle T_3$};
		\node [style=none] (6) at (0.5, -0.75) {};
		\node [style=none] (7) at (1.25, -0.75) {};
		\node [style=none] (8) at (1, -0.25) {$\scriptstyle T_4$};
		\node [style=none] (9) at (6.75, 0.75) {};
		\node [style=none] (10) at (7.5, 0.75) {};
		\node [style=none] (11) at (7, 1.25) {$\scriptstyle T_7$};
		\node [style=small-square] (12) at (0, -0.75) {$\scriptstyle U_A$};
		\node [style=none] (13) at (-7.5, -0.75) {};
		\node [style=none] (14) at (-6.75, -0.75) {};
		\node [style=none] (15) at (6.75, -0.75) {};
		\node [style=none] (16) at (7.5, -0.75) {};
		\node [style=none] (17) at (-7, -0.25) {$\scriptstyle C_0$};
		\node [style=none] (18) at (7, -0.25) {$\scriptstyle C_7$};
		\node [style=none] (19) at (-8, -1.25) {};
		\node [style=none] (20) at (-8, 1.25) {};
		\node [style=none] (21) at (-7.5, -0.75) {};
		\node [style=none] (22) at (-7.5, 0.75) {};
		\node [style=none] (23) at (-8, 0) {$\phi$};
		\node [style=none] (24) at (-7.5, 0.75) {};
		\node [style=none] (25) at (7.5, 0.75) {};
		\node [style=none] (26) at (7.5, -0.75) {};
		\node [style=none] (27) at (8, 1.25) {};
		\node [style=none] (28) at (8, -1.25) {};
		\node [style=none] (29) at (7.5, -0.75) {};
		\node [style=none] (30) at (8, 0) {$\psi$};
		\node [style=none] (31) at (7.5, -0.75) {};
		\node [style=none] (32) at (7.5, -0.75) {};
		\node [style=none] (33) at (-7.5, 1.25) {};
		\node [style=none] (34) at (-7.5, -1.25) {};
		\node [style=none] (35) at (7.5, -1.25) {};
		\node [style=none] (36) at (7.5, 1.25) {};
		\node [style=none] (37) at (-6.5, 2) {};
		\node [style=none] (38) at (-6.5, -1.5) {};
		\node [style=none] (39) at (-1.5, 2) {};
		\node [style=none] (40) at (-1.5, -1.5) {};
		\node [style=none] (41) at (1.5, 2) {};
		\node [style=none] (42) at (1.5, -1.5) {};
		\node [style=none] (43) at (6.5, 2) {};
		\node [style=none] (44) at (6.5, -1.5) {};
		\node [style=none] (45) at (-6.5, -2) {$t_0$};
		\node [style=none] (46) at (-1.5, -2) {$t_3$};
		\node [style=none] (47) at (1.5, -2) {$t_4$};
		\node [style=none] (48) at (6.5, -2) {$t_7$};
		\node [style=none] (49) at (-5.25, 1.5) {};
		\node [style=none] (50) at (-2.75, 1.5) {};
		\node [style=none] (51) at (-5.25, -1.5) {};
		\node [style=none] (52) at (-2.75, -1.5) {};
		\node [style=none] (53) at (2.75, 1.5) {};
		\node [style=none] (54) at (5.25, 1.5) {};
		\node [style=none] (55) at (2.75, -1.5) {};
		\node [style=none] (56) at (5.25, -1.5) {};
		\node [style=none] (57) at (-2.75, 0.75) {};
		\node [style=none] (58) at (2.75, 0.75) {};
		\node [style=none] (59) at (1.75, -0.75) {};
		\node [style=none] (60) at (2.75, -0.75) {};
		\node [style=none] (61) at (-2.75, -0.75) {};
		\node [style=none] (62) at (-1.75, -0.75) {};
		\node [style=none] (63) at (-6.25, 0.75) {};
		\node [style=none] (64) at (-5.25, 0.75) {};
		\node [style=none] (65) at (5.25, 0.75) {};
		\node [style=none] (66) at (6.25, 0.75) {};
		\node [style=none] (67) at (5.25, -0.75) {};
		\node [style=none] (68) at (6.25, -0.75) {};
		\node [style=none] (69) at (-6.25, -0.75) {};
		\node [style=none] (70) at (-5.25, -0.75) {};
		\node [style=none] (71) at (-4, 0) {};
		\node [style=none] (72) at (4, 0) {};
		\node [style=none] (73) at (-4, 0) {$\scriptstyle\Phi_A(U_B)$};
		\node [style=none] (74) at (4, 0) {$\scriptstyle\Pi_A(U_B)$};
		\node [style=none] (75) at (2.25, -0.25) {};
		\node [style=none] (76) at (-2.25, -0.25) {$\scriptstyle T_3$};
		\node [style=none] (77) at (2.25, -0.25) {$\scriptstyle T_4$};
		\node [style=none] (78) at (0, 1.25) {$\scriptstyle E$};
		\node [style=none] (79) at (5.75, 1.25) {$\scriptstyle T_7$};
		\node [style=none] (80) at (-5.75, 1.25) {$\scriptstyle T_0$};
		\node [style=none] (81) at (-5.75, -0.25) {$\scriptstyle C_0$};
		\node [style=none] (82) at (5.75, -0.25) {$\scriptstyle C_7$};
	\end{pgfonlayer}
	\begin{pgfonlayer}{edgelayer}
		\draw [style=syst-wire] (0.center) to (1.center);
		\draw [style=syst-wire] (3.center) to (4.center);
		\draw [style=syst-wire] (6.center) to (7.center);
		\draw [style=syst-wire] (9.center) to (10.center);
		\draw [style=syst-wire] (13.center) to (14.center);
		\draw [style=syst-wire] (15.center) to (16.center);
		\draw [style=box-wire, bend right=90] (20.center) to (19.center);
		\draw [style=box-wire] (20.center) to (33.center);
		\draw [style=box-wire] (33.center) to (34.center);
		\draw [style=box-wire] (34.center) to (19.center);
		\draw [style=box-wire, bend left=90] (27.center) to (28.center);
		\draw [style=box-wire] (27.center) to (36.center);
		\draw [style=box-wire] (36.center) to (35.center);
		\draw [style=box-wire] (35.center) to (28.center);
		\draw [style=contr-wire] (38.center) to (37.center);
		\draw [style=contr-wire] (40.center) to (39.center);
		\draw [style=contr-wire] (42.center) to (41.center);
		\draw [style=contr-wire] (44.center) to (43.center);
		\draw [style=box-wire] (49.center) to (50.center);
		\draw [style=box-wire] (50.center) to (52.center);
		\draw [style=box-wire] (52.center) to (51.center);
		\draw [style=box-wire] (49.center) to (51.center);
		\draw [style=box-wire] (53.center) to (54.center);
		\draw [style=box-wire] (54.center) to (56.center);
		\draw [style=box-wire] (56.center) to (55.center);
		\draw [style=box-wire] (53.center) to (55.center);
		\draw [style=syst-wire] (57.center) to (58.center);
		\draw [style=syst-wire] (59.center) to (60.center);
		\draw [style=syst-wire] (61.center) to (62.center);
		\draw [style=syst-wire] (63.center) to (64.center);
		\draw [style=syst-wire] (65.center) to (66.center);
		\draw [style=syst-wire] (67.center) to (68.center);
		\draw [style=syst-wire] (69.center) to (70.center);
	\end{pgfonlayer}
\end{tikzpicture}

%% file: causalframe_A.tikz
\begin{tikzpicture}
	\begin{pgfonlayer}{nodelayer}
		\node [style=none] (0) at (-4.25, 1.5) {};
		\node [style=none] (1) at (-1.75, 1.5) {};
		\node [style=none] (2) at (-4.25, -1.5) {};
		\node [style=none] (3) at (-1.75, -1.5) {};
		\node [style=none] (4) at (1.75, 1.5) {};
		\node [style=none] (5) at (4.25, 1.5) {};
		\node [style=none] (6) at (1.75, -1.5) {};
		\node [style=none] (7) at (4.25, -1.5) {};
		\node [style=none] (8) at (-1.75, 0.75) {};
		\node [style=none] (9) at (1.75, 0.75) {};
		\node [style=none] (10) at (0.75, -0.75) {};
		\node [style=none] (11) at (1.75, -0.75) {};
		\node [style=none] (12) at (-1.75, -0.75) {};
		\node [style=none] (13) at (-0.75, -0.75) {};
		\node [style=none] (14) at (-5.25, 0.75) {};
		\node [style=none] (15) at (-4.25, 0.75) {};
		\node [style=none] (16) at (4.25, 0.75) {};
		\node [style=none] (17) at (5.25, 0.75) {};
		\node [style=none] (18) at (4.25, -0.75) {};
		\node [style=none] (19) at (5.25, -0.75) {};
		\node [style=none] (20) at (-5.25, -0.75) {};
		\node [style=none] (21) at (-4.25, -0.75) {};
		\node [style=none] (22) at (-3, 0) {};
		\node [style=none] (23) at (3, 0) {};
		\node [style=none] (24) at (-3, 0) {$\scriptstyle\Pi_A(U_B)$};
		\node [style=none] (25) at (3, 0) {$\scriptstyle\Phi_A(U_B)$};
		\node [style=none] (27) at (1.25, -0.25) {};
		\node [style=none] (28) at (-1.25, -0.25) {$\scriptstyle A_I$};
		\node [style=none] (29) at (1.25, -0.25) {$\scriptstyle A_O$};
		\node [style=none] (30) at (1.25, 1.25) {$\scriptstyle E_A$};
		\node [style=none] (30) at (-1.25, 1.25) {$\scriptstyle E_A$};
		\node [style=none] (31) at (4.75, 1.25) {$\scriptstyle F_S$};
		\node [style=none] (32) at (-4.75, 1.25) {$\scriptstyle P_S$};
		\node [style=none] (33) at (-4.75, -0.25) {$\scriptstyle P_C$};
		\node [style=none] (34) at (4.75, -0.25) {$\scriptstyle F_C$};
	\end{pgfonlayer}
	\begin{pgfonlayer}{edgelayer}
		\draw [style=box-wire] (0.center) to (1.center);
		\draw [style=box-wire] (1.center) to (3.center);
		\draw [style=box-wire] (3.center) to (2.center);
		\draw [style=box-wire] (0.center) to (2.center);
		\draw [style=box-wire] (4.center) to (5.center);
		\draw [style=box-wire] (5.center) to (7.center);
		\draw [style=box-wire] (7.center) to (6.center);
		\draw [style=box-wire] (4.center) to (6.center);
		\draw [style=syst-wire] (8.center) to (9.center);
		\draw [style=syst-wire] (10.center) to (11.center);
		\draw [style=syst-wire] (12.center) to (13.center);
		\draw [style=syst-wire] (14.center) to (15.center);
		\draw [style=syst-wire] (16.center) to (17.center);
		\draw [style=syst-wire] (18.center) to (19.center);
		\draw [style=syst-wire] (20.center) to (21.center);
	\end{pgfonlayer}
\end{tikzpicture}

%% file: gate_A.tikz
\begin{tikzpicture}
	\begin{pgfonlayer}{nodelayer}
		\node [style=small-square] (0) at (0, 0) {$\scriptstyle U_A$};
		\node [style=none] (1) at (0.5, 0) {};
		\node [style=none] (2) at (1.5, 0) {};
		\node [style=none] (3) at (-0.5, 0) {};
		\node [style=none] (4) at (-1.5, 0) {};
		\node [style=none] (5) at (-1.2, 0.5) {$\scriptstyle A_I$};
		\node [style=none] (6) at (1.2, 0.5) {$\scriptstyle A_O$};
	\end{pgfonlayer}
	\begin{pgfonlayer}{edgelayer}
		\draw [style=syst-wire] (1.center) to (2.center);
		\draw [style=syst-wire] (4.center) to (3.center);
	\end{pgfonlayer}
\end{tikzpicture}

%% file: state.tikz
\begin{tikzpicture}
	\begin{pgfonlayer}{nodelayer}
		\node [style=none] (5) at (0, 0) {};
		\node [style=none] (6) at (0.5, 0.5) {$\scriptstyle P_S$};
		\node [style=none] (7) at (1, 0) {};
		\node [style=none] (8) at (0, 0.5) {};
		\node [style=none] (9) at (-0.75, 0.5) {};
		\node [style=none] (10) at (-0.75, -0.5) {};
		\node [style=none] (11) at (0, -0.5) {};
		\node [style=none] (12) at (-0.75, 0.5) {};
		\node [style=none] (13) at (-0.5, 0) {};
		\node [style=none] (14) at (-0.5, 0) {$\phi$};
	\end{pgfonlayer}
	\begin{pgfonlayer}{edgelayer}
		\draw [style=syst-wire] (7.center) to (5.center);
		\draw [style=box-wire] (8.center) to (12.center);
		\draw [style=box-wire] (8.center) to (11.center);
		\draw [style=box-wire] (11.center) to (10.center);
		\draw [style=box-wire, bend right=270, looseness=1.25] (10.center) to (12.center);
	\end{pgfonlayer}
\end{tikzpicture}

%% file: measure.tikz
\begin{tikzpicture}
	\begin{pgfonlayer}{nodelayer}
		\node [style=none] (0) at (0.75, 0.5) {};
		\node [style=none] (1) at (0.75, -0.5) {};
		\node [style=none] (2) at (0, -0.5) {};
		\node [style=none] (3) at (0, 0.5) {};
		\node [style=none] (4) at (0.75, 0.5) {};
		\node [style=none] (5) at (0, 0) {};
		\node [style=none] (6) at (-1, 0) {};
		\node [style=none] (7) at (0.5, 0) {};
		\node [style=none] (8) at (0.5, 0) {$\psi$};
		\node [style=none] (9) at (-0.5, 0.5) {};
		\node [style=none] (10) at (-0.5, 0.5) {$\scriptstyle F_S$};
	\end{pgfonlayer}
	\begin{pgfonlayer}{edgelayer}
		\draw [style=box-wire] (3.center) to (4.center);
		\draw [style=box-wire, bend left=90] (4.center) to (1.center);
		\draw [style=box-wire] (1.center) to (2.center);
		\draw [style=box-wire] (2.center) to (3.center);
		\draw [style=syst-wire] (6.center) to (5.center);
	\end{pgfonlayer}
\end{tikzpicture}

%% file: transformation_switch.tikz
\begin{tikzpicture}
	\begin{pgfonlayer}{nodelayer}
		\node [style=none] (1) at (-0.75, 0) {};
		\node [style=none] (2) at (0.75, 0) {};
		\node [style=none] (3) at (-0.75, -1.5) {};
		\node [style=none] (4) at (0.75, -1.5) {};
		\node [style=none] (5) at (-1, 3.25) {};
		\node [style=none] (6) at (1, 3.25) {};
		\node [style=none] (7) at (-1, 0.25) {};
		\node [style=none] (8) at (1, 0.25) {};
		\node [style=none] (15) at (2.5, -2.25) {};
		\node [style=none] (17) at (1, 2.75) {};
		\node [style=none] (18) at (2.5, 2.75) {};
		\node [style=none] (19) at (1, 1.75) {};
		\node [style=none] (20) at (2.5, 1.75) {};
		\node [style=none] (21) at (1, 0.75) {};
		\node [style=none] (22) at (2.5, 0.75) {};
		\node [style=none] (23) at (-4.5, 2.75) {};
		\node [style=none] (24) at (-1, 2.75) {};
		\node [style=none] (25) at (-2.5, 1.75) {};
		\node [style=none] (26) at (-1, 1.75) {};
		\node [style=none] (27) at (-2.5, 0.75) {};
		\node [style=none] (28) at (-1, 0.75) {};
		\node [style=none] (29) at (0.75, -0.75) {};
		\node [style=none] (30) at (2.5, -0.75) {};
		\node [style=none] (31) at (-2.5, -0.75) {};
		\node [style=none] (32) at (-0.75, -0.75) {};
		\node [style=none] (33) at (1.75, -0.25) {$\scriptstyle A_O$};
		\node [style=none] (34) at (1.75, -0.25) {};
		\node [style=none] (35) at (1.75, 1.25) {};
		\node [style=none] (36) at (1.75, 1.25) {$\scriptstyle A_I$};
		\node [style=none] (37) at (1.75, 2.25) {$\scriptstyle F_S$};
		\node [style=none] (38) at (1.75, 3.25) {$\scriptstyle F_C$};
		\node [style=none] (39) at (0, 1.75) {};
		\node [style=none] (40) at (0, 1.75) {$\scriptscriptstyle W(U_B)$};
		\node [style=none] (41) at (0, -0.75) {$\scriptscriptstyle U_A$};
		\node [style=none] (50) at (-1.75, 3.25) {$\scriptstyle P_C$};
		\node [style=none] (51) at (-1.75, 2.25) {$\scriptstyle P_S$};
		\node [style=none] (52) at (-1.75, 1.25) {$\scriptstyle A_O$};
		\node [style=none] (53) at (-1.75, -0.25) {$\scriptstyle A_I$};
		\node [style=none] (54) at (2.5, -2.75) {};
		\node [style=none] (55) at (2.5, 3.25) {};
		\node [style=none] (56) at (2.5, 3.25) {};
		\node [style=none] (57) at (4.5, -2.75) {};
		\node [style=none] (58) at (4.5, 3.25) {};
		\node [style=none] (59) at (3.5, 0.5) {};
		\node [style=none] (60) at (3.5, 0.5) {};
		\node [style=none] (61) at (3.5, 0.5) {$\scriptscriptstyle J_{A\to B}$};
		\node [style=none] (67) at (-4.5, -4) {};
		\node [style=none] (68) at (-4.5, 5) {};
		\node [style=none] (69) at (-4.5, 5) {};
		\node [style=none] (70) at (-2.5, -4) {};
		\node [style=none] (71) at (-2.5, 4) {};
		\node [style=none] (72) at (-3.5, 0.5) {};
		\node [style=none] (73) at (-3.5, 0.5) {};
		\node [style=none] (74) at (-3.5, 0.5) {$\scriptscriptstyle J_{A\to B}^\dagger$};
		\node [style=none] (75) at (-6, 2.75) {};
		\node [style=none] (76) at (-4.5, 2.75) {};
		\node [style=none] (77) at (-6, 1.5) {};
		\node [style=none] (78) at (-4.5, 1.5) {};
		\node [style=none] (79) at (-6, -1) {};
		\node [style=none] (80) at (-4.5, -1) {};
		\node [style=none] (81) at (-5.25, 2) {$\scriptstyle P_S'$};
		\node [style=none] (82) at (-5.25, -0.5) {$\scriptstyle B_O$};
		\node [style=none] (83) at (-6, -2.25) {};
		\node [style=none] (84) at (-4.5, -2.25) {};
		\node [style=none] (85) at (-5.25, -1.75) {$\scriptstyle B_I$};
		\node [style=none] (86) at (4.5, 2.75) {};
		\node [style=none] (87) at (6, 2.75) {};
		\node [style=none] (88) at (4.5, 1.5) {};
		\node [style=none] (89) at (9.5, 1.5) {};
		\node [style=none] (90) at (4.5, 0.25) {};
		\node [style=none] (91) at (9.5, 0.25) {};
		\node [style=none] (92) at (8.75, 2) {$\scriptstyle P_S'$};
		\node [style=none] (93) at (8.75, 0.75) {$\scriptstyle F_S'$};
		\node [style=none] (94) at (4.5, -1) {};
		\node [style=none] (95) at (9.5, -1) {};
		\node [style=none] (96) at (8.75, -0.5) {$\scriptstyle B_I$};
		\node [style=none] (97) at (-5.25, 3.25) {};
		\node [style=none] (98) at (-5.25, 3.25) {$\scriptstyle P_C$};
		\node [style=none] (99) at (-6, 0.25) {};
		\node [style=none] (100) at (-4.5, 0.25) {};
		\node [style=none] (101) at (-5.25, 0.75) {$\scriptstyle F_S'$};
		\node [style=none] (102) at (4.5, -2.25) {};
		\node [style=none] (103) at (9.5, -2.25) {};
		\node [style=none] (104) at (8.75, -1.75) {$\scriptstyle B_O$};
		\node [style=none] (105) at (5.25, 3.25) {$\scriptstyle F_C$};
		\node [style=none] (106) at (0.75, -2.25) {};
		\node [style=none] (107) at (1.75, -1.75) {$\scriptstyle P_S$};
		\node [style=none] (108) at (2.5, -2.25) {};
		\node [style=none] (109) at (0.75, -1.75) {};
		\node [style=none] (110) at (0, -1.75) {};
		\node [style=none] (111) at (0, -2.75) {};
		\node [style=none] (112) at (0.75, -2.75) {};
		\node [style=none] (113) at (0, -1.75) {};
		\node [style=none] (114) at (0.25, -2.25) {};
		\node [style=none] (115) at (0.25, -2.25) {$\phi$};
		\node [style=none] (116) at (0, -3) {};
		\node [style=none] (117) at (0, -4) {};
		\node [style=none] (118) at (-0.75, -4) {};
		\node [style=none] (119) at (-0.75, -3) {};
		\node [style=none] (120) at (0, -3) {};
		\node [style=none] (121) at (-0.75, -3.5) {};
		\node [style=none] (122) at (-2.5, -3.5) {};
		\node [style=none] (123) at (-0.25, -3.5) {};
		\node [style=none] (124) at (-0.25, -3.5) {$\psi$};
		\node [style=none] (125) at (-1.75, -3) {};
		\node [style=none] (126) at (-1.75, -3) {$\scriptstyle F_S$};
		\node [style=none] (127) at (6, 4) {};
		\node [style=none] (128) at (6, 2.25) {};
		\node [style=none] (129) at (6, 2.25) {};
		\node [style=none] (130) at (8, 2.25) {};
		\node [style=none] (131) at (8, 5) {};
		\node [style=none] (132) at (8, 2.75) {};
		\node [style=none] (133) at (9.5, 2.75) {};
		\node [style=none] (134) at (8.75, 3.25) {$\scriptstyle F_C$};
	\end{pgfonlayer}
	\begin{pgfonlayer}{edgelayer}
		\draw [style=box-wire] (1.center) to (2.center);
		\draw [style=box-wire] (2.center) to (4.center);
		\draw [style=box-wire] (4.center) to (3.center);
		\draw [style=box-wire] (3.center) to (1.center);
		\draw [style=box-wire] (5.center) to (6.center);
		\draw [style=box-wire] (6.center) to (8.center);
		\draw [style=box-wire] (8.center) to (7.center);
		\draw [style=box-wire] (7.center) to (5.center);
		\draw [style=syst-wire] (17.center) to (18.center);
		\draw [style=syst-wire] (19.center) to (20.center);
		\draw [style=syst-wire] (21.center) to (22.center);
		\draw [style=syst-wire] (23.center) to (24.center);
		\draw [style=syst-wire] (25.center) to (26.center);
		\draw [style=syst-wire] (27.center) to (28.center);
		\draw [style=syst-wire] (29.center) to (30.center);
		\draw [style=syst-wire] (31.center) to (32.center);
		\draw [style=box-wire] (15.center) to (20.center);
		\draw [style=box-wire] (54.center) to (15.center);
		\draw [style=box-wire] (20.center) to (56.center);
		\draw [style=box-wire] (54.center) to (57.center);
		\draw [style=box-wire] (57.center) to (58.center);
		\draw [style=box-wire] (56.center) to (58.center);
		\draw [style=box-wire] (67.center) to (70.center);
		\draw [style=box-wire] (70.center) to (71.center);
		\draw [style=box-wire] (69.center) to (67.center);
		\draw [style=syst-wire] (75.center) to (76.center);
		\draw [style=syst-wire] (77.center) to (78.center);
		\draw [style=syst-wire] (79.center) to (80.center);
		\draw [style=syst-wire] (83.center) to (84.center);
		\draw [style=syst-wire] (86.center) to (87.center);
		\draw [style=syst-wire] (88.center) to (89.center);
		\draw [style=syst-wire] (90.center) to (91.center);
		\draw [style=syst-wire] (94.center) to (95.center);
		\draw [style=syst-wire] (99.center) to (100.center);
		\draw [style=syst-wire] (102.center) to (103.center);
		\draw [style=syst-wire] (108.center) to (106.center);
		\draw [style=box-wire] (109.center) to (113.center);
		\draw [style=box-wire] (109.center) to (112.center);
		\draw [style=box-wire] (112.center) to (111.center);
		\draw [style=box-wire, bend right=270, looseness=1.25] (111.center) to (113.center);
		\draw [style=box-wire] (119.center) to (120.center);
		\draw [style=box-wire, bend left=90] (120.center) to (117.center);
		\draw [style=box-wire] (117.center) to (118.center);
		\draw [style=box-wire] (118.center) to (119.center);
		\draw [style=syst-wire] (122.center) to (121.center);
		\draw [style=box-wire] (129.center) to (130.center);
		\draw [style=box-wire] (127.center) to (129.center);
		\draw [style=box-wire] (127.center) to (71.center);
		\draw [style=box-wire] (69.center) to (131.center);
		\draw [style=box-wire] (131.center) to (130.center);
		\draw [style=syst-wire] (132.center) to (133.center);
	\end{pgfonlayer}
\end{tikzpicture}

%% file: transformed_AtoB.tikz
\begin{tikzpicture}
	\begin{pgfonlayer}{nodelayer}
		\node [style=none] (0) at (-0.75, -1.75) {};
		\node [style=none] (1) at (0.75, -1.75) {};
		\node [style=none] (2) at (-0.75, -3.25) {};
		\node [style=none] (3) at (0.75, -3.25) {};
		\node [style=none] (4) at (0.75, -2.5) {};
		\node [style=none] (5) at (2.25, -2.5) {};
		\node [style=none] (6) at (-2.25, -2.5) {};
		\node [style=none] (7) at (-0.75, -2.5) {};
		\node [style=none] (10) at (0, -2.5) {$\scriptscriptstyle U_B$};
		\node [style=none] (11) at (-1.5, -2) {$\scriptstyle B_I$};
		\node [style=none] (12) at (-1.5, -1) {};
		\node [style=none] (13) at (1.5, -1) {};
		\node [style=none] (14) at (-1.5, 3.75) {};
		\node [style=none] (15) at (1.5, 3.75) {};
		\node [style=none] (16) at (-1.5, 3.25) {};
		\node [style=none] (17) at (-3, 3.25) {};
		\node [style=none] (18) at (-1.5, 3.25) {};
		\node [style=none] (19) at (-3, 2) {};
		\node [style=none] (20) at (-1.5, 2) {};
		\node [style=none] (21) at (-3, -0.5) {};
		\node [style=none] (22) at (-1.5, -0.5) {};
		\node [style=none] (23) at (-2.25, 2.5) {$\scriptstyle P_S'$};
		\node [style=none] (24) at (-2.25, 0) {$\scriptstyle B_O$};
		\node [style=none] (25) at (-2.25, 3.75) {};
		\node [style=none] (26) at (-2.25, 3.75) {$\scriptstyle P_C$};
		\node [style=none] (27) at (-3, 0.75) {};
		\node [style=none] (28) at (-1.5, 0.75) {};
		\node [style=none] (29) at (-2.25, 1.25) {$\scriptstyle F_S'$};
		\node [style=none] (30) at (3, 2) {};
		\node [style=none] (31) at (3, 0.75) {};
		\node [style=none] (32) at (2.25, 2.5) {$\scriptstyle P_S'$};
		\node [style=none] (33) at (2.25, 1.25) {$\scriptstyle F_S'$};
		\node [style=none] (34) at (3, -0.5) {};
		\node [style=none] (35) at (2.25, 0) {$\scriptstyle B_I$};
		\node [style=none] (36) at (1.5, 2) {};
		\node [style=none] (37) at (1.5, 3.25) {};
		\node [style=none] (38) at (3, 3.25) {};
		\node [style=none] (39) at (2.25, 3.75) {$\scriptstyle F_C$};
		\node [style=none] (40) at (1.5, 0.75) {};
		\node [style=none] (41) at (1.5, -0.5) {};
		\node [style=none] (42) at (0, 1.5) {$\scriptstyle \widetilde W$};
		\node [style=none] (43) at (1.5, -2) {$\scriptstyle B_O$};
	\end{pgfonlayer}
	\begin{pgfonlayer}{edgelayer}
		\draw [style=box-wire] (0.center) to (1.center);
		\draw [style=box-wire] (1.center) to (3.center);
		\draw [style=box-wire] (3.center) to (2.center);
		\draw [style=box-wire] (2.center) to (0.center);
		\draw [style=syst-wire] (4.center) to (5.center);
		\draw [style=syst-wire] (6.center) to (7.center);
		\draw [style=box-wire] (12.center) to (13.center);
		\draw [style=box-wire] (13.center) to (15.center);
		\draw [style=box-wire] (15.center) to (14.center);
		\draw [style=box-wire] (14.center) to (12.center);
		\draw [style=syst-wire] (17.center) to (18.center);
		\draw [style=syst-wire] (19.center) to (20.center);
		\draw [style=syst-wire] (21.center) to (22.center);
		\draw [style=syst-wire] (27.center) to (28.center);
		\draw [style=syst-wire] (37.center) to (38.center);
		\draw [style=syst-wire] (36.center) to (30.center);
		\draw [style=syst-wire] (40.center) to (31.center);
		\draw [style=syst-wire] (41.center) to (34.center);
	\end{pgfonlayer}
\end{tikzpicture}

%% file: timeframe_B.tikz
\begin{tikzpicture}
	\begin{pgfonlayer}{nodelayer}
		\node [style=none] (0) at (-5.25, 1.5) {};
		\node [style=none] (1) at (-1.75, 1.5) {};
		\node [style=none] (2) at (-5.25, -1.5) {};
		\node [style=none] (3) at (-1.75, -1.5) {};
		\node [style=none] (4) at (1.75, 1.5) {};
		\node [style=none] (5) at (5.25, 1.5) {};
		\node [style=none] (6) at (1.75, -1.5) {};
		\node [style=none] (7) at (5.25, -1.5) {};
		\node [style=none] (8) at (-1.75, 0.75) {};
		\node [style=none] (9) at (1.75, 0.75) {};
		\node [style=none] (10) at (0.75, -0.75) {};
		\node [style=none] (11) at (1.75, -0.75) {};
		\node [style=none] (12) at (-1.75, -0.75) {};
		\node [style=none] (13) at (-0.75, -0.75) {};
		\node [style=none] (14) at (-6.25, 1) {};
		\node [style=none] (15) at (-5.25, 1) {};
		\node [style=none] (16) at (5.25, 1) {};
		\node [style=none] (17) at (6.25, 1) {};
		\node [style=none] (18) at (5.25, -1) {};
		\node [style=none] (19) at (6.25, -1) {};
		\node [style=none] (20) at (-6.25, -1) {};
		\node [style=none] (21) at (-5.25, -1) {};
		\node [style=none] (22) at (-3.5, 0) {};
		\node [style=none] (23) at (3.5, 0) {};
		\node [style=none] (24) at (-3.5, 0) {$\scriptstyle\Pi_B(U_A, \phi , \psi)$};
		\node [style=none] (25) at (3.5, 0) {$\scriptstyle\Phi_B(U_A, \phi, \psi)$};
		\node [style=none] (26) at (1.25, -0.25) {};
		\node [style=none] (27) at (-1.25, -0.25) {$\scriptstyle B_I$};
		\node [style=none] (28) at (1.25, -0.25) {$\scriptstyle B_O$};
		\node [style=none] (29) at (0, 1.25) {$\scriptstyle E$};
		\node [style=none] (30) at (5.75, 1.5) {$\scriptstyle P_S'$};
		\node [style=none] (31) at (-5.75, 1.5) {$\scriptstyle P_S'$};
		\node [style=none] (32) at (-5.75, -0.5) {$\scriptstyle P_C$};
		\node [style=none] (33) at (5.75, -0.5) {$\scriptstyle F_C$};
		\node [style=none] (34) at (-5.25, 0) {};
		\node [style=none] (35) at (-6.25, 0) {};
		\node [style=none] (36) at (-5.75, 0.5) {$\scriptstyle F_S'$};
		\node [style=none] (37) at (6.25, 0) {};
		\node [style=none] (38) at (5.25, 0) {};
		\node [style=none] (39) at (5.75, 0.5) {$\scriptstyle F_S'$};
	\end{pgfonlayer}
	\begin{pgfonlayer}{edgelayer}
		\draw [style=box-wire] (0.center) to (1.center);
		\draw [style=box-wire] (1.center) to (3.center);
		\draw [style=box-wire] (3.center) to (2.center);
		\draw [style=box-wire] (0.center) to (2.center);
		\draw [style=box-wire] (4.center) to (5.center);
		\draw [style=box-wire] (5.center) to (7.center);
		\draw [style=box-wire] (7.center) to (6.center);
		\draw [style=box-wire] (4.center) to (6.center);
		\draw [style=syst-wire] (8.center) to (9.center);
		\draw [style=syst-wire] (10.center) to (11.center);
		\draw [style=syst-wire] (12.center) to (13.center);
		\draw [style=syst-wire] (14.center) to (15.center);
		\draw [style=syst-wire] (16.center) to (17.center);
		\draw [style=syst-wire] (18.center) to (19.center);
		\draw [style=syst-wire] (20.center) to (21.center);
		\draw [style=syst-wire] (35.center) to (34.center);
		\draw [style=syst-wire] (38.center) to (37.center);
	\end{pgfonlayer}
\end{tikzpicture}

%% file: spacetime_process.tikz
\begin{tikzpicture}
	\begin{pgfonlayer}{nodelayer}
		\node [style=none] (1) at (-7.5, -1) {};
		\node [style=none] (2) at (-4.5, -1) {};
		\node [style=none] (3) at (-7.5, -1.5) {};
		\node [style=none] (4) at (-9.5, -1.5) {};
		\node [style=none] (5) at (-9.5, 1.5) {};
		\node [style=none] (6) at (-7.5, 1.5) {};
		\node [style=none] (7) at (-7.5, 1) {};
		\node [style=none] (8) at (-6.5, 1) {};
		\node [style=none] (9) at (-7.5, 0) {};
		\node [style=none] (10) at (-4.5, 0) {};
		\node [style=none] (11) at (-2.5, -1) {};
		\node [style=none] (12) at (-2.5, -1.5) {};
		\node [style=none] (13) at (-4.5, -1.5) {};
		\node [style=none] (14) at (-4.5, 1.5) {};
		\node [style=none] (15) at (-2.5, 1.5) {};
		\node [style=none] (16) at (-2.5, 1) {};
		\node [style=none] (17) at (-2.5, 0) {};
		\node [style=none] (18) at (-5.5, 1) {};
		\node [style=none] (19) at (-4.5, 1) {};
		\node [style=none] (20) at (-2.5, -1) {};
		\node [style=none] (21) at (0.5, -1) {};
		\node [style=none] (22) at (-2.5, 0) {};
		\node [style=none] (23) at (0.5, 0) {};
		\node [style=none] (24) at (2.5, -1) {};
		\node [style=none] (25) at (5.5, -1) {};
		\node [style=none] (26) at (2.5, -1.5) {};
		\node [style=none] (27) at (0.5, -1.5) {};
		\node [style=none] (28) at (0.5, 1.5) {};
		\node [style=none] (29) at (2.5, 1.5) {};
		\node [style=none] (30) at (2.5, 1) {};
		\node [style=none] (31) at (3.5, 1) {};
		\node [style=none] (32) at (2.5, 0) {};
		\node [style=none] (33) at (5.5, 0) {};
		\node [style=none] (34) at (7.5, -1) {};
		\node [style=none] (35) at (7.5, -1.5) {};
		\node [style=none] (36) at (5.5, -1.5) {};
		\node [style=none] (37) at (5.5, 1.5) {};
		\node [style=none] (38) at (7.5, 1.5) {};
		\node [style=none] (39) at (7.5, 1) {};
		\node [style=none] (40) at (7.5, 0) {};
		\node [style=none] (41) at (4.5, 1) {};
		\node [style=none] (42) at (5.5, 1) {};
		\node [style=none] (43) at (7.5, -1) {};
		\node [style=none] (44) at (7.5, 0) {};
		\node [style=none] (45) at (-2.5, 1) {};
		\node [style=none] (46) at (-1.5, 1) {};
		\node [style=none] (47) at (-0.5, 1) {};
		\node [style=none] (48) at (0.5, 1) {};
		\node [style=none] (49) at (7.5, 1) {};
		\node [style=none] (50) at (8.5, 1) {};
		\node [style=none] (51) at (7.5, 0) {};
		\node [style=none] (52) at (8.5, 0) {};
		\node [style=none] (53) at (7.5, -1) {};
		\node [style=none] (54) at (8.5, -1) {};
		\node [style=none] (56) at (-10.5, 1) {};
		\node [style=none] (57) at (-10.5, 0) {};
		\node [style=none] (59) at (-10.5, 0) {};
		\node [style=none] (60) at (-10.5, 1) {};
		\node [style=none] (61) at (-9.5, 1) {};
		\node [style=none] (62) at (-10.5, 0) {};
		\node [style=none] (63) at (-9.5, 0) {};
		\node [style=none] (65) at (-9.5, -1) {};
		\node [style=none] (66) at (-10, 1.5) {$\scriptstyle R_0$};
		\node [style=none] (67) at (-7, 1.5) {$\scriptstyle T_1$};
		\node [style=none] (68) at (-5, 1.5) {$\scriptstyle T_2$};
		\node [style=none] (69) at (-2, 1.5) {$\scriptstyle T_3$};
		\node [style=none] (70) at (0, 1.5) {$\scriptstyle T_4$};
		\node [style=none] (71) at (3, 1.5) {$\scriptstyle T_5$};
		\node [style=none] (72) at (5, 1.5) {$\scriptstyle T_6$};
		\node [style=none] (73) at (8, 1.5) {$\scriptstyle R_7$};
		\node [style=none] (74) at (-10, 0.5) {$\scriptstyle S_0$};
		\node [style=none] (75) at (8.25, 0.5) {$\scriptstyle S_7$};
		\node [style=none] (80) at (-10, -0.5) {$\scriptstyle C_0$};
		\node [style=none] (85) at (-8.5, 0) {$\set T^{0\to 1}$};
		\node [style=none] (86) at (-3.5, 0) {$\set T^{2\to 3}$};
		\node [style=none] (87) at (1.5, 0) {$\set T^{4\to 5}$};
		\node [style=none] (88) at (6.5, 0) {$\set T^{6\to 7}$};
		\node [style=none] (89) at (8.25, -0.5) {$\scriptstyle C_7$};
		\node [style=none] (90) at (-7, 0.5) {$\scriptstyle S_1$};
		\node [style=none] (91) at (-2, 0.5) {$\scriptstyle S_3$};
		\node [style=none] (92) at (-5, 0.5) {$\scriptstyle S_2$};
		\node [style=none] (93) at (0, 0.5) {$\scriptstyle S_4$};
		\node [style=none] (94) at (3, 0.5) {$\scriptstyle S_5$};
		\node [style=none] (95) at (5, 0.5) {$\scriptstyle S_6$};
		\node [style=none] (96) at (-7, -0.5) {$\scriptstyle C_1$};
		\node [style=none] (97) at (-5, -0.5) {$\scriptstyle C_2$};
		\node [style=none] (98) at (-2, -0.5) {$\scriptstyle C_3$};
		\node [style=none] (99) at (0, -0.5) {$\scriptstyle C_4$};
		\node [style=none] (100) at (3, -0.5) {$\scriptstyle C_5$};
		\node [style=none] (101) at (5, -0.5) {$\scriptstyle C_6$};
		\node [style=none] (102) at (-10.5, -1) {};
		\node [style=none] (103) at (-10.5, -1) {};
		\node [style=none] (104) at (-10.5, -1) {};
		\node [style=none] (105) at (-9.5, -1) {};
	\end{pgfonlayer}
	\begin{pgfonlayer}{edgelayer}
		\draw [style=syst-wire] (2.center) to (1.center);
		\draw [style=box-wire] (6.center) to (5.center);
		\draw [style=box-wire] (5.center) to (4.center);
		\draw [style=box-wire] (4.center) to (3.center);
		\draw [style=box-wire] (3.center) to (6.center);
		\draw [style=syst-wire] (8.center) to (7.center);
		\draw [style=syst-wire] (10.center) to (9.center);
		\draw [style=box-wire] (15.center) to (14.center);
		\draw [style=box-wire] (14.center) to (13.center);
		\draw [style=box-wire] (13.center) to (12.center);
		\draw [style=box-wire] (12.center) to (15.center);
		\draw [style=syst-wire] (19.center) to (18.center);
		\draw [style=syst-wire] (21.center) to (20.center);
		\draw [style=syst-wire] (23.center) to (22.center);
		\draw [style=syst-wire] (25.center) to (24.center);
		\draw [style=box-wire] (29.center) to (28.center);
		\draw [style=box-wire] (28.center) to (27.center);
		\draw [style=box-wire] (27.center) to (26.center);
		\draw [style=box-wire] (26.center) to (29.center);
		\draw [style=syst-wire] (31.center) to (30.center);
		\draw [style=syst-wire] (33.center) to (32.center);
		\draw [style=box-wire] (38.center) to (37.center);
		\draw [style=box-wire] (37.center) to (36.center);
		\draw [style=box-wire] (36.center) to (35.center);
		\draw [style=box-wire] (35.center) to (38.center);
		\draw [style=syst-wire] (42.center) to (41.center);
		\draw [style=syst-wire] (46.center) to (45.center);
		\draw [style=syst-wire] (48.center) to (47.center);
		\draw [style=syst-wire] (50.center) to (49.center);
		\draw [style=syst-wire] (52.center) to (51.center);
		\draw [style=syst-wire] (54.center) to (53.center);
		\draw [style=syst-wire] (61.center) to (60.center);
		\draw [style=syst-wire] (63.center) to (62.center);
		\draw [style=syst-wire] (105.center) to (104.center);
	\end{pgfonlayer}
\end{tikzpicture}

%% file: timecircuit_fragment_B.tikz
\begin{tikzpicture}
	\begin{pgfonlayer}{nodelayer}
		\node [style=none] (0) at (-4, 2.5) {};
		\node [style=none] (1) at (-2, 2.5) {};
		\node [style=none] (2) at (-2, 0.5) {};
		\node [style=none] (3) at (-4, 0.5) {};
		\node [style=none] (4) at (2, 2.5) {};
		\node [style=none] (5) at (4, 2.5) {};
		\node [style=none] (6) at (4, 0.5) {};
		\node [style=none] (7) at (2, 0.5) {};
		\node [style=none] (8) at (-2, 1.5) {};
		\node [style=none] (9) at (2, 1.5) {};
		\node [style=none] (10) at (-3, 0.5) {};
		\node [style=none] (11) at (3, 0.5) {};
		\node [style=none] (12) at (-3, -1.5) {};
		\node [style=none] (13) at (3, -1.5) {};
		\node [style=none] (14) at (-5, 1.5) {};
		\node [style=none] (15) at (-5, -1.5) {};
		\node [style=none] (16) at (5, -1.5) {};
		\node [style=none] (17) at (5, 1.5) {};
		\node [style=none] (18) at (1, 1.5) {};
		\node [style=none] (19) at (-1, 1.5) {};
		\node [style=none] (20) at (-4, 1.5) {};
		\node [style=none] (21) at (4, 1.5) {};
		\node [style=new style 0] (22) at (-3, -1.5) {};
		\node [style=new style 1] (23) at (3, -1.5) {};
		\node [style=none] (24) at (4.5, -1) {$\scriptstyle C_0$};
		\node [style=none] (25) at (-4.5, -1) {$\scriptstyle C_0$};
		\node [style=none] (26) at (-4.75, 2) {$\scriptstyle T_1$};
		\node [style=none] (27) at (-1.25, 2) {$\scriptstyle T_2$};
		\node [style=none] (28) at (4.75, 2) {$\scriptstyle T_6$};
		\node [style=none] (29) at (1.25, 2) {$\scriptstyle T_5$};
		\node [style=none] (30) at (-3, 1.5) {$\scriptstyle U_B$};
		\node [style=none] (31) at (3, 1.5) {$\scriptstyle U_B$};
	\end{pgfonlayer}
	\begin{pgfonlayer}{edgelayer}
		\draw [style=box-wire] (0.center) to (1.center);
		\draw [style=box-wire] (1.center) to (2.center);
		\draw [style=box-wire] (2.center) to (3.center);
		\draw [style=box-wire] (3.center) to (0.center);
		\draw [style=box-wire] (4.center) to (5.center);
		\draw [style=box-wire] (5.center) to (6.center);
		\draw [style=box-wire] (6.center) to (7.center);
		\draw [style=box-wire] (7.center) to (4.center);
		\draw [style=box-wire] (14.center) to (20.center);
		\draw [style=box-wire] (8.center) to (19.center);
		\draw [style=box-wire] (18.center) to (9.center);
		\draw [style=box-wire] (21.center) to (17.center);
		\draw [style=box-wire] (15.center) to (16.center);
		\draw [style=syst-wire] (14.center) to (20.center);
		\draw [style=syst-wire] (8.center) to (19.center);
		\draw [style=syst-wire] (18.center) to (9.center);
		\draw [style=syst-wire] (21.center) to (17.center);
		\draw [style=syst-wire] (15.center) to (16.center);
		\draw [style=syst-wire] (11.center) to (23);
		\draw [style=syst-wire] (10.center) to (22);
	\end{pgfonlayer}
\end{tikzpicture}

%% file: spacetime_gate_A.tikz
\begin{tikzpicture}
	\begin{pgfonlayer}{nodelayer}
		\node [style=small-square] (0) at (0, 0) {$\scriptstyle U_A$};
		\node [style=none] (1) at (0.5, 0) {};
		\node [style=none] (2) at (1.5, 0) {};
		\node [style=none] (3) at (-0.5, 0) {};
		\node [style=none] (4) at (-1.5, 0) {};
		\node [style=none] (5) at (-1.2, 0.5) {$\scriptstyle T_3$};
		\node [style=none] (6) at (1.2, 0.5) {$\scriptstyle T_4$};
	\end{pgfonlayer}
	\begin{pgfonlayer}{edgelayer}
		\draw [style=syst-wire] (1.center) to (2.center);
		\draw [style=syst-wire] (4.center) to (3.center);
	\end{pgfonlayer}
\end{tikzpicture}

%% file: transformed_fragments_AB.tikz
\begin{tikzpicture}
	\begin{pgfonlayer}{nodelayer}
		\node [style=none] (0) at (-3.5, 0.5) {};
		\node [style=none] (1) at (-1.5, 0.5) {};
		\node [style=none] (2) at (-1.5, -1.5) {};
		\node [style=none] (3) at (-3.5, -1.5) {};
		\node [style=none] (4) at (1.5, 0.5) {};
		\node [style=none] (5) at (3.5, 0.5) {};
		\node [style=none] (6) at (3.5, -1.5) {};
		\node [style=none] (7) at (1.5, -1.5) {};
		\node [style=none] (8) at (-1.5, -0.5) {};
		\node [style=none] (9) at (1.5, -0.5) {};
		\node [style=none] (10) at (-2.5, -1.5) {};
		\node [style=none] (11) at (2.5, -1.5) {};
		\node [style=none] (12) at (-2.5, -3.5) {};
		\node [style=none] (13) at (2.5, -3.5) {};
		\node [style=none] (14) at (-4.5, -0.5) {};
		\node [style=none] (15) at (-6.5, -3.5) {};
		\node [style=none] (16) at (4.5, -3.5) {};
		\node [style=none] (17) at (4.5, -0.5) {};
		\node [style=none] (18) at (0.5, -0.5) {};
		\node [style=none] (19) at (-0.5, -0.5) {};
		\node [style=none] (20) at (-3.5, -0.5) {};
		\node [style=none] (21) at (3.5, -0.5) {};
		\node [style=new style 0] (22) at (-2.5, -3.5) {};
		\node [style=new style 1] (23) at (2.5, -3.5) {};
		\node [style=none] (24) at (4, -3) {$\scriptstyle C_0$};
		\node [style=none] (25) at (-6, -3) {$\scriptstyle C_0$};
		\node [style=none] (26) at (-4, 0) {$\scriptstyle T_1$};
		\node [style=none] (27) at (-1, 0) {$\scriptstyle T_2$};
		\node [style=none] (28) at (4, 0) {$\scriptstyle T_6$};
		\node [style=none] (29) at (1, 0) {$\scriptstyle T_5$};
		\node [style=none] (30) at (-2.5, -0.5) {$\scriptstyle U_B$};
		\node [style=none] (31) at (2.5, -0.5) {$\scriptstyle U_B$};
		\node [style=none] (32) at (4.5, 1) {};
		\node [style=none] (33) at (4.5, -4) {};
		\node [style=none] (34) at (5.5, -4) {};
		\node [style=none] (35) at (5.5, 4) {};
		\node [style=none] (36) at (0.5, 1) {};
		\node [style=none] (37) at (-0.5, 1) {};
		\node [style=none] (38) at (-0.5, -1) {};
		\node [style=none] (39) at (0.5, -1) {};
		\node [style=none] (40) at (-4.5, 1) {};
		\node [style=none] (41) at (-4.5, -1) {};
		\node [style=none] (42) at (-5.5, -1) {};
		\node [style=none] (43) at (-5.5, 4) {};
		\node [style=none] (44) at (5.5, -3.5) {};
		\node [style=none] (45) at (6.5, -3.5) {};
		\node [style=none] (46) at (5.5, 1.5) {};
		\node [style=none] (47) at (5.5, 3.5) {};
		\node [style=none] (48) at (6.5, 1.5) {};
		\node [style=none] (49) at (6.5, 3.5) {};
		\node [style=none] (50) at (-6.5, -3.5) {};
		\node [style=none] (53) at (-6.5, 3.5) {};
		\node [style=none] (54) at (-5.5, -1) {};
		\node [style=none] (55) at (-5.5, 3.5) {};
		\node [style=none] (56) at (-6, 2) {$\scriptstyle \widetilde T_3$};
		\node [style=none] (58) at (4.5, 4) {};
		\node [style=none] (59) at (-4.5, 4) {};
		\node [style=none] (60) at (-4.5, 2) {};
		\node [style=none] (61) at (4.5, 2) {};
		\node [style=none] (62) at (-1, 4.5) {};
		\node [style=none] (63) at (1, 4.5) {};
		\node [style=none] (64) at (1, 2.5) {};
		\node [style=none] (65) at (-1, 2.5) {};
		\node [style=none] (66) at (1, 3.5) {};
		\node [style=none] (67) at (0, 2.5) {};
		\node [style=none] (68) at (-1, 3.5) {};
		\node [style=none] (69) at (0, 3.5) {$\scriptstyle U_A$};
		\node [style=none] (70) at (-4.5, 3.5) {};
		\node [style=none] (71) at (4.5, 3.5) {};
		\node [style=none] (72) at (-6.5, 1.5) {};
		\node [style=none] (73) at (-5.5, 1.5) {};
		\node [style=none] (74) at (6, 2) {$\scriptstyle \widetilde T_4$};
		\node [style=none] (75) at (-6, 4) {$\scriptstyle \widetilde T_5$};
		\node [style=none] (76) at (6, 4) {$\scriptstyle \widetilde T_6$};
		\node [style=none] (77) at (6, -3) {$\scriptstyle C_0$};
		\node [style=none] (78) at (0, 1.5) {$\scriptstyle S_{A\to B}$};
		\node [style=none] (79) at (-6, 0) {$\scriptstyle \widetilde T_1$};
		\node [style=none] (80) at (-6.5, -0.5) {};
		\node [style=none] (81) at (-5.5, -0.5) {};
		\node [style=none] (82) at (5.5, -0.5) {};
		\node [style=none] (83) at (6.5, -0.5) {};
		\node [style=none] (84) at (6, 0) {$\scriptstyle \widetilde T_2$};
		\node [style=none] (85) at (-2.5, 4) {$\scriptstyle T_3$};
		\node [style=none] (86) at (2.5, 4) {$\scriptstyle T_4$};
	\end{pgfonlayer}
	\begin{pgfonlayer}{edgelayer}
		\draw [style=box-wire] (0.center) to (1.center);
		\draw [style=box-wire] (1.center) to (2.center);
		\draw [style=box-wire] (2.center) to (3.center);
		\draw [style=box-wire] (3.center) to (0.center);
		\draw [style=box-wire] (4.center) to (5.center);
		\draw [style=box-wire] (5.center) to (6.center);
		\draw [style=box-wire] (6.center) to (7.center);
		\draw [style=box-wire] (7.center) to (4.center);
		\draw [style=box-wire] (14.center) to (20.center);
		\draw [style=box-wire] (8.center) to (19.center);
		\draw [style=box-wire] (18.center) to (9.center);
		\draw [style=box-wire] (21.center) to (17.center);
		\draw [style=box-wire] (15.center) to (16.center);
		\draw [style=syst-wire] (14.center) to (20.center);
		\draw [style=syst-wire] (8.center) to (19.center);
		\draw [style=syst-wire] (18.center) to (9.center);
		\draw [style=syst-wire] (21.center) to (17.center);
		\draw [style=syst-wire] (15.center) to (16.center);
		\draw [style=syst-wire] (11.center) to (23);
		\draw [style=syst-wire] (10.center) to (22);
		\draw [style=box-wire] (33.center) to (32.center);
		\draw [style=box-wire] (37.center) to (38.center);
		\draw [style=box-wire] (38.center) to (39.center);
		\draw [style=box-wire] (39.center) to (36.center);
		\draw [style=box-wire] (36.center) to (32.center);
		\draw [style=box-wire] (33.center) to (34.center);
		\draw [style=box-wire] (34.center) to (35.center);
		\draw [style=box-wire] (37.center) to (40.center);
		\draw [style=box-wire] (40.center) to (41.center);
		\draw [style=box-wire] (41.center) to (42.center);
		\draw [style=box-wire] (42.center) to (43.center);
		\draw [style=syst-wire] (47.center) to (49.center);
		\draw [style=syst-wire] (46.center) to (48.center);
		\draw [style=syst-wire] (44.center) to (45.center);
		\draw [style=syst-wire] (53.center) to (55.center);
		\draw [style=box-wire] (60.center) to (61.center);
		\draw [style=box-wire] (61.center) to (58.center);
		\draw [style=box-wire] (35.center) to (58.center);
		\draw [style=box-wire] (43.center) to (59.center);
		\draw [style=box-wire] (59.center) to (60.center);
		\draw [style=box-wire] (62.center) to (63.center);
		\draw [style=box-wire] (63.center) to (64.center);
		\draw [style=box-wire] (64.center) to (65.center);
		\draw [style=box-wire] (65.center) to (62.center);
		\draw [style=syst-wire] (70.center) to (68.center);
		\draw [style=syst-wire] (66.center) to (71.center);
		\draw [style=syst-wire] (72.center) to (73.center);
		\draw [style=syst-wire] (80.center) to (81.center);
		\draw [style=syst-wire] (82.center) to (83.center);
	\end{pgfonlayer}
\end{tikzpicture}

%% file: transformed_fragments_AB_1.tikz
\begin{tikzpicture}
	\begin{pgfonlayer}{nodelayer}
		\node [style=none] (0) at (-3.5, 3) {};
		\node [style=none] (1) at (-1.5, 3) {};
		\node [style=none] (2) at (-1.5, 1) {};
		\node [style=none] (3) at (-3.5, 1) {};
		\node [style=none] (4) at (1.5, 3) {};
		\node [style=none] (5) at (3.5, 3) {};
		\node [style=none] (6) at (3.5, 1) {};
		\node [style=none] (7) at (1.5, 1) {};
		\node [style=none] (8) at (-1.5, 2) {};
		\node [style=none] (9) at (1.5, 2) {};
		\node [style=none] (10) at (-2.5, 1) {};
		\node [style=none] (11) at (2.5, 1) {};
		\node [style=none] (12) at (-2.5, 0) {};
		\node [style=none] (13) at (2.5, 0) {};
		\node [style=none] (14) at (-4.5, 2) {};
		\node [style=none] (15) at (-4.5, 0) {};
		\node [style=none] (16) at (4.5, 0) {};
		\node [style=none] (17) at (4.5, 2) {};
		\node [style=none] (18) at (0.5, 2) {};
		\node [style=none] (19) at (-0.5, 2) {};
		\node [style=none] (20) at (-3.5, 2) {};
		\node [style=none] (21) at (3.5, 2) {};
		\node [style=new style 0] (22) at (-2.5, 0) {};
		\node [style=new style 1] (23) at (2.5, 0) {};
		\node [style=none] (24) at (4, 0.5) {$\scriptstyle C_0$};
		\node [style=none] (25) at (-4, 0.5) {$\scriptstyle C_0$};
		\node [style=none] (26) at (-4.25, 2.5) {$\scriptstyle \widetilde T_1$};
		\node [style=none] (27) at (-0.75, 2.5) {$\scriptstyle \widetilde T_2$};
		\node [style=none] (28) at (4.25, 2.5) {$\scriptstyle \widetilde T_6$};
		\node [style=none] (29) at (0.75, 2.5) {$\scriptstyle \widetilde T_5$};
		\node [style=none] (30) at (-2.5, 2) {$\scriptstyle U_A$};
		\node [style=none] (31) at (2.5, 2) {$\scriptstyle U_A$};
		\node [style=none] (32) at (-1, -1) {};
		\node [style=none] (33) at (1, -1) {};
		\node [style=none] (34) at (1, -3) {};
		\node [style=none] (35) at (-1, -3) {};
		\node [style=none] (36) at (-1, -2) {};
		\node [style=none] (37) at (0, -3) {};
		\node [style=none] (38) at (2, -2) {};
		\node [style=none] (39) at (-2, -2) {};
		\node [style=none] (40) at (1, -2) {};
		\node [style=none] (41) at (1.75, -1.5) {$\scriptstyle \widetilde T_4$};
		\node [style=none] (42) at (-1.75, -1.5) {$\scriptstyle \widetilde T_3$};
		\node [style=none] (43) at (0, -2) {$\scriptstyle U_B$};
	\end{pgfonlayer}
	\begin{pgfonlayer}{edgelayer}
		\draw [style=box-wire] (0.center) to (1.center);
		\draw [style=box-wire] (1.center) to (2.center);
		\draw [style=box-wire] (2.center) to (3.center);
		\draw [style=box-wire] (3.center) to (0.center);
		\draw [style=box-wire] (4.center) to (5.center);
		\draw [style=box-wire] (5.center) to (6.center);
		\draw [style=box-wire] (6.center) to (7.center);
		\draw [style=box-wire] (7.center) to (4.center);
		\draw [style=box-wire] (14.center) to (20.center);
		\draw [style=box-wire] (8.center) to (19.center);
		\draw [style=box-wire] (18.center) to (9.center);
		\draw [style=box-wire] (21.center) to (17.center);
		\draw [style=box-wire] (15.center) to (16.center);
		\draw [style=syst-wire] (14.center) to (20.center);
		\draw [style=syst-wire] (8.center) to (19.center);
		\draw [style=syst-wire] (18.center) to (9.center);
		\draw [style=syst-wire] (21.center) to (17.center);
		\draw [style=syst-wire] (15.center) to (16.center);
		\draw [style=syst-wire] (11.center) to (23);
		\draw [style=syst-wire] (10.center) to (22);
		\draw [style=box-wire] (32.center) to (33.center);
		\draw [style=box-wire] (33.center) to (34.center);
		\draw [style=box-wire] (34.center) to (35.center);
		\draw [style=box-wire] (35.center) to (32.center);
		\draw [style=box-wire] (39.center) to (36.center);
		\draw [style=box-wire] (40.center) to (38.center);
		\draw [style=syst-wire] (39.center) to (36.center);
		\draw [style=syst-wire] (40.center) to (38.center);
	\end{pgfonlayer}
\end{tikzpicture}

%% file: transformed_background.tikz
\begin{tikzpicture}
	\begin{pgfonlayer}{nodelayer}
		\node [style=none] (0) at (-6.5, -2) {};
		\node [style=none] (1) at (-3.5, -2) {};
		\node [style=none] (2) at (-6.5, -2.5) {};
		\node [style=none] (3) at (-8.5, -2.5) {};
		\node [style=none] (4) at (-8.5, 0.5) {};
		\node [style=none] (5) at (-6.5, 0.5) {};
		\node [style=none] (6) at (-6.5, 0) {};
		\node [style=none] (7) at (-5.5, 0) {};
		\node [style=none] (8) at (-6.5, -1) {};
		\node [style=none] (9) at (-3.5, -1) {};
		\node [style=none] (10) at (-1.5, -2) {};
		\node [style=none] (11) at (-1.5, -2.5) {};
		\node [style=none] (12) at (-3.5, -2.5) {};
		\node [style=none] (13) at (-3.5, 0.5) {};
		\node [style=none] (14) at (-1.5, 0.5) {};
		\node [style=none] (15) at (-1.5, 0) {};
		\node [style=none] (16) at (-1.5, -1) {};
		\node [style=none] (17) at (-4.5, 0) {};
		\node [style=none] (18) at (-3.5, 0) {};
		\node [style=none] (19) at (-1.5, -2) {};
		\node [style=none] (20) at (1.5, -2) {};
		\node [style=none] (21) at (-1.5, -1) {};
		\node [style=none] (22) at (1.5, -1) {};
		\node [style=none] (23) at (3.5, -2) {};
		\node [style=none] (24) at (6.5, -2) {};
		\node [style=none] (25) at (3.5, -2.5) {};
		\node [style=none] (26) at (1.5, -2.5) {};
		\node [style=none] (27) at (1.5, 0.5) {};
		\node [style=none] (28) at (3.5, 0.5) {};
		\node [style=none] (29) at (3.5, 0) {};
		\node [style=none] (30) at (4.5, 0) {};
		\node [style=none] (31) at (3.5, -1) {};
		\node [style=none] (32) at (6.5, -1) {};
		\node [style=none] (33) at (8.5, -2) {};
		\node [style=none] (34) at (8.5, -2.5) {};
		\node [style=none] (35) at (6.5, -2.5) {};
		\node [style=none] (36) at (6.5, 0.5) {};
		\node [style=none] (37) at (8.5, 0.5) {};
		\node [style=none] (38) at (8.5, 0) {};
		\node [style=none] (39) at (8.5, -1) {};
		\node [style=none] (40) at (5.5, 0) {};
		\node [style=none] (41) at (6.5, 0) {};
		\node [style=none] (42) at (8.5, -2) {};
		\node [style=none] (43) at (8.5, -1) {};
		\node [style=none] (44) at (-1.5, 0) {};
		\node [style=none] (45) at (-0.5, 0) {};
		\node [style=none] (46) at (0.5, 0) {};
		\node [style=none] (47) at (1.5, 0) {};
		\node [style=none] (48) at (8.5, 0) {};
		\node [style=none] (49) at (9.5, 0) {};
		\node [style=none] (50) at (8.5, -1) {};
		\node [style=none] (51) at (9.5, -1) {};
		\node [style=none] (52) at (8.5, -2) {};
		\node [style=none] (53) at (9.5, -2) {};
		\node [style=none] (54) at (-11.5, 0) {};
		\node [style=none] (55) at (-11.5, -1) {};
		\node [style=none] (56) at (-11.5, -1) {};
		\node [style=none] (57) at (-11.5, 0) {};
		\node [style=none] (58) at (-8.5, 0) {};
		\node [style=none] (59) at (-11.5, -1) {};
		\node [style=none] (60) at (-8.5, -1) {};
		\node [style=none] (61) at (-8.5, -2) {};
		\node [style=none] (62) at (-11, 0.5) {$\scriptstyle R_0$};
		\node [style=none] (63) at (-6, 0.5) {$\scriptstyle T_1$};
		\node [style=none] (64) at (-4, 0.5) {$\scriptstyle T_2$};
		\node [style=none] (65) at (-1, 0.5) {$\scriptstyle T_3$};
		\node [style=none] (66) at (1, 0.5) {$\scriptstyle T_4$};
		\node [style=none] (67) at (4, 0.5) {$\scriptstyle T_5$};
		\node [style=none] (68) at (6, 0.5) {$\scriptstyle T_6$};
		\node [style=none] (69) at (9, 0.5) {$\scriptstyle R_7$};
		\node [style=none] (70) at (-11, -0.5) {$\scriptstyle S_0$};
		\node [style=none] (71) at (9, -0.5) {$\scriptstyle S_7$};
		\node [style=none] (73) at (-7.5, -1) {$\set T^{0\to 1}$};
		\node [style=none] (74) at (-2.5, -1) {$\set T^{2\to 3}$};
		\node [style=none] (75) at (2.5, -1) {$\set T^{4\to 5}$};
		\node [style=none] (76) at (7.5, -1) {$\set T^{6\to 7}$};
		\node [style=none] (77) at (9, -1.5) {$\scriptstyle C_7$};
		\node [style=none] (78) at (-6, -0.5) {$\scriptstyle S_1$};
		\node [style=none] (79) at (-1, -0.5) {$\scriptstyle S_3$};
		\node [style=none] (80) at (-4, -0.5) {$\scriptstyle S_2$};
		\node [style=none] (81) at (1, -0.5) {$\scriptstyle S_4$};
		\node [style=none] (82) at (4, -0.5) {$\scriptstyle S_5$};
		\node [style=none] (83) at (6, -0.5) {$\scriptstyle S_6$};
		\node [style=none] (84) at (-6, -1.5) {$\scriptstyle C_1$};
		\node [style=none] (85) at (-4, -1.5) {$\scriptstyle C_2$};
		\node [style=none] (86) at (-1, -1.5) {$\scriptstyle C_3$};
		\node [style=none] (87) at (1, -1.5) {$\scriptstyle C_4$};
		\node [style=none] (88) at (4, -1.5) {$\scriptstyle C_5$};
		\node [style=none] (89) at (6, -1.5) {$\scriptstyle C_6$};
		\node [style=none] (90) at (-9.5, -2) {};
		\node [style=none] (91) at (-9.5, -2) {};
		\node [style=none] (92) at (-9.5, -2) {};
		\node [style=none] (93) at (-8.5, -2) {};
		\node [style=none] (94) at (-9.5, 1) {};
		\node [style=none] (95) at (-10.5, -2.5) {};
		\node [style=none] (96) at (-9.5, -2.5) {};
		\node [style=none] (97) at (-10.5, 3.25) {};
		\node [style=none] (98) at (-5.5, 1) {};
		\node [style=none] (99) at (-4.5, 1) {};
		\node [style=none] (100) at (-5.5, -0.5) {};
		\node [style=none] (101) at (-4.5, -0.5) {};
		\node [style=none] (102) at (-0.5, 1) {};
		\node [style=none] (103) at (0.5, 1) {};
		\node [style=none] (104) at (-0.5, -0.5) {};
		\node [style=none] (105) at (0.5, -0.5) {};
		\node [style=none] (106) at (4.5, -0.5) {};
		\node [style=none] (107) at (5.5, -0.5) {};
		\node [style=none] (108) at (4.5, 1) {};
		\node [style=none] (109) at (5.5, 1) {};
		\node [style=none] (110) at (0, 1.5) {};
		\node [style=none] (111) at (-2.5, 1.5) {$\scriptstyle S_{A\to B}^{\dagger}$};
		\node [style=none] (112) at (-11.5, -2) {};
		\node [style=none] (113) at (-11.5, -2) {};
		\node [style=none] (114) at (-10.5, -2) {};
		\node [style=none] (115) at (-11, -1.5) {$\scriptstyle C_0$};
		\node [style=none] (116) at (-9.5, 3.25) {};
		\node [style=none] (117) at (-9.5, 2) {};
		\node [style=none] (118) at (-9.5, 3.25) {};
		\node [style=none] (119) at (-5.5, 2) {};
		\node [style=none] (120) at (-4.5, 2) {};
		\node [style=none] (121) at (-0.5, 2) {};
		\node [style=none] (122) at (0.5, 2) {};
		\node [style=none] (123) at (4.5, 2) {};
		\node [style=none] (124) at (5.5, 1) {};
		\node [style=none] (125) at (-5.5, 3.25) {};
		\node [style=none] (126) at (-4.5, 3.25) {};
		\node [style=none] (127) at (-0.5, 3.25) {};
		\node [style=none] (128) at (0.5, 3.25) {};
		\node [style=none] (129) at (4.5, 3.25) {};
		\node [style=none] (130) at (5.5, 3.25) {};
		\node [style=none] (131) at (-9.5, 2.75) {};
		\node [style=none] (132) at (-9.5, 2.75) {};
		\node [style=none] (133) at (-8.5, 2.75) {};
		\node [style=none] (134) at (-6.5, 2.75) {};
		\node [style=none] (135) at (-6.5, 2.75) {};
		\node [style=none] (136) at (-5.5, 2.75) {};
		\node [style=none] (137) at (-4.5, 2.75) {};
		\node [style=none] (138) at (-4.5, 2.75) {};
		\node [style=none] (139) at (-3.5, 2.75) {};
		\node [style=none] (140) at (-1.5, 2.75) {};
		\node [style=none] (141) at (-1.5, 2.75) {};
		\node [style=none] (142) at (-0.5, 2.75) {};
		\node [style=none] (143) at (0.5, 2.75) {};
		\node [style=none] (144) at (0.5, 2.75) {};
		\node [style=none] (145) at (1.5, 2.75) {};
		\node [style=none] (146) at (3.5, 2.75) {};
		\node [style=none] (147) at (3.5, 2.75) {};
		\node [style=none] (148) at (4.5, 2.75) {};
		\node [style=none] (149) at (-9, 3.25) {$\scriptstyle \widetilde T_1$};
		\node [style=none] (150) at (-6, 3.25) {$\scriptstyle \widetilde T_2$};
		\node [style=none] (151) at (-4, 3.25) {$\scriptstyle \widetilde T_3$};
		\node [style=none] (152) at (-1, 3.25) {$\scriptstyle \widetilde T_4$};
		\node [style=none] (153) at (1, 3.25) {$\scriptstyle \widetilde T_5$};
		\node [style=none] (154) at (4, 3.25) {$\scriptstyle \widetilde T_6$};
	\end{pgfonlayer}
	\begin{pgfonlayer}{edgelayer}
		\draw [style=syst-wire] (1.center) to (0.center);
		\draw [style=box-wire] (5.center) to (4.center);
		\draw [style=box-wire] (4.center) to (3.center);
		\draw [style=box-wire] (3.center) to (2.center);
		\draw [style=box-wire] (2.center) to (5.center);
		\draw [style=syst-wire] (7.center) to (6.center);
		\draw [style=syst-wire] (9.center) to (8.center);
		\draw [style=box-wire] (14.center) to (13.center);
		\draw [style=box-wire] (13.center) to (12.center);
		\draw [style=box-wire] (12.center) to (11.center);
		\draw [style=box-wire] (11.center) to (14.center);
		\draw [style=syst-wire] (18.center) to (17.center);
		\draw [style=syst-wire] (20.center) to (19.center);
		\draw [style=syst-wire] (22.center) to (21.center);
		\draw [style=syst-wire] (24.center) to (23.center);
		\draw [style=box-wire] (28.center) to (27.center);
		\draw [style=box-wire] (27.center) to (26.center);
		\draw [style=box-wire] (26.center) to (25.center);
		\draw [style=box-wire] (25.center) to (28.center);
		\draw [style=syst-wire] (30.center) to (29.center);
		\draw [style=syst-wire] (32.center) to (31.center);
		\draw [style=box-wire] (37.center) to (36.center);
		\draw [style=box-wire] (36.center) to (35.center);
		\draw [style=box-wire] (35.center) to (34.center);
		\draw [style=box-wire] (34.center) to (37.center);
		\draw [style=syst-wire] (41.center) to (40.center);
		\draw [style=syst-wire] (45.center) to (44.center);
		\draw [style=syst-wire] (47.center) to (46.center);
		\draw [style=syst-wire] (49.center) to (48.center);
		\draw [style=syst-wire] (51.center) to (50.center);
		\draw [style=syst-wire] (53.center) to (52.center);
		\draw [style=syst-wire] (58.center) to (57.center);
		\draw [style=syst-wire] (60.center) to (59.center);
		\draw [style=syst-wire] (93.center) to (92.center);
		\draw [style=box-wire] (95.center) to (97.center);
		\draw [style=box-wire] (96.center) to (95.center);
		\draw [style=box-wire] (96.center) to (94.center);
		\draw [style=box-wire] (100.center) to (101.center);
		\draw [style=box-wire] (101.center) to (99.center);
		\draw [style=box-wire] (100.center) to (98.center);
		\draw [style=box-wire] (104.center) to (102.center);
		\draw [style=box-wire] (104.center) to (105.center);
		\draw [style=box-wire] (105.center) to (103.center);
		\draw [style=box-wire] (106.center) to (107.center);
		\draw [style=box-wire] (107.center) to (109.center);
		\draw [style=box-wire] (106.center) to (108.center);
		\draw [style=box-wire] (94.center) to (98.center);
		\draw [style=box-wire] (99.center) to (102.center);
		\draw [style=box-wire] (103.center) to (108.center);
		\draw [style=syst-wire] (114.center) to (113.center);
		\draw [style=box-wire] (118.center) to (97.center);
		\draw [style=box-wire] (118.center) to (117.center);
		\draw [style=box-wire] (117.center) to (119.center);
		\draw [style=box-wire] (119.center) to (125.center);
		\draw [style=box-wire] (125.center) to (126.center);
		\draw [style=box-wire] (126.center) to (120.center);
		\draw [style=box-wire] (120.center) to (121.center);
		\draw [style=box-wire] (121.center) to (127.center);
		\draw [style=box-wire] (127.center) to (128.center);
		\draw [style=box-wire] (128.center) to (122.center);
		\draw [style=box-wire] (122.center) to (123.center);
		\draw [style=box-wire] (123.center) to (129.center);
		\draw [style=box-wire] (129.center) to (130.center);
		\draw [style=box-wire] (130.center) to (124.center);
		\draw [style=syst-wire] (133.center) to (132.center);
		\draw [style=syst-wire] (136.center) to (135.center);
		\draw [style=syst-wire] (139.center) to (138.center);
		\draw [style=syst-wire] (142.center) to (141.center);
		\draw [style=syst-wire] (145.center) to (144.center);
		\draw [style=syst-wire] (148.center) to (147.center);
	\end{pgfonlayer}
\end{tikzpicture}

%% file: spacetime_process_B.tikz
\begin{tikzpicture}
	\begin{pgfonlayer}{nodelayer}
		\node [style=none] (0) at (-6.5, -1) {};
		\node [style=none] (1) at (-3.5, -1) {};
		\node [style=none] (2) at (-6.5, -1.5) {};
		\node [style=none] (3) at (-8.5, -1.5) {};
		\node [style=none] (4) at (-8.5, 1.5) {};
		\node [style=none] (5) at (-6.5, 1.5) {};
		\node [style=none] (6) at (-6.5, 1) {};
		\node [style=none] (7) at (-5.5, 1) {};
		\node [style=none] (8) at (-6.5, 0) {};
		\node [style=none] (9) at (-3.5, 0) {};
		\node [style=none] (10) at (-1.5, -1) {};
		\node [style=none] (11) at (-1.5, -1.5) {};
		\node [style=none] (12) at (-3.5, -1.5) {};
		\node [style=none] (13) at (-3.5, 1.5) {};
		\node [style=none] (14) at (-1.5, 1.5) {};
		\node [style=none] (15) at (-1.5, 1) {};
		\node [style=none] (16) at (-1.5, 0) {};
		\node [style=none] (17) at (-4.5, 1) {};
		\node [style=none] (18) at (-3.5, 1) {};
		\node [style=none] (19) at (-1.5, -1) {};
		\node [style=none] (20) at (1.5, -1) {};
		\node [style=none] (21) at (-1.5, 0) {};
		\node [style=none] (22) at (1.5, 0) {};
		\node [style=none] (23) at (3.5, -1) {};
		\node [style=none] (24) at (6.5, -1) {};
		\node [style=none] (25) at (3.5, -1.5) {};
		\node [style=none] (26) at (1.5, -1.5) {};
		\node [style=none] (27) at (1.5, 1.5) {};
		\node [style=none] (28) at (3.5, 1.5) {};
		\node [style=none] (29) at (3.5, 1) {};
		\node [style=none] (30) at (4.5, 1) {};
		\node [style=none] (31) at (3.5, 0) {};
		\node [style=none] (32) at (6.5, 0) {};
		\node [style=none] (33) at (8.5, -1) {};
		\node [style=none] (34) at (8.5, -1.5) {};
		\node [style=none] (35) at (6.5, -1.5) {};
		\node [style=none] (36) at (6.5, 1.5) {};
		\node [style=none] (37) at (8.5, 1.5) {};
		\node [style=none] (38) at (8.5, 1) {};
		\node [style=none] (39) at (8.5, 0) {};
		\node [style=none] (40) at (5.5, 1) {};
		\node [style=none] (41) at (6.5, 1) {};
		\node [style=none] (42) at (8.5, -1) {};
		\node [style=none] (43) at (8.5, 0) {};
		\node [style=none] (44) at (-1.5, 1) {};
		\node [style=none] (45) at (-0.5, 1) {};
		\node [style=none] (46) at (0.5, 1) {};
		\node [style=none] (47) at (1.5, 1) {};
		\node [style=none] (48) at (8.5, 1) {};
		\node [style=none] (49) at (9.5, 1) {};
		\node [style=none] (50) at (8.5, 0) {};
		\node [style=none] (51) at (9.5, 0) {};
		\node [style=none] (52) at (8.5, -1) {};
		\node [style=none] (53) at (9.5, -1) {};
		\node [style=none] (54) at (-9.5, 1) {};
		\node [style=none] (55) at (-9.5, 0) {};
		\node [style=none] (56) at (-9.5, 0) {};
		\node [style=none] (57) at (-9.5, 1) {};
		\node [style=none] (58) at (-8.5, 1) {};
		\node [style=none] (59) at (-9.5, 0) {};
		\node [style=none] (60) at (-8.5, 0) {};
		\node [style=none] (61) at (-8.5, -1) {};
		\node [style=none] (62) at (-9, 1.5) {$\scriptstyle R_0$};
		\node [style=none] (63) at (4, 1.5) {$\scriptstyle \widetilde T_5$};
		\node [style=none] (64) at (6, 1.5) {$\scriptstyle \widetilde T_6$};
		\node [style=none] (65) at (-1, 1.5) {$\scriptstyle \widetilde T_3$};
		\node [style=none] (66) at (1, 1.5) {$\scriptstyle \widetilde T_4$};
		\node [style=none] (67) at (-6, 1.5) {$\scriptstyle \widetilde T_1$};
		\node [style=none] (68) at (-4, 1.5) {$\scriptstyle \widetilde T_2$};
		\node [style=none] (69) at (9, 1.5) {$\scriptstyle R_7$};
		\node [style=none] (70) at (-9, 0.5) {$\scriptstyle S_0$};
		\node [style=none] (71) at (9.25, 0.5) {$\scriptstyle S_7$};
		\node [style=none] (72) at (-9, -0.5) {$\scriptstyle C_0$};
		\node [style=none] (73) at (-7.5, 0) {$\widetilde{\set T}^{0\to 1}$};
		\node [style=none] (74) at (-2.5, 0) {$\widetilde{\set T}^{2\to 3}$};
		\node [style=none] (75) at (2.5, 0) {$\widetilde{\set T}^{4\to 5}$};
		\node [style=none] (76) at (7.5, 0) {$\widetilde{\set T}^{6\to 7}$};
		\node [style=none] (77) at (9.25, -0.5) {$\scriptstyle C_7$};
		\node [style=none] (78) at (-6, 0.5) {$\scriptstyle S_1$};
		\node [style=none] (79) at (-1, 0.5) {$\scriptstyle S_3$};
		\node [style=none] (80) at (-4, 0.5) {$\scriptstyle S_2$};
		\node [style=none] (81) at (1, 0.5) {$\scriptstyle S_4$};
		\node [style=none] (82) at (4, 0.5) {$\scriptstyle S_5$};
		\node [style=none] (83) at (6, 0.5) {$\scriptstyle S_6$};
		\node [style=none] (84) at (-6, -0.5) {$\scriptstyle C_1$};
		\node [style=none] (85) at (-4, -0.5) {$\scriptstyle C_2$};
		\node [style=none] (86) at (-1, -0.5) {$\scriptstyle C_3$};
		\node [style=none] (87) at (1, -0.5) {$\scriptstyle C_4$};
		\node [style=none] (88) at (4, -0.5) {$\scriptstyle C_5$};
		\node [style=none] (89) at (6, -0.5) {$\scriptstyle C_6$};
		\node [style=none] (90) at (-9.5, -1) {};
		\node [style=none] (91) at (-9.5, -1) {};
		\node [style=none] (92) at (-9.5, -1) {};
		\node [style=none] (93) at (-8.5, -1) {};
	\end{pgfonlayer}
	\begin{pgfonlayer}{edgelayer}
		\draw [style=syst-wire] (1.center) to (0.center);
		\draw [style=box-wire] (5.center) to (4.center);
		\draw [style=box-wire] (4.center) to (3.center);
		\draw [style=box-wire] (3.center) to (2.center);
		\draw [style=box-wire] (2.center) to (5.center);
		\draw [style=syst-wire] (7.center) to (6.center);
		\draw [style=syst-wire] (9.center) to (8.center);
		\draw [style=box-wire] (14.center) to (13.center);
		\draw [style=box-wire] (13.center) to (12.center);
		\draw [style=box-wire] (12.center) to (11.center);
		\draw [style=box-wire] (11.center) to (14.center);
		\draw [style=syst-wire] (18.center) to (17.center);
		\draw [style=syst-wire] (20.center) to (19.center);
		\draw [style=syst-wire] (22.center) to (21.center);
		\draw [style=syst-wire] (24.center) to (23.center);
		\draw [style=box-wire] (28.center) to (27.center);
		\draw [style=box-wire] (27.center) to (26.center);
		\draw [style=box-wire] (26.center) to (25.center);
		\draw [style=box-wire] (25.center) to (28.center);
		\draw [style=syst-wire] (30.center) to (29.center);
		\draw [style=syst-wire] (32.center) to (31.center);
		\draw [style=box-wire] (37.center) to (36.center);
		\draw [style=box-wire] (36.center) to (35.center);
		\draw [style=box-wire] (35.center) to (34.center);
		\draw [style=box-wire] (34.center) to (37.center);
		\draw [style=syst-wire] (41.center) to (40.center);
		\draw [style=syst-wire] (45.center) to (44.center);
		\draw [style=syst-wire] (47.center) to (46.center);
		\draw [style=syst-wire] (49.center) to (48.center);
		\draw [style=syst-wire] (51.center) to (50.center);
		\draw [style=syst-wire] (53.center) to (52.center);
		\draw [style=syst-wire] (58.center) to (57.center);
		\draw [style=syst-wire] (60.center) to (59.center);
		\draw [style=syst-wire] (93.center) to (92.center);
	\end{pgfonlayer}
\end{tikzpicture}